\documentclass{aa}

\usepackage{graphicx}
\usepackage{txfonts}
\usepackage{multirow}
\usepackage{natbib}

\usepackage{caption}

\usepackage{hyperref}
\hypersetup{linkcolor=blue,citecolor=blue,colorlinks=true}

\newcommand*{\sedsize}{0.47\textwidth}

\usepackage{makecell}

\newcommand{\eq}{Eq.~}

\newcommand{\Fig}{Fig.~}
\newcommand{\Figs}{Figs.~}

\newcommand{\Sec}{Section~}

\newcommand{\App}{Appendix~}

\newcommand{\Tab}{Tab.~}

\begin{document}

    \title{The Spectra of IceCube Neutrino (SIN) candidate sources - V. Modeling and interpretation of multiwavelength and neutrino data}
    \titlerunning{SIN candidate sources - V}
   \author{X. Rodrigues 
          \inst{1,2}
          \and
          M. Karl
            \inst{3,1}
          \and
            P. Padovani
            \inst{1}
          \and
            P. Giommi
            \inst{4,5,6}
          \and
            S. Paiano
            \inst{7,8}
          \and
            R. Falomo
            \inst{9}
          \and
            M. Petropoulou
            \inst{10}
          \and
            F. Oikonomou
            \inst{11}
            }

    \institute{European Southern Observatory, Karl-Schwarzschild-Straße 2, 85748 Garching bei M\"unchen, Germany
                \and
                Excellence Cluster ORIGINS, Boltzmannstr. 2, D-85748 Garching bei M\"unchen, Germany
                \and
                Technische Universit\"at M{\"u}nchen, TUM School of Natural Sciences, Physics Department, 
James-Frank-Str. 1, D-85748 Garching bei M{\"u}nchen, Germany
            \and
            Institute for Advanced Study, Technische Universität M\"unchen,
Lichtenbergstrasse 2a, D-85748 Garching bei M\"unchen, Germany
            \and
            Center for Astrophysics and Space Science (CASS), New York University Abu Dhabi, PO Box 129188 Abu Dhabi, United Arab Emirates
            \and    
            Associated with INAF, Osservatorio Astronomico di Brera, via Brera, 28, I-20121 Milano, Italy
            \and
            INAF - IASF Palermo, via Ugo La Malfa, 153, I-90146, Palermo, Italy
            \and
            INAF - IASF Milano, via Corti 12, I-20133, Milano, Italy
            \and
            INAF - Osservatorio Astronomico di Padova, vicolo dell’Osservatorio 5, I-35122, Padova, Italy
            \and
            Department of Physics, National and Kapodistrian University of Athens, University Campus Zografos, GR 15783, Athens, Greece
            \and
            Institutt for fysikk, Norwegian University of Science and Technology (NTNU), 7491 Trondheim, Norway
    }

   \date{Submitted 2 May 2024 | Accepted 3 June 2024}

 
\abstract
{A correlation has been reported between the arrival directions of high-energy IceCube events and $\gamma$-ray blazars classified as intermediate- and high-synchrotron-peaked BL Lacs. Subsequent studies have investigated the optical properties of these sources, compiled and analyzed public multiwavelength data, and constrained their individual neutrino emission based on public IceCube point-source data.}  
 {We provide a theoretical interpretation of public multiwavelength and neutrino point source data for the 32 BL~Lac objects in the sample previously associated with an IceCube alert event. We combined the individual source results to draw conclusions regarding the multimesssenger properties of the sample and the required power in relativistic protons.}
 {We performed particle interaction modeling using open-source numerical simulation software. We constrained the model parameters using a novel and unique approach that simultaneously describes the host galaxy contribution, the observed synchrotron peak properties, the average multiwavelength fluxes, and, where possible, the IceCube point source constraints.}
 {We show that a single-zone leptohadronic model can describe the multiwavelength broadband fluxes from all 32 IceCube candidates. In some cases, the model suggests that hadronic emission may contribute a considerable fraction of the $\gamma$-ray flux. The required power in relativistic protons ranges from a few percent to a factor of ten of the Eddington luminosity, which is energetically less demanding compared to other leptohadronic blazar models in recent literature. The model can describe the 68\% confidence level IceCube flux for a large fraction of the masquerading BL~Lacs in the sample, including TXS~0506+056; whereas, for true BL Lacs, the model predicts a low neutrino flux in the IceCube sensitivity range. Physically, this distinction is due to the presence of photons from broad line emission in masquerading BL~Lacs, which increase the efficiency of hadronic interactions. The predicted neutrino flux peaks between a few petaelectronvolt and 100~PeV and scales positively with the flux in the gigaelectronvolt, megaelectronvolt, X-ray, and optical bands. Based on these results, we provide a list of the brightest neutrino emitters, which can be used for future searches targeting the 10-100~PeV regime.}
 {}


   \maketitle


\section{Introduction}
\label{sec:intro}

The IceCube Neutrino Observatory, located in the geographic South Pole, has detected a cosmic flux of neutrinos with energies up to a few petaelectronvolt \citep[PeV,][]{IceCube_2013,IceCube:2015qii,IceCube:2015gsk,IceCube:2013cdw}. Over a decade after its discovery, the origin of the bulk of this neutrino flux remains unclear   \citep[see e.g.,][]{Stecker:1991vm,Halzen:2002pg,Petropoulou:2015upa,Hooper:2016jls,Murase:2019vdl}.

Multiple astrophysical sources have been associated with IceCube events at different confidence levels~\citep[e.g.,][]{2014MNRAS.443..474P,2018Sci...361..147I,icfermi,vanVelzen:2020cwu,Franckowiak:2020qrq,Sahakyan:2022nbz,doi:10.1126/science.abg3395f}.
One such event was a muon neutrino with an estimated energy of 290~TeV, detected in 2017 from the direction of source TXS~0506+056 during a half-year-long $\gamma$-ray flare. This spatial and temporal coincidence led to an association between the event and the source at the $3.5~\sigma$ confidence level~\citep{icfermi}. TXS~0506+056 is a blazar, or an active galactic nucleus (AGN) displaying a relativistic jet that points close to the line of sight. Observationally, this blazar is classified as a BL Lac due to the apparent lack of broad lines in its optical spectrum. The observed synchrotron emission, originating in electrons accelerated in the relativistic jet, peaks at $\sim10^{15}~\mathrm{Hz}$, placing the source between the category of intermediate- and high-synchrotron-peaked BL Lac (commonly referred to as IBL and HBL respectively, collectively denoted here as IHBL). Upon further analysis of the source's spectrum, \citet{Padovani2019} have shown that it is in fact a masquerading BL Lac, where broad line emission is intrinsically present but is outshone by the high-frequency synchrotron continuum.
This indicates that the supermassive black hole of TXS~0506+056 is surrounded by a broad line region (BLR) of dense and rapidly rotating gas, as in the case of flat-spectrum radio quasars (FSRQs). 

Under certain conditions, broad line and thermal emission from a BLR can serve as an interaction target for protons or other nuclei that may be accelerated in the jet together with electrons. This effect can boost the emission of high-energy neutrinos and other secondary particles produced in hadronic interactions, as supported by theoretical studies of TXS 0506+056~\citep[e.g.][]{Reimer:2018vvw,Rodrigues:2018tku,Petropoulou:2019zqp}, other masquerading BL Lacs~\citep[e.g.][]{Petropoulou2020,Sahakyan:2022nbz,deClairfontaine:2023pgo}, and FSRQs~\citep[e.g.][]{Murase:2014foa,Rodrigues:2020fbu,Oikonomou:2021akf,Rodrigues:2023vbv}. At the same time, those studies generally indicate that the interactions of
protons of about petaelectronvolt energies lead to abundant cascade emission in the X-ray range, which means that the maximum flux of $100~\mathrm{TeV}-\mathrm{PeV}$ neutrinos that can be expected theoretically is limited by the observed X-ray flux.


Upon investigating the sample of public IceCube high-energy neutrino track events, \citet[][henceforth G20]{Giommi:2020hbx} have reported a spatial correlation with the population of $\gamma$-ray-detected IHBLs with a significance of 3.2 $\sigma$ (post-trial). For low-synchrotron-peaked BL Lacs (LBLs), no significant correlation was found. G20 then identified 47 IHBLs in the sample coincident with IceCube events. 
Based on the properties of the IceCube sample, this means an expected $16\pm4$ potential neutrino sources.

This prompted a sequence of follow-up analyses within an umbrella project titled Spectra of IceCube Neutrino (SIN) candidate sources. \citet[][henceforth referred to as Paper I]{Paiano:2021zpc} performed optical spectroscopy of 17 of the G20 blazars and used those data to estimate or set lower limits on the source redshift. \citet[][Paper II]{Padovani:2021kjr} utilized the spectroscopic data to characterize the blazars in the sample as either true or masquerading BL Lacs, in the cases where that characterization was possible. \citet[][Paper III]{Paiano:2023nsw} extended the spectroscopy campaign and the corresponding analysis to the entire sample. Most recently, \citet{Karl:2023huw}, referred to henceforth as Paper IV \citep[see also][]{Karl_2024}, investigated the blazar light curves from infrared to $\gamma$ rays, compiled public multiwavelength data, and derived constraints on the neutrino flux from each source. The neutrino flux estimate was computed based on public IceCube data and assuming a power-law neutrino spectrum. Paper IV also provided an update on the original BL Lac sample by applying the revised criteria for astrophysical neutrino identification recently adopted in the recent IceCat-I catalog \citep{blaufuss2019generation,abbasi2023icecat1}. 

For this work, we leveraged the wealth of data on the G20 sample, which includes TXS~0506+056 and other known masquerading BL Lacs, to test a common theoretical multimesssenger framework of IHBLs. This framework is based on time-dependent numerical simulations of relativistic protons and electrons radiating in a single dissipation region in the relativistic jet. We utilized a novel fitting method that describes not only the available multiwavelength data from each source, but also the observed synchrotron peak frequency and flux, the host galaxy emission (obtained through optical decomposition when possible), the derived accretion disk and BLR luminosity, and the 68\% confidence level IceCube point-source fluxes. The latter fluxes were derived by comparing public IceCube data from each source with Monte Carlo simulations using a spectral shape typical of p$\gamma$ interactions as our signal assumption. We show that this can result in different neutrino flux  estimates compared to the commonly used assumption of a power-law signal. The predicted neutrino spectrum peaks above the petaelectronvolt range for all sources, and the flux scales approximately linearly with the $\gamma$-ray flux. The model can describe the IceCube flux limits for most masquerading BL Lacs, including TXS~0506+056, while this is challenging for true BL Lacs due to their low neutrino efficiency in the IceCube sensitivity range. The results suggest a dissipation region in the jets of masquerading BL Lacs lying at a distance to the supermassive black hole between one and three times the radius of the BLR.

The paper is organized as follows: in \Sec\ref{sec:methods} we present the methods underlying the preparation of the multiwavelength data, the modeling framework, the novel method for the IceCube point source flux estimation, and the optimization method utilized to constrain the model parameters. In \Sec\ref{sec:results} we present the results of the modeling of the sample, highlighting the differences between the predictions for masquerading and non-masquerading BL Lacs. In \Sec\ref{sec:discussion} we contextualize the results in the landscape of previous leptohadronic models, discuss statistical trends that can be derived at the sample level, and address potential pathways for expanding this method in the future. We present our conclusions in \Sec\ref{sec:conclusion}.

\begin{table*}[htpb!]
\setlength{\tabcolsep}{1.7pt}
\caption{List of blazars in the sample and information on their associated IceCube alert.}
\begin{center}
\begin{tabular}{lllllr}
\hline\hline
{\it Fermi}-LAT catalog designation & Associated source & Source classification\hspace{2mm} & IceCube real-time alert \hspace{2mm} & Alert type\hspace{2mm} & $E / \mathrm{TeV}$ \\
\hline
4FGL J0158.8+0101    & 5BZU J0158+0101         & Masquerading  & Diffuse (II.1) & -- & 480$^{*}$ \\
4FGL J0244.7+1316    & CRATESJ024445+132002    & Undetermined     & IC161103A  & Bronze & 85   \\
4FGL J0239.5+1326    & 3HSP J023927.2+13273    & Undetermined     & IC161103A  & Bronze & 85      \\
4FGL J0224.2+1616    & VOU J022411+161500      & Undetermined     & IC111216A  & Gold   & 891   \\
4FGL J0232.8+2018    & 3HSP J023248.5+20171    & True BL Lac   & IC111216A  & Gold   & 891   \\
4FGL J0344.4+3432    & 3HSP J034424.9+34301    & Undetermined     & IC150831A  & Gold   & 181   \\
4FGL J0509.4+0542    & TXS 0506+056            & Masquerading  & IC170922A  & Gold   & 264   \\
3FGL J0627.9-1517    & 3HSP J062753.3-15195    & True BL Lac   & IC170321A  & Gold   & 231   \\
4FGL J0649.5-3139    & 3HSP J064933.6-31392    & Undetermined     & IC140721A  & Gold   & 157   \\
4FGL J0854.0+2753    & 3HSP J085410.1+27542    & True BL Lac   & IC150904A  & Gold   & 302   \\
4FGL J0946.2+0104    & 3HSP J094620.2+01045    & True BL Lac   & IC190819A  & Bronze & 113   \\
4FGL J0955.1+3551    & 3HSP J095507.9+35510    & True BL Lac   & IC200107A  & --   & --  \\
4FGL J1003.4+0205    & 3HSP J100326.6+02045    & Undetermined     & IC190819A  & Bronze & 113   \\
4FGL J1055.7-1807    & VOU J105603-180929      & Undetermined     & IC171015A  & Gold   & 72    \\
4FGL J1117.0+2013    & 3HSP J111706.2+20140    & Masquerading  & IC130408A  & Gold   & 65    \\
4FGL J1124.0+2045    & 3HSP J112405.3+20455    & Undetermined     & IC130408A  & Gold   & 65    \\
4FGL J1124.9+2143    & 3HSP J112503.6+21430    & Undetermined     & IC130408A  & Gold   & 65    \\
3FGL J1258.4+2123    & 3HSP J125821.5+21235    & Undetermined     & IC151017A  & Gold   & 321   \\
4FGL J1258.7-0452    & 3HSP J125848.0-04474    & True BL Lac   & IC150926A  & Gold   & 216   \\
4FGL J1300.0+1753    & 3HSP J130008.5+17553    & Undetermined     & IC151017A  & Gold   & 321   \\
4FGL J1314.7+2348    & 5BZB J1314+2348         & Masquerading  & IC151017A  & Gold   & 321   \\
4FGL J1321.9+3219    & 5BZB J1322+3216         & Masquerading  & IC120515A  & Gold   & 194   \\
4FGL J1507.3-3710    & VOU J150720-370902      & Masquerading  & IC181014A  & Bronze & 62      \\
4FGL J1528.4+2004    & 3HSP J152835.7+20042    & Masquerading  & Diffuse (II.10)   & --     & 420$^{*}$  \\
4FGL J1533.2+1855    & 3HSP J153311.2+18542    & Undetermined     & Diffuse (II.10)   & --     & 420$^{*}$  \\
4FGL J1554.2+2008    & 3HSP J155424.1+20112    & Undetermined     & Diffuse (II.10)   & --     & 420$^{*}$ \\
4FGL J1808.2+3500    & CRATESJ180812+350104    & Masquerading  & IC110610A  & Gold   & 294   \\
4FGL J1808.8+3522    & 3HSP J180849.7+35204    & True BL Lac   & IC110610A  & Gold   & 294   \\
4FGL J2030.5+2235    & 3HSP J203031.6+22343    & Undetermined     & Diffuse (II.4)   &  --      & 200$^{*}$  \\
4FGL J2030.9+1935    & 3HSP J203057.1+19361    & Masquerading  & Diffuse (II.4) & -- & 200$^{*}$ \\
4FGL J2133.1+2529    & 3HSP J213314.3+25285    & Undetermined     & IC150714A  & Gold   & 439   \\
4FGL J2223.3+0102    & 3HSP J222329.5+01022    & Undetermined     & IC140114A  & Bronze & 54      \\
4FGL J2227.9+0036    & 5BZB J2227+0037         & Masquerading  & IC140114A  & Bronze & 54      \\
4FGL J2326.2+0113    & CRATESJ232625+011147    & Masquerading  & IC160510A  & Gold   & 208   \\
\hline\hline
\multicolumn{6}{p{\linewidth}}{$^*$High-energy tracks detected earlier than IceCube alerts listed in IceCat-1 are labeled as ``Diffuse'' and taken from \citet[Table 8]{Abbasi_2022}. The respective energies marked with asterisks are the reconstructed muon energies instead of the reconstructed neutrino energies. The event IC200107A was not part of the alert stream but was issued separately in \url{https://gcn.gsfc.nasa.gov/gcn/gcn3/26655.gcn3}.}\\
\end{tabular}
\end{center}
\label{tab:sample}
\end{table*}

\section{Methods} 
\label{sec:methods}

In Paper IV, a multiwavelength spectral energy distribution (SED) of each source in the G20 sample was compiled based on public data from multiwavelength experiments, using the Open Universe VOU-Blazars tool~\citep{Chang:2019cdu}. We now wish to analyze these multiwavelength fluxes as well as the IceCube data, using a self-consistent physical model of each source. We adopted a novel model-building approach based on five observational constraints: \textit{1)} the luminosity of the accretion disk and BLR, which may play a crucial role in particle interactions in masquerading BL Lacs; 
\textit{2)} the contribution of the host galaxy to the optical spectrum, which plays a role on the overall SED fit; 
\textit{3)} the frequency of the synchrotron peak emission, which must be correctly described by the interaction model; \textit{4)} spectral fluxes in radio, infrared, optical, ultraviolet, $\gamma$ rays and, where available, X-rays; and \textit{5)} IceCube point-source fluxes, using a self-consistent likelihood minimization based on the spectral shape predicted by the model rather than assuming a power-law spectrum. 

In the following we describe the procedure underlying each of these components. We also describe the assumptions behind the modeling of the radiation processes taking place in the relativistic jet.

\subsection{Accretion disk and broad line region emission}
\label{sec:disk}

Additionally to the nonthermal emission from the relativistic jet that we wish to model, the optical emission from blazars can have contributions from the AGN core region as well as from the host galaxy. In this section we discuss the procedure used to model these components.

In masquerading BL Lacs, the thermal emission from the accretion disk of the supermassive black hole, so-called big blue bump, is outshone by the jet emission, as is the atomic emission from the BLR surrounding the central engine. Even though these components cannot be directly observed in the optical SED, they can play a fundamental role in the emission of $\gamma$ rays through inverse Compton scattering, the development of electromagnetic cascades in the jet, and the production of high-energy neutrinos, as explained in \Sec\ref{sec:model_BLR}.

Since the accretion disk luminosity cannot be directly determined in these sources, an indirect approach has to be used instead. This was done in previous papers in this series by using 
relationships between the accretion-related bolometric luminosity, $L_{\rm bol}$, and $L_{\rm [\ion{O}{II}]}$ and $L_{\rm [\ion{O}{III}]}$ (\citealt{Punsly:2011}; see also \citealt{Padovani2019}
for more details.)

Papers II and III made use of these indirect methods to estimate a value or an upper limit to the bolometric luminosity for each blazar in the G20 sample with [\ion{O}{II}] or [\ion{O}{III}] information. That result was used as one of the four criteria in determining the nature of each source as a masquerading BL Lac or a true BL Lac. In most cases, however, there was
simply not enough information to make a decision either way. 

In this paper, we consider disk luminosity values derived in two different ways, following 
\citet{Padovani2019}. Namely: 1. we assume $\langle L_{\rm bol}/L_{\rm disk} 
\rangle \approx 2$, which is consistent with typical quasar SEDs \citep[e.g.,]
[]{Richards:2006}. Hence $L_{\rm disk,bol} = 0.5 \times L_{\rm bol}$; 2. we derive the 
narrow line luminosity (NLR) from $L_{\rm NLR} = 3 \times (3 \times L_{\rm 
\ion{O}{II}} + 1.5 \times L_{\rm \ion{O}{III}})$ 
\citep{Rawlings:1991}, from which we get $L_{\rm BLR}$ assuming $L_{\rm BLR}/L_{\rm NLR} 
\sim 10$, typical of FSRQs \citep{Gu:2009}. It then follows that 
$L_{\rm disk,NLR} = 10 \times L_{\rm BLR}$, for a standard covering 
factor $\sim 10\%$. In case both estimates were available, we took the logarithmic mean of the two as our best value; this required 
both $L_{\rm \ion{O}{III}}$ and $L_{\rm \ion{O}{II}}$, as otherwise 
$L_{\rm disk,NLR}$ could not be derived. When only $L_{\rm disk,bol}$
was present we used $10^{\log L_{\rm disk,bol} - 0.25}$ as our best estimate,
since $\langle \log L_{\rm disk,bol} - \log L_{\rm disk,NLR} \rangle \sim 0.5$. When only upper limits were available on $L_{\rm \ion{O}{III}}$ and/or $L_{\rm \ion{O}{II}}$ upper limits on $L_{\rm disk}$ were derived. Lower limits on the uncertainties of
$L_{\rm disk}$ are given as 0.5 $\times (\log L_{\rm disk,bol} - \log L_{\rm disk,NLR})$ dex or 0.25 
when only $L_{\rm disk,bol}$ is available. Since the derivation of the thermal, accretion-related bolometric
luminosity is not that easy for BL Lacs and requires a number of assumptions and correlations, as described above, the real uncertainties are very likely larger. 

By adopting upper limits as the assumed disk luminosity, we may, of course, overestimate the real disk luminosity for those sources. However, this does not affect the model fitting because the energy density of broad line photons is independent of the disk luminosity, owing to geometric assumptions that are detailed further in \Sec\ref{sec:model_BLR}.

We then assumed a spectrum based on a template that includes a multi-temperature continuum from the accretion disk as well as emission lines from the BLR~\citep{SDSS:2001ros}.
For each blazar, we normalized this spectrum to the disk luminosity value and compare it to the SED presented in Paper IV. We show an example of this in \Fig\ref{fig:optical_fit_example}. The gray points represent the SED from Paper IV in the infrared and optical range, and the blue curve represents the spectrum of the accretion disk, the BLR, and a dust torus, which is the component below $10^{14}~\mathrm{Hz}$. The blue band represents the uncertainty range on the normalization, obtained through the method described earlier in this section.

In the example shown in \Fig\ref{fig:optical_fit_example}, we can see that the spectrum lies below the broadband SED, which suggests that the jet emission dominates the observed fluxes. We, therefore, accept this disk luminosity value as a valid assumption.
In the cases where the assumed disk luminosity leads to a BLR spectrum that overshoots the SED, we reevaluated the original assumption by reducing the disk luminosity so that the SED data are not violated (cf.~\Fig\ref{fig:app_host} in \App\ref{app:host}).

The result of this procedure is summarized in the left panel of \Fig\ref{fig:disc_and_host} for all blazars in our sample.
The short vertical lines represent the value deduced by the method described above, and the  horizontal lines show the corresponding uncertainty range. The final disk luminosity values are shown as circles; for most sources, this corresponds to the deduced value, while for three sources, the estimate had to be reduced after accounting for the multiwavelength SED fluxes, in all cases by a factor smaller than three. 

Blazars of undetermined nature, that is, which cannot conclusively be determined to be masquerading or true BL~Lacs are shown in \Fig\ref{fig:disc_and_host} in gray. These sources are treated a priori the same way as masquerading BL Lacs (red), while for true BL Lacs (blue) we did not account for disk or BLR emission. For reference, the optical SEDs of all sources in the sample are shown in \App\ref{app:host} in \Fig\ref{fig:app_host}, together with the assumed disk, BLR, and dust torus spectra resulting from this procedure. 

\subsection{Host galaxy emission}
\label{sec:host}

The results of the optical spectroscopy campaign of the G20 sample were presented in Papers I -- III. This resulted in a redshift determination and the decomposition of the optical spectrum into a host galaxy contribution, assumed to follow the giant elliptical template adopted from \citet{Mannucci:2001qa}, and a nonthermal contribution from the relativistic jet, assumed to follow a power-law spectrum in the frequency range of the analysis.

\begin{figure}
\includegraphics[width=\linewidth,trim={0mm 17mm 0mm 0mm},clip]{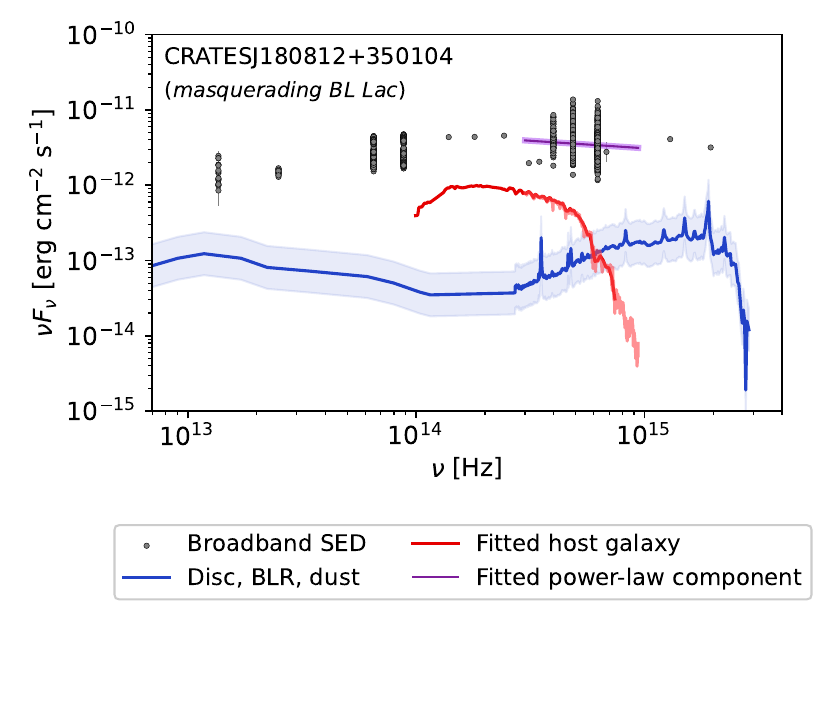}

\caption{Infrared and optical spectrum of a masquerading BL Lac from the sample, given in the observer's frame. The gray points represent the broadband SED data (Paper IV). In red and purple, we show the result of the spectral decomposition following the optical spectroscopy campaign of the G20 sample (Paper III). In blue we show the template spectrum adapted from ~\citep{SDSS:2001ros} and the respective error range, which comes from the uncertainty on the disk luminosity (cf.~\Fig\ref{fig:disc_and_host}). At lower frequencies we see the infrared emission from a dust torus, and toward higher frequencies a thermal continuum from the accretion disk and broad lines from the BLR. Each of these three components is seen with a different relativistic boost in the rest frame of the relativistic jet, as explained in \Sec\ref{sec:disk} and \App\ref{app:dust}.}
\label{fig:optical_fit_example}
\end{figure}

In \Fig\ref{fig:optical_fit_example} we show an example of the derived host galaxy spectrum, following the spectral decomposition based on optical spectroscopy of the source, as described in detail in Papers I and III. In \Fig\ref{fig:app_host} (\App\ref{app:host}), we show the host spectrum for the remaining sources in the sample. In all cases, this component is compatible with the SED compiled from the public multiwavelength data in Paper IV; however, we did not use the best-fit spectral index to constrain the model, because \textit{a)} the model does not necessarily predict an exact power law in this frequency range, and \textit{b)} the variability of the nonthermal emission can easily lead to a time-dependent change in the actual spectral index, which cannot be captured by the current approach. 

\begin{figure*}
\includegraphics[width=\linewidth]{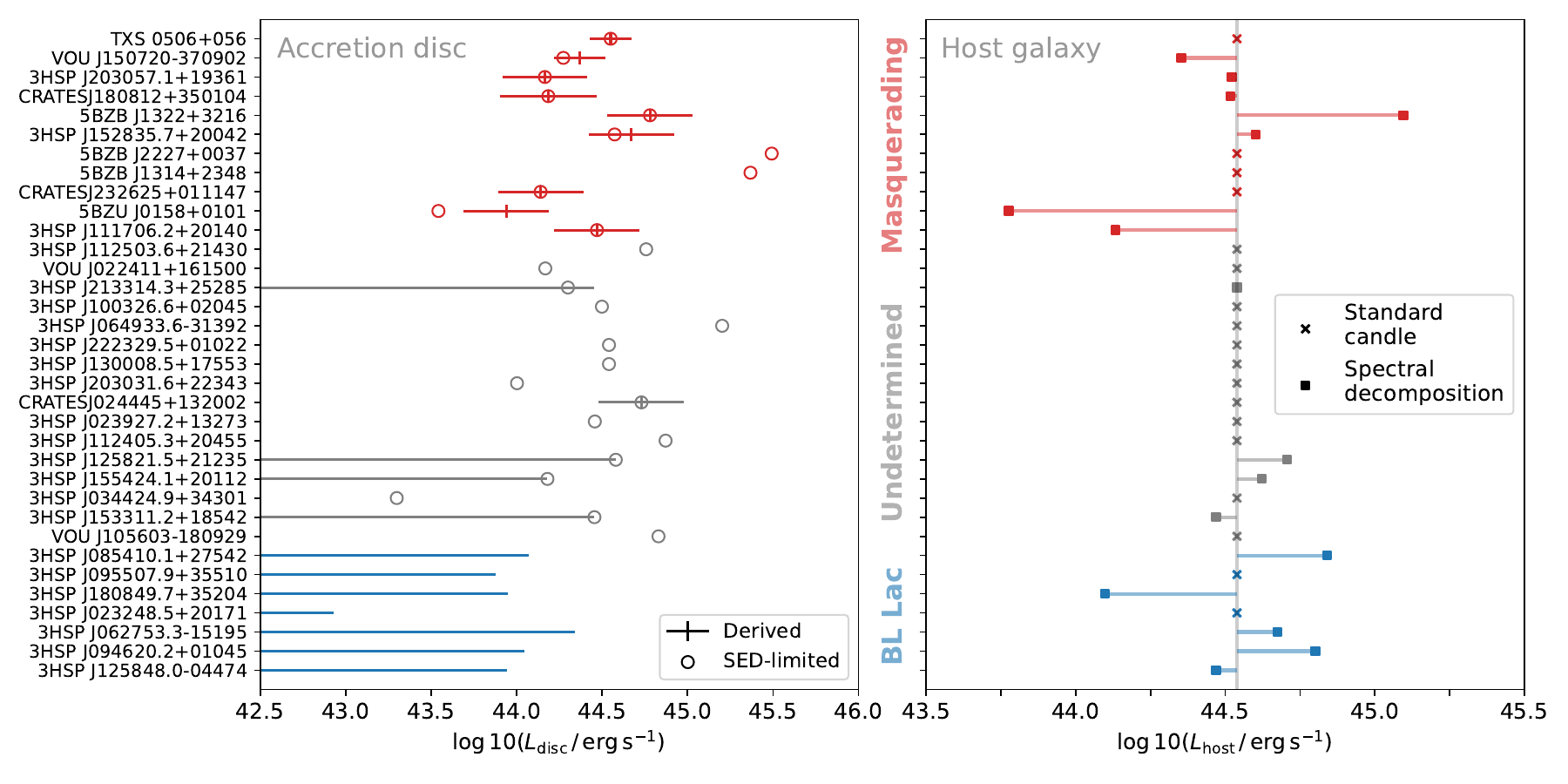}
\caption{Constraints on the luminosity of the accretion disk and host galaxy for all blazars in the sample (masquerading BL Lacs in red, true BL Lacs in blue, and undetermined in gray). \textit{Left:} accretion disk luminosity constraints. The vertical line markers show the estimate derived in Section \ref{sec:disk} and the horizontal line their respective uncertainties. In the case of true BL Lacs, as well as four of the blazars of undetermined nature, only upper limits are available. In the cases where there is no vertical line marker, Papers II and III either report no value of the bolometric luminosity or only an upper limit. The circles show the final value adopted, based on an additional comparison with the optical SED, as discussed in \Sec\ref{sec:disk}. We can see that the two estimates only disagree for three objects: VOU~J150720-370902, 3HSP~J152835.7+20042, and 5BZU~J0158+0101. 
\textit{Right:} Estimated host galaxy luminosity following the procedure explained in \Sec\ref{sec:host}. The estimates marked with a square result from optical spectrography, as reported in Paper III  (see \App\ref{app:host} for further details)}
\label{fig:disc_and_host}
\end{figure*}

In the right panel of \Fig\ref{fig:disc_and_host}, we show the resulting host galaxy luminosity for all blazars in the sample. The host luminosity is denoted with a square in the cases where the above procedure was adopted. The crosses, on the other hand, indicate those blazars for which no spectroscopic data were available. In those cases, we assume that the host galaxy is a typical giant elliptical with a standard-candle luminosity of $3.5\times10^{44}~\rm{erg}\,\rm{s}^{-1}$~\citep[vertical gray line][]{}. We can see that in the cases where spectroscopy was performed, the derived luminosity differs at most by a factor of three compared to the standard candle assumption.

\subsection{Synchrotron peak}
\label{sec:synchrotron_peak}

Additionally to describing of the multiwavelength and neutrino fluxes, we also wish to accurately determine the synchrotron peak frequency for each source. As explained in \Sec\ref{sec:optimization}, using data on the synchrotron emission also allows to maximally constrain the parameters of the electrons and the jet at an early stage of the optimization process, before including the complexity of hadronic processes.

\begin{figure}
\includegraphics[width=\linewidth,trim={0mm 5mm 0mm 3mm},clip]{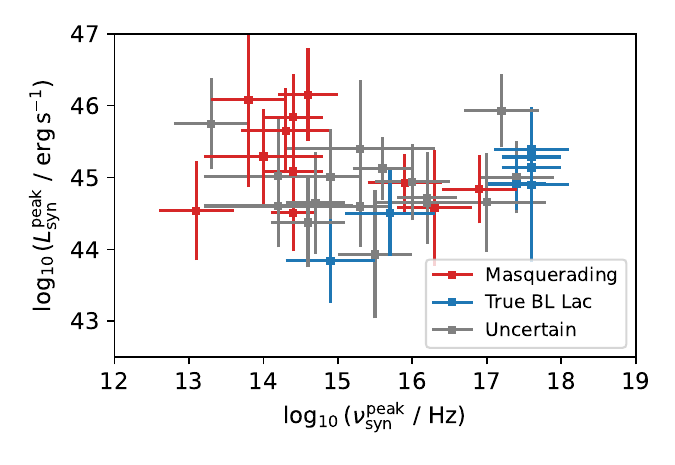}

\caption{Estimated synchrotron peak luminosity of the BL Lacs in the sample, as a function of the respective synchrotron peak frequency, adopted from Paper IV~\citep{Karl:2023huw,Karl_2024}, shown here in the comoving frame of the source.}
\label{fig:synchrotron_peak}
\end{figure}

For each source, we estimated a window in photon frequency and flux where the synchrotron peak predicted by the model is allowed to fall. The synchrotron peak frequency of each BL Lac in the comoving source frame is provided in Table 1 of Paper IV, accompanied by the respective uncertainty, $\nu_\mathrm{syn}^\mathrm{peak}\pm\sigma_{\nu_\mathrm{syn}^\mathrm{peak}}$. Those estimates were obtained using the open-source tool BlaST~\citep{Glauch:2022xth}, which utilizes a neural network to estimate the synchrotron peak based on the entire multiwavelength SED. These values and respective uncertainties are shown in \Fig\ref{fig:synchrotron_peak}, along the x-axis. 

On the y-axis, \Fig\ref{fig:synchrotron_peak} shows the estimated synchrotron peak luminosity for each blazar, which we used as well to constrain the electron parameters in the model. To estimate this, we took the multiwavelength SED of each source, provided in Paper IV, and considered the range of fluxes observed in the infrared through the X-ray bands. This procedure is exemplified in \Fig\ref{fig:binning_example} for one of the sources. The multiwavelength fluxes are shown as gray data points; we can ignore for the moment the gray triangles, which represent upper limits, and the orange boxes, which result from the binning procedure described in the next section. The vertical green band represents the estimated synchrotron peak frequency range, as provided in Paper IV. To constrain the synchrotron peak flux, we utilized the multiwavelength SED, and consider the range of detected fluxes within the peak frequency range defined above. When there are no available data in that range, we considered the SED data falling within one order of magnitude of the determined peak synchrotron frequency.

The above procedure results in a double constraint that can be represented as a rectangular window in frequency and flux, as exemplified in \Fig\ref{fig:binning_example}. Here, the green shaded area represents the 1$\sigma$ range of the synchrotron peak frequency. Based on the multiwavelength SED, we then additionally defined the range of synchrotron peak fluxes, resulting in the dashed red rectangle. This constrain is also shown for the remaining sources in \Fig\ref{fig:app_components_1}, in \App\ref{app:components}.

\subsection{Multiwavelength data binning}
\label{sec:binning}

\begin{figure}[htpb!]
\includegraphics[width=\linewidth,trim={0 6mm 0 0}, clip]{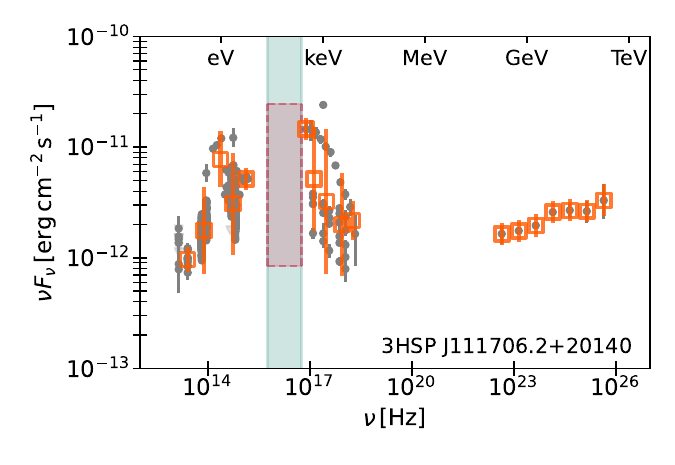}
\caption{Result of the binning procedure for one source in the sample, used as input for the model optimization. As gray data points we show the observations compiled in Paper IV; the upper limits are shown as downward triangles. The orange squares represent the binned fluxes while the orange error bars show the uncertainty in each binned flux. This uncertainty reflects both the intrinsic uncertainty in the data as well as the spread in the flux measurements within the same frequency bin. The red rectangle represents the area of acceptable synchrotron peaks based on the data, following the procedure described in  \Sec\ref{sec:synchrotron_peak}.}
\label{fig:binning_example}
\end{figure}

The multiwavelength data we wish to describe are generally nonsimultaneous. The $\gamma$-ray data between $\sim100$~MeV and $\sim100$~GeV consist of a time-averaged spectrum from the \textit{Fermi} Large Area Telescope (LAT), as analyzed in Paper IV. At other wavelengths, the data originate from nonsimultaneous observations and often exhibit large flux variability. We therefore started by binning the data by averaging the fluxes in logarithmic frequency bins and estimating the corresponding uncertainties.

We divided the frequency range into bins of width $\log_{10}(\nu)=0.4$ for all sources, as exemplified in \Fig\ref{fig:binning_example}. For each bin $i$ that contains observations, we calculated the weighted average of the logarithmic fluxes falling within the bin,
\begin{equation}
  \langle \log_{10}\nu F_{\nu} \rangle_{i} \,=\, \sum_{j\,\mathrm{in\,bin},i} \,\frac{\log_{10}\nu F_{\nu,j}}{\sigma_{\log_{10}(\nu {F_\nu},j)}},
  \label{eq:binning}
\end{equation}
where $\nu F_{\nu,j}$ are the individual observed fluxes in frequency bin $i$ and $\sigma_{\log_{10}(\nu {F_\nu},j)}$ the respective error, calculated on the logarithmic flux. To estimate the uncertainty in each bin $i$, we considered the maximum value of the observational error values $\sigma_{\log_{10}(\nu {F_\nu},j)}$ within that bin and the spread in the values of $\log_{10}(\nu F_{\nu,j})$, which may indicate variability of the source in that frequency band. We took the maximum value between these two as the uncertainty on the binned flux, $\sigma_{\log, i}$.

As we can see in the example of \Fig\ref{fig:binning_example}, through this method the original nonsimultaneous data, shown in gray, are binned in logarithmic frequency, as shown in orange. The corresponding error bars in each bin capture the variability of the original data within that bin. Above 100~MeV, where the data are time-averaged, we can see that each point occupies its own logarithmic frequency bin, which means the original data are unaffected by this procedure. In this case, the final error bars simply reflect the intrinsic error bars of the LAT fluxes.

We tested the goodness of a modeled SED against the binned multiwavelength data by means of a logarithmic chi-squared, $\chi^2_\mathrm{log}$:
\begin{equation}
  \chi^2_\mathrm{log} \,=\, \sum_{i} \,\frac{[\langle \log_{10}\nu F_\nu\rangle_i - \log_{10}(\nu \mathcal{F}_{\nu,i})]^2}{\sigma_{\log, i}^2},
  \label{eq:logchi}
\end{equation}
where $\langle \log_{10}\nu F_\nu\rangle_i$ is a binned flux as defined in \eq~(\ref{eq:binning}), and $\nu \mathcal{F}_{\nu,i}$ is the value of the SED predicted by the model interpolated at the central frequency of bin $i$. For radio data below 300~GHz we considered the respective flux values only as upper limits. That is because the compact dissipation region responsible for the high-energy emission is necessarily optically thick to low-frequency radio emission, owing to synchrotron self-absorption~\citep[cf.][]{Rodrigues:2023vbv}. That means that for data points $i$ for which $\nu_i<300\,\mathrm{GHz}$, the respective terms are only accounted for in the sum of \eq(\ref{eq:logchi}) if the model overshoots the observation, that is, if $\langle \log_{10}\nu F_\nu\rangle_i < \log_{10}(\nu\mathcal{F}_{\nu,i})$.

Compared to the minimization of a linear variable, such as the conventional $\chi^2$, the advantage of a logarithmic approach is the more equal inclusion of data points lying at different orders of magnitude in flux. On the other hand, as discussed later, the value of this variable cannot be easily interpreted in absolute terms, but only as a point of comparison between solutions.

\subsection{Leptohadronic jet model}
\label{sec:model_jet}

The core element of this analysis is the multimessenger modeling of the sample, which consists of a numerical simulation of the electromagnetic and hadronic interactions of protons accelerated in the relativistic jet. For each blazar, the best-fit solution of the model should be able to describe both the multiwavelength SED and the data resulting from the IceCube point source analysis and the IceCube alert stream, in a fully self-consistent manner.

The particle interactions are simulated using the open-source software AM$^3$ \citep{Klinger:2023zzv}. This is a time-dependent numerical framework that solves the coupled partial differential equations describing the evolution of the energy spectrum of a homogeneous population of electrons, protons and photons, as well as all the secondary particles produced in their interactions. 

We assume that electrons and protons are accelerated in the relativistic jet, and consider the radiative interactions taking place in a single zone, referred to as the dissipation region. For simplicity, we model the dissipation region as a spherical blob of radius $R_\mathrm{b}^\prime$\footnote{Throughout this work, primed quantities will refer to the rest frame of the relativistic jet, while quantities that are not primed will refer either to the host galaxy frame or the observer's frame, as indicated explicitly in the text.}. The dissipation region is assumed to be permeated by a homogeneous and isotropic magnetic field of strength $B^\prime$, and to move along the jet with a bulk Lorentz factor $\Gamma_\mathrm{b}$. The majority of the radiation is relativistically beamed into an angle $\theta_\mathrm{beam}\sim1/\Gamma_\mathrm{b}$. We conservatively assume that the observation angle is $\theta_\mathrm{obs}=\theta_\mathrm{beam}$, which means that the emission is boosted with a Doppler factor $\delta_\mathrm{D}=\Gamma_\mathrm{b}$ (see, e.g., Appendix A of \citealt{Urry1995}).

We assume the spectrum of accelerated electrons and protons can be described by power laws with spectral indices $p_\mathrm{e}$ and $p_\mathrm{p}$, respectively, up to maximum Lorentz factors $\gamma_\mathrm{e}^{\prime \mathrm{max}}$ and $\gamma_\mathrm{p}^{\prime \mathrm{max}}$. For simplicity, we generally\footnote{For most sources, the model is not sensitive to the minimum electron Lorentz factor $\gamma_\mathrm{e}^{\prime\mathrm{min}}$, owing to the fact that we did not attempt to describe radio data below 300~GHz, as explained in \Sec\ref{sec:binning}. In the case of source TXS 0506+056, 300~GHz data suggest a value of $\gamma_\mathrm{e}^{\prime\mathrm{min}}=300$ in our best-fit scenario. We therefore adopted this value in the final result. For all other sources, the data do not constrain the $\gamma_\mathrm{e}^{\prime\mathrm{min}}$ parameter. In those cases, we adopted a value of $\gamma_\mathrm{e}^{\prime\mathrm{min}}=100$.} fixed the minimum Lorentz factor of both species to $\gamma_\mathrm{e}^{\prime \mathrm{min}}=\gamma_\mathrm{p}^{\prime \mathrm{min}}=100$. The normalization constants of the distributions are determined by the total electron and proton injection powers, $L_\mathrm{e}^{\prime}$ and $L_\mathrm{p}^{\prime}$.

Since constraining the proton luminosity is challenging due to a high level of degeneracy in the leptohadronic solutions, we limited a priori the possible values of $L_\mathrm{p}^{\prime}$. As explained in \Sec\ref{sec:optimization}, we required that the physical proton luminosity\footnote{\label{footnote:physical_luminosity}The physical proton luminosity is defined here as the power in protons accelerated in the jet given in the rest frame of the supermassive black hole, given by $L_\mathrm{p}^{\mathrm{phys}}=L_\mathrm{p}^{\prime}\,\Gamma_\mathrm{b}^2/2$.} does not exceed the Eddington luminosity of the source by a factor larger than ten, $L_\mathrm{p}^{\mathrm{phys}}<10\,L_\mathrm{Edd}$. In terms of the proton luminosity in the rest frame of the jet, this translates into the constraint $L_\mathrm{p}^\prime<20\,L_\mathrm{Edd}/\Gamma_\mathrm{b}$.

\subsection{External radiation in masquerading BL Lacs}
\label{sec:model_BLR}

In the case of blazars that have either been identified as masquerading BL Lacs or whose nature cannot be determined, we consider the presence of a BLR and a dust torus surrounding the central engine. Although in IHBLs the radiation from these elements is swamped by the jet emission in the observer's frame, their energy density in the jet frame can, under certain conditions, be relativistically boosted into the jet frame, in which case they play an important role in particle interactions in the jet.

We start by describing the BLR treatment. The radius of the BLR is assumed to scale with the square-root of the accretion disk luminosity, $R_\mathrm{BLR}=10^{17}(L_\mathrm{disk}/10^{45}\,\mathrm{erg/s})^{0.5}\,\mathrm{cm}$\citep{Cleary:2006pe,Ghisellini:2008zp}. This is also the prescription followed by previous blazar models~\citep[e.g.,][]{Murase:2014foa,Rodrigues:2023vbv}. The isotropic component of the BLR photon fields can be boosted into the jet frame and potentially contribute as targets to electromagnetic and hadronic interactions. The strongest of these components is broad line emission. We considered explicitly the hydrogen and helium Ly$\alpha$ lines\footnote{The hydrogen and helium spectral lines are modeled as Gaussian distributions with peak energies in the black hole rest frame of 10.2~eV and 40.8~eV, respectively, and a relative width of 5\% of the peak energy.}, with a total luminosity of $f_\mathrm{cov}^\mathrm{BLR}L_\mathrm{disk}$, where we assumed a BLR covering factor of $f_\mathrm{cov}^\mathrm{BLR}=0.1$~\citep{Greene:2005nj}. A fraction of the thermal emission from the accretion disk,  modeled here as a multi-temperature template spectrum \citep{SDSS:2001ros}, is also isotropized in the BLR thanks to Thomson scattering. The total luminosity of this isotropized field is  $\tau_\mathrm{T}\,L_\mathrm{disk}$, where we assumed an optical thickness to Thomson scattering of $\tau_\mathrm{T}=0.01$~\citep{Blandford:1995yf}.

Because of the relation for the BLR radius given above, the photon energy density inside the BLR does not depend on the disk luminosity~\citep{Ghisellini:2009wa}:
\begin{equation}
u_\mathrm{BLR}=\frac{f_\mathrm{cov}^\mathrm{BLR}\,L_\mathrm{disk}}{4\pi c\,R_\mathrm{BLR}^2}=2.7\times10^{-2} \left(\frac{f_\mathrm{cov}^\mathrm{BLR}}{0.1}\right)~\mathrm{erg~cm^{-3}}.
\label{eq:u_BLR}
\end{equation}
The effective BLR photon density in the rest frame of the dissipation region, which is a key element of our masquerading BL Lac model, depends on the ratio between the dissipation radius, which is the distance to the central engine and the BLR radius, that is, $R_\mathrm{diss}/R_\mathrm{BLR}$. The absolute value of $R_\mathrm{diss}$, on the other hand, cannot be directly constrained in this model. For $R_\mathrm{diss}/R_\mathrm{BLR}\approx1$, the dissipation region is located approximately on the BLR. The local photon density is then given by \eq(\ref{eq:u_BLR}). The relative Doppler factor between the BLR and the dissipation region is $\delta^\mathrm{rel}_\mathrm{BLR}\approx\Gamma_\mathrm{b}$, which means that the photon energy density in the jet frame is given by $u_\mathrm{BLR}^\prime=\Gamma^2 u_\mathrm{BLR}$, and the photon energy by $E_\gamma^\prime=\Gamma_\mathrm{b}E_\gamma$. 
We did not test scenarios involving $R_\mathrm{diss}<R_\mathrm{BLR}$, since in that case the $\gamma$ rays emitted by the jet would be strongly attenuated, in contradiction with observations~\citep{Costamante:2018anp}. While neutrino production could in principle occur deep inside the BLR, co-explaining neutrino and $\gamma$-ray emission in such a scenario would require a more complex framework, such as a multiple-zone model.

Outside the BLR, the local photon density drops. Additionally to the decrease of the photon energy density by a geometric factor  $R_\mathrm{diss}^{-2}$, the relative Doppler factor also decreases, because the photons impinge increasingly from behind. As an approximation, we followed the prescription by \citet{Ghisellini:2009wa} (see Eq~20 of that reference and subsequent discussion). That is, we considered only the photons impinging tangentially from the edges of the BLR, which have the highest relative Doppler factor, and ignore the photons impinging more from behind, which are more drastically deboosted. In the case where $R_\mathrm{diss}/R_\mathrm{BLR}\gtrsim1$, these tangentially impinging photons can play a significant role in high-energy interactions; for values of $R_\mathrm{diss}/R_\mathrm{BLR}\gtrsim4$, the radiation zone is too distant to be significantly affected by the BLR, and the masquerading model effectively defaults to that of a true BL Lac \citep[see][where a similar assumption is considered in the modeling of a FSRQ catalog]{Rodrigues:2023vbv}.

The dust surrounding the central AGN region is heated by the accretion disk emission~\citep[e.g.,][]{1987ApJ...320..537B}. The infrared photons emitted by the warm dust can also play a role as a target for particle interactions. At the same time, the geometry and size of the dust torus is a topic of debate and exhibits large variation among sources, without a clear scaling with the disk luminosity~\citep{2011A&A...527A.121K,Burtscher:2013aza}. This makes it challenging to constrain the contribution of the thermal emission from the dust as targets for radiative interactions taking place in the jet. We therefore neglected this contribution in our treatment, and discuss its possible effects a posteriori. As we show in \App\ref{app:dust}, the effect of this dust emission is negligible for most sources in the sample. For four sources, it may lead to a slight enhancement in photo-pair production, resulting in an increase in the predicted megaelectronvolt $\gamma$-ray flux by a factor up to 1.8. Given the current absence of data in this band, it is therefore challenging to constrain the dust contribution to our leptohadronic IHBL model. 

\subsection{Reanalyzing the IceCube point source data}
\label{sec:icecube}

To constrain neutrino emission from each blazar, we utilized public IceCube data encompassing 10
years of through-going muon tracks \citep{https://doi.org/10.21234/cpkq-k003}. For each blazar, we derived the point source neutrino flux by estimating the likelihood of a signal component clustering around the source, and comparing it to the null hypothesis of pure atmospheric background.

As detailed in Appendix A of Paper IV, we performed simulations of a signal neutrino flux using the open-source software SkyLLH\footnote{https://github.com/icecube/skyllh} \citep{Bellenghi:20230u}. In that work, the simulated neutrino signal was assumed to have an energy probability density function (pdf) given by a power law distribution, that is, $\propto E^{-\gamma}$. In this work, we adopted a more self-consistent treatment and assume instead that the simulated signal follows a pdf given by a spectral shape typical of p$\gamma$ interactions. Specifically, we considered an average neutrino spectral shape provided by a recent blazar sample modeling study~\citep{Rodrigues:2023vbv} and adopt it as a template for the signal energy pdf. Unlike a power-law spectrum, a neutrino spectrum emitted in photohadronic interactions is highly peaked in a $\nu F_\nu$ representation, as our modeling results also show (\Sec\ref{sec:results}). In \App\ref{app:icecube} we provide a description of the adopted spectral shape, lay out the details of this procedure, and compare it to the commonly used power-law assumption.

We then compared the simulation results to the public IceCube point source data from each blazar. Following the procedure by \citet{Feldman_1998}, we calculated the 68\% confidence level intervals on the neutrino flux from each blazar. The fact that the signal spectral shape is fixed means that \textit{a)} the derived flux range depends only on the peak energy $E_\nu^{\mathrm{peak}}$; and \textit{b)} the integrated flux of signal neutrinos is proportional to the differential flux at the peak, $\nu F_\nu^{\mathrm{peak}}$ (assuming that the neutrino spectral shape does not drastically differ from source to source, an assumption whose accuracy is quantified in \App\ref{app:icecube}). Therefore, for each value of the neutrino peak energy, $E_\nu^{\mathrm{peak}}$, the 68\% confidence results on a given source can be translated into an allowed range of muon neutrino peak flux levels $\nu F_\nu^{\mathrm{peak}}$. Whenever the lower bound on the neutrino flux is compatible with zero for a given value of $E_\nu^{\mathrm{peak}}$, we only considered the 68\% upper limit. As we show in \Sec\ref{sec:results}, we found a flux incompatible with 0 on the 68\% level in 12 out of 32 cases. In comparison, in Paper IV we found this for eleven sources, with CRATESJ232625+011147 behaving differently. These neutrino constraints are combined with the multiwavelength data discussed previously to constrain the leptohadronic model parameters of each blazar, as described in the following section.

\subsection{Multimessenger model optimization}
\label{sec:optimization}

We now turn to the optimization of the leptohadronic model based on the multiwavelength and neutrino data from each source. Rather than optimizing the model parameters by means of a direct global minimization \citep[see e.g.,][]{Rodrigues:2023vbv}, in this work we divided the procedure into four steps. This ensures that the model captures individual observational characteristics: \textit{1)} the synchrotron peak frequency and flux; \textit{2)} the effect of external fields on the $\gamma$-ray emission in the case of masquerading BL Lacs; \textit{3)} the full multiwavelength SED; \textit{4)} limits on the IceCube neutrino point source flux.

\begin{table*}[htbp!]
    \caption{Workflow of the leptohadronic model optimization method.}
    \begin{tabular}{llll}
    \hline
    \hline
        & Step 1 & Step 2 & Step 3 \\
    \hline
    \makecell[l]{Processes\\included}&
        \begin{tabular}{@{}l@{}}Electron synchrotron \\ \\ \\ \\ \\ \\ \end{tabular} & 
        \begin{tabular}{@{}l@{}} Electron synchrotron\\ Inverse Compton \\ Photon-photon pair production \\ \\ \\ \\ \end{tabular} & 
        \begin{tabular}{@{}l@{}} Electron synchrotron\\ Inverse Compton\\Photon-photon pair productions \\ Photo-pion production \\ Proton photo-pair production \\ Proton synchrotron \end{tabular} \\
    \hline
        
    \makecell[l]{Observational\\constraints}& 
        \begin{tabular}{@{}l@{}} $\bullet$ Describe synchrotron peak$^\ast$ \\ \end{tabular} & 
        \begin{tabular}{@{}l@{}} $\bullet$ Not overshoot LAT\end{tabular} & 
        \begin{tabular}{@{}l@{}}$\bullet$ Minimize binned $\chi^2_{\mathrm{log}}$ of predicted SED \\  $\bullet$ Neutrino flux consistent with IceCube\end{tabular} \\
    \hline
    \multirow{6}{*}{Parameters$^\dagger$} & $R_\mathrm{b}^\prime$ ($10^{15.0}$ - $10^{16.5}$ cm) & $R_\mathrm{diss}$ (1.0-4.0$\,R_\mathrm{BLR}$) & $L^\prime_\mathrm{p}$ ($10\,L^\prime_\mathrm{e}$ - $20\,L_\mathrm{Edd}/\Gamma_\mathrm{b}$) \\ 
            & $B^\prime$ ($10^{-1.5}$ - $10^{1.0}$ G) & &  $\gamma_\mathrm{p}^{\prime\,\mathrm{max}}$ ($10^{4.0}$ - $10^{9.0}$) \\ 
            &$\Gamma_\mathrm{b}$  ($10^{0.5}$ - $10^{1.5}$) & & $p_\mathrm{p}$ (1.0 - 3.0)\\
            & $L^\prime_\mathrm{e}$ ($10^{39}$-$10^{44}$ erg s$^{-1}$) & & \\ 
            & $\gamma_\mathrm{e}^{\prime\,\mathrm{max}}$  ($10^{3.0}$ - $10^{7.0}$) & & \\ 
            & $p_\mathrm{e}$  (1.0 - 3.0) & & \\    
      \hline
      \hline
    \end{tabular}
    $^\ast$The synchrotron peak frequency and flux are required to lie within the derived boundaries, as shown in \Fig\ref{fig:synchrotron_peak}, according to the method described in \Sec\ref{sec:synchrotron_peak}.
    
    $^\dagger$The parameters whose boundary values are indicated in the format $10^x$ were scanned in logarithmic space due to their search range spanning several orders of magnitude. The definition of each parameter, as well as the total number of parameter sets tested at each step, are detailed in~\Sec\ref{sec:optimization}.
    \label{tab:steps}
\end{table*}

We start by constraining the parameters of each source based on the observed synchrotron peak. For that, we considered a uniform six-dimensional parameter space describing the electron population and the relativistic jet. These six parameters are listed in the first column of \Tab\ref{tab:steps} together with their respective boundary values. We simulated $10^6$ different parameter sets, which translates to ten bins per parameter. The parameters whose limits are listed in the form $10^x$ were searched in a logarithmic grid, which at these stage are all except the electron spectral index $p_\mathrm{e}$. We calculated the synchrotron emission for each of the grid points and compare the peak flux and frequency with the limits derived in \Sec\ref{sec:synchrotron_peak}, as summarized in \Fig\ref{fig:synchrotron_peak}. We calculated the emission numerically using AM$^3$, including synchrotron emission, cooling, and synchrotron self-absorption, and excluding all remaining processes. This ensures that the cooling of the accelerated electrons is self-consistently accounted for in the steady state emission. Neglecting inverse Compton losses at this stage maximizes the efficiency of the calculation without significantly affecting the results. This is because in all blazars in the sample, and indeed in IHBLs in general, the Compton dominance is lower than unity. By requiring the predicted synchrotron peak flux and frequency to fall within the derived boundaries as explained above, we excluded 95-99\% of the original six-dimensional parameter space hypervolume.

In a second step, we included the effect of inverse Compton scattering on the high-energy fluxes. In the case of masquerading BL Lacs and BL Lacs of undetermined nature, we also included the dissipation radius parameter, $R_\mathrm{diss}/R_\mathrm{BLR}$ (see middle column of \Tab\ref{tab:steps}), which regulates the extent of external Compton scattering as explained in \Sec\ref{sec:model_BLR}. For each source, we performed $5\times10^4$ purely leptonic simulations, now accounting for inverse Compton emission and cooling, and estimate the emitted multiwavelength fluxes. We excluded a given parameter set only if the leptonic emission overshoots any of the {\it Fermi}-LAT data points by more than $1\sigma$. This process eliminates an additional 90-98\% of the remaining parameter space of the jet and the electron population. The allowed solutions correctly describe the synchrotron peak emission and do not overshoot the $\gamma$-ray fluxes. We did not make any requirement on the minimum $\gamma$-ray flux from leptonic processes, thus leaving open the possibility for hadronic emission to contribute to, or even dominate, the LAT spectrum. This methodology substantially differs from the type of leptohadronic modeling approaches employed in recent literature on other IceCube blazar candidates, as reviewed in \Sec\ref{sec:intro}, where the LAT spectrum is typically dominated by primary electron emission.

As a final step, we included proton interactions and optimize the parameters of the full leptohadronic model. 
Additionally to the parameters describing the electron distribution and the source geometry, already highly constrained in the two previous steps, we considered three additional parameters characterizing a proton distribution, as listed in \Tab\ref{tab:steps}. We simulated approximately $2\times10^{4}$ parameter sets per source, now accounting for all electromagnetic and hadronic interaction channels including cascade emission, and compute the steady-state  multiwavelength fluxes and the corresponding steady-state neutrino spectrum.

As shown in \Tab\ref{tab:steps}, we set an upper boundary on the proton physical luminosity, $L^\mathrm{phys}_\mathrm{p}=L^\prime_\mathrm{p}\,\Gamma_\mathrm{b}^2/2<10\,L_\mathrm{Edd}$. The Eddington luminosity of each source is calculated based on the black hole mass derived in Papers II and III. This imposes a limit on the super-Eddington accretion rate of the resulting solutions. On the other end of the spectrum, we limited the search to $L^\prime_\mathrm{p}>10\,L^\prime_\mathrm{e}$, since below this level hadronic processes will not significantly contribute to the SED and we would therefore obtain solutions compatible with a purely leptonic scenario (cf. leptohadronic literature reviewed in \Sec\ref{sec:intro}).

After performing the full leptohadronic model on the grid, we excluded parameter sets for which the resulting muon neutrino spectrum peaks outside the range allowed by the IceCube data. In practice, this step is only relevant in the cases where the IceCube data provide a lower limit on the neutrino flux at the 68\% confidence level. If, for a given source, we found no solutions that satisfy the IceCube lower limits within the acceptable proton power range, we then relaxed this condition, which means that the model fails to comply with the IceCube data given the other conditions imposed. It is worth noting that the neutrino flux limits are given at the 68\% confidence level, so it is natural to expect that the results do not comply with these limits for a fraction of the sources.

For each solution that obeys the IceCube limits (or for the entire pool of leptohadronic solutions in the cases where no solution obeys the IceCube constraints or only upper limits exist), we performed a local minimization using iminuit~\citep{iminuit}. We used as cost function for the minimization the logarithmic chi-squared $\chi^2_\mathrm{log}$ given in \eq(\ref{eq:logchi}), which evaluates the relative goodness of fit of the predicted SED to the binned multiwavelength data. The optimized leptohadronic parameter set with the lowest value of $\chi^2_\mathrm{log}$ is selected as the best-fit result.

\section{Results}
\label{sec:results}

We now present the model results and discuss the corresponding multimesssenger predictions. In \Fig\ref{fig:all_masquerading} we show the predicted multiwavelength and neutrino fluxes for the 11 masquerading BL Lacs in the G20 sample. The total photon flux resulting from the best-fit parameters is shown as a black curve. In purple we show the leptonic component of the emission, and in light green the hadronic component. The muon neutrino spectrum, computed self-consistently, is shown as a dark green curve. The best-fit parameter values leading to these results are provided in \Tab\ref{tab:parameters} in \App\ref{app:parameters}. Along with the best-fit result, we show as colored bands the respective 1$\sigma$ uncertainty range. In the lower row of \Tab\ref{tab:parameters} we provide the average uncertainty of each model parameter underlying these $1\sigma$ error bands on the predicted flux. The gray data points are the public multiwavelength flux data analyzed in Paper IV. Finally, we show in green the constraints resulting from the IceCube data analysis introduced in \Sec\ref{sec:icecube} and further detailed in \App\ref{app:icecube}, where we considered a neutrino signal spectrum that is peaked in $\nu F_\nu$ as predicted by photohadronic models. Specifically, the green region shows the allowed energy and flux values of the peak of the muon neutrino spectrum in the observer's frame, in order for the spectrum to be consistent with the point source data at the 68\% confidence level. 

\begin{figure*}[htpb!]
\centering

\includegraphics[width=\sedsize,trim={4mm 5mm 4mm 4mm}, clip]{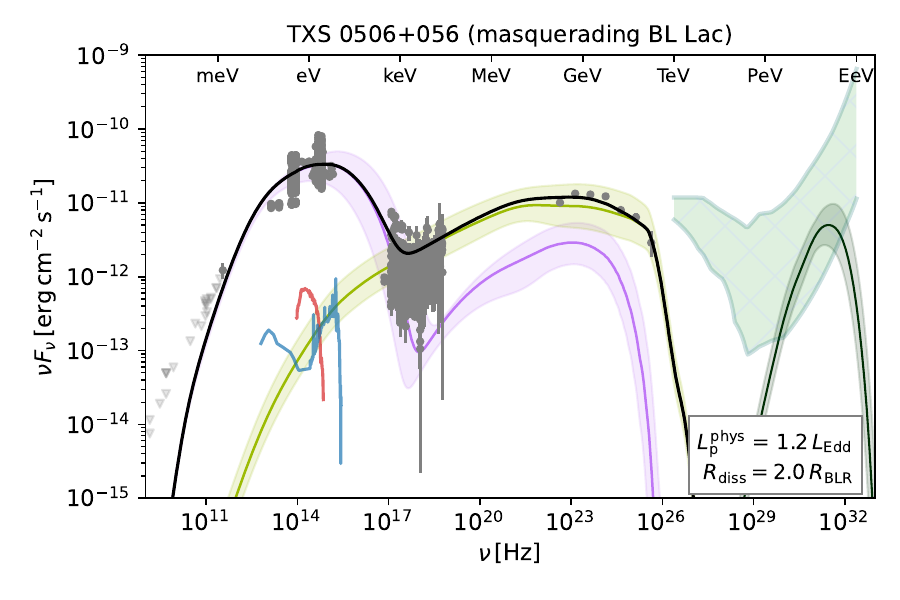}\includegraphics[width=\sedsize,trim={4mm 5mm 4mm 4mm}, clip]{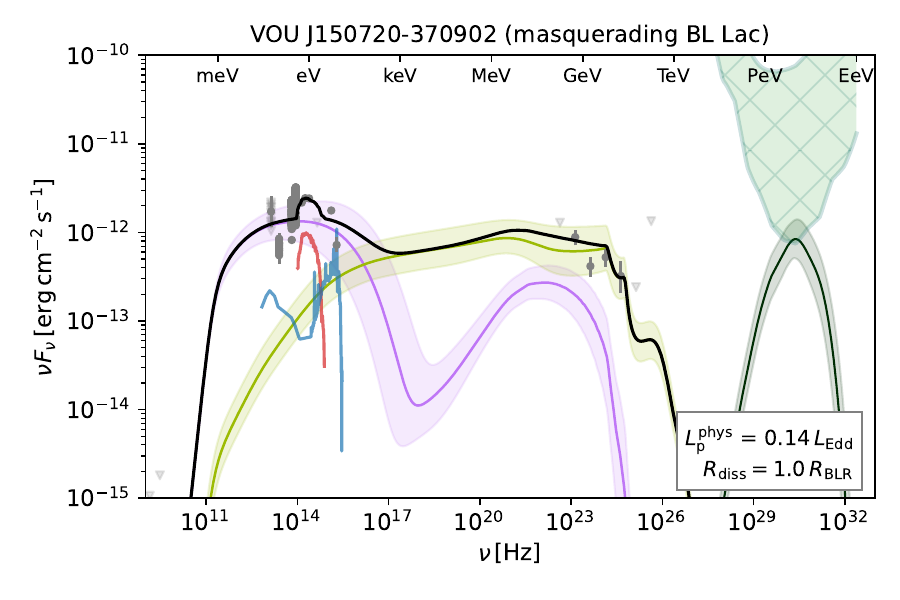}

\includegraphics[width=\sedsize,trim={4mm 5mm 4mm 4mm}, clip]{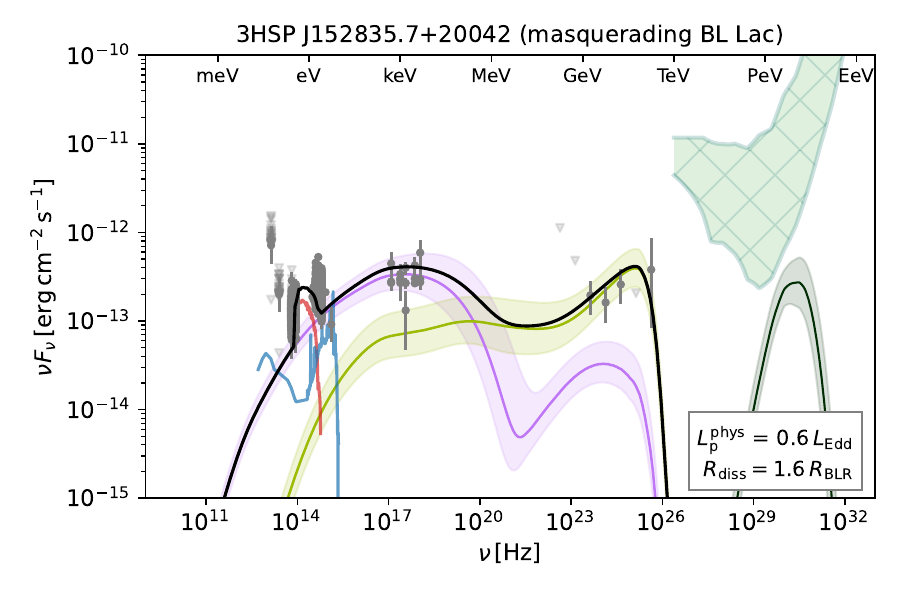}\includegraphics[width=\sedsize,trim={4mm 5mm 4mm 4mm}, clip]{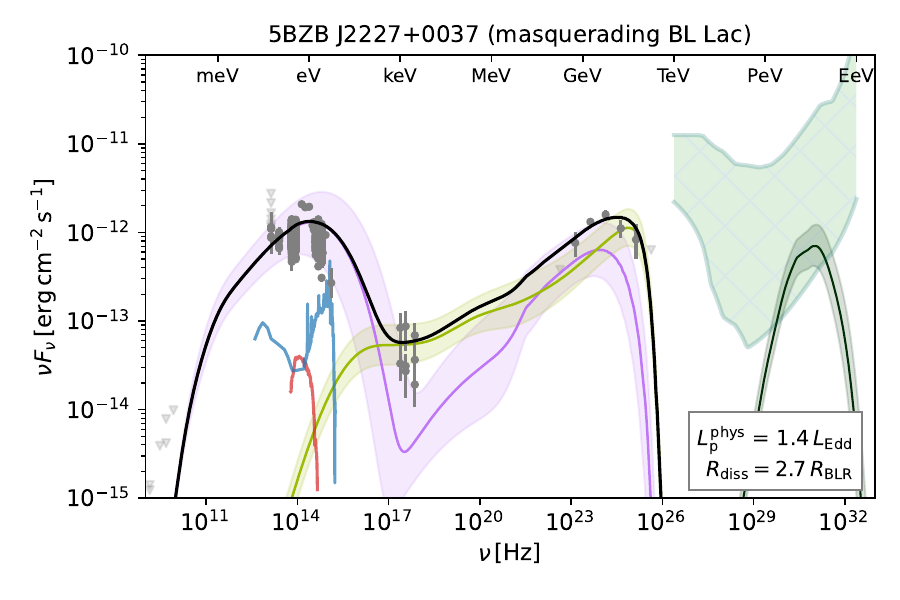}

\includegraphics[width=\sedsize,trim={4mm 5mm 4mm 4mm}, clip]{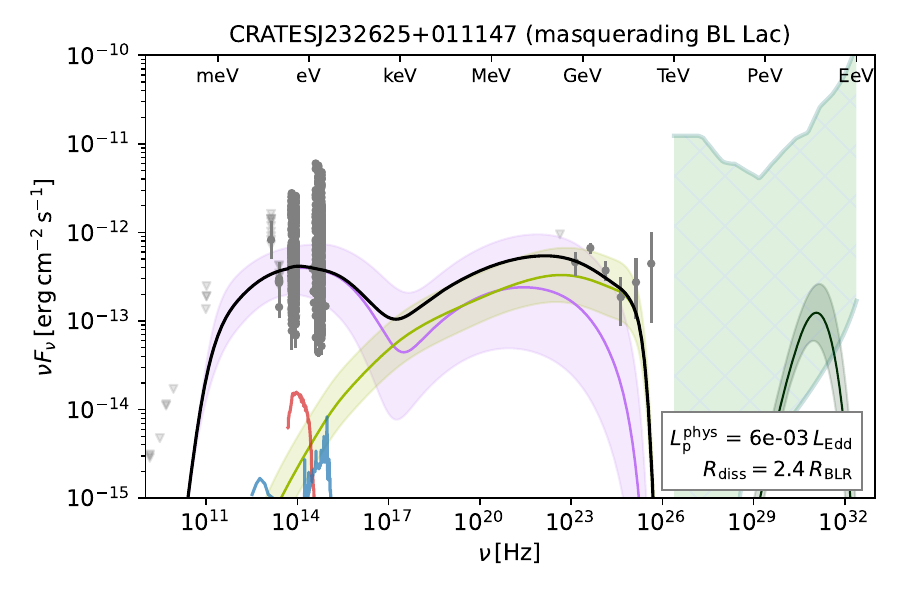}\includegraphics[width=\sedsize,trim={4mm 5mm 4mm 4mm}, clip]{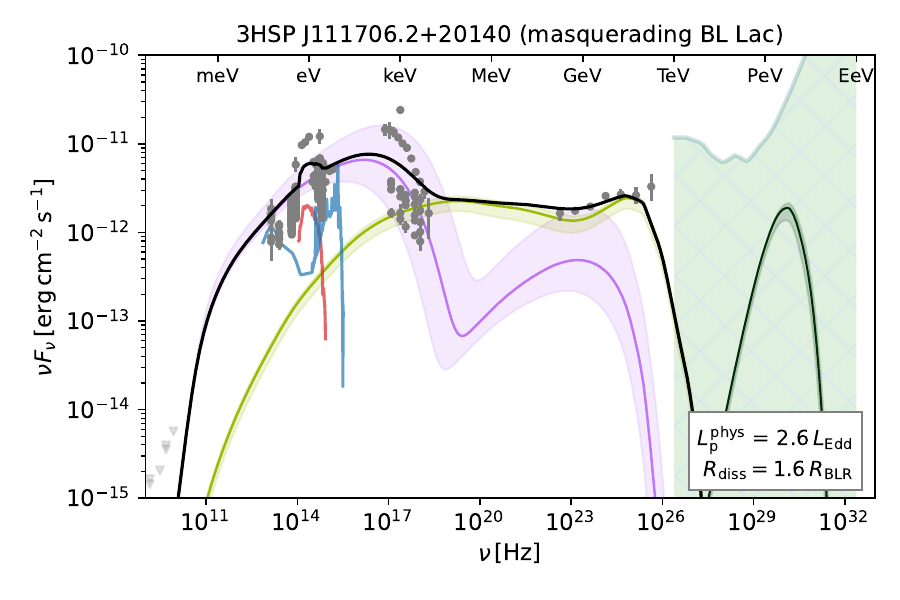}

\includegraphics[width=\sedsize,trim={4mm 5mm 4mm 4mm}, clip]{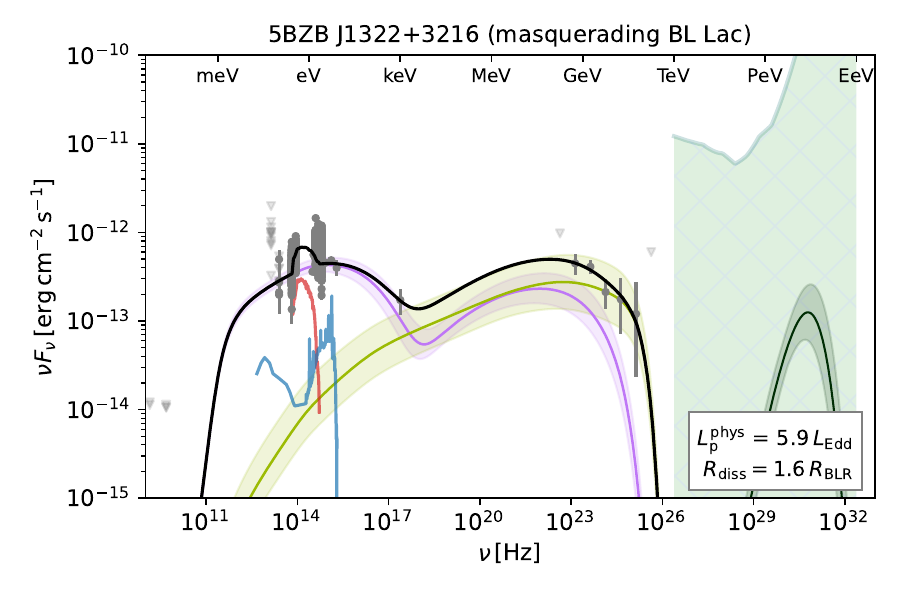}\includegraphics[width=\sedsize,trim={4mm 5mm 4mm 4mm}, clip]{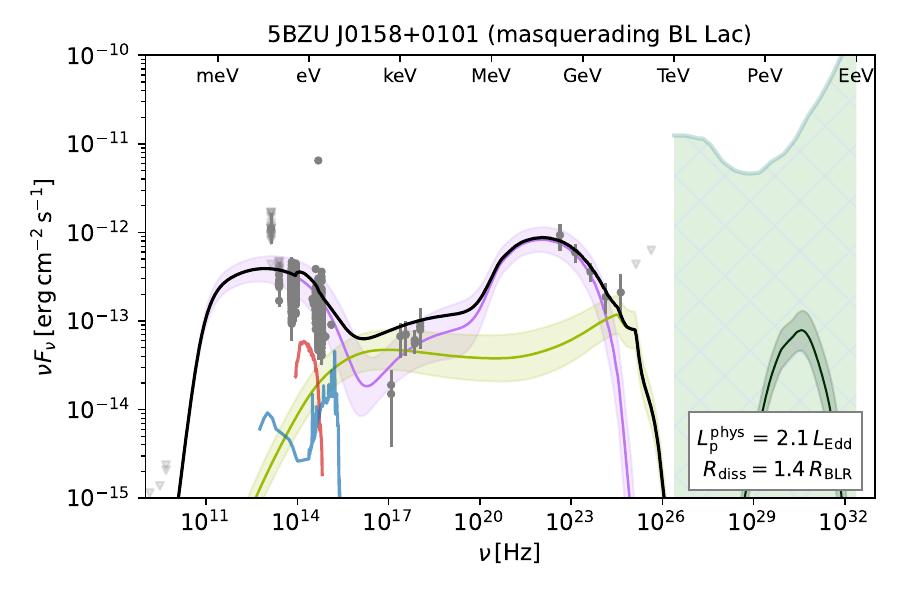}

\includegraphics[height=9mm,trim={0 0 0 0}, clip]{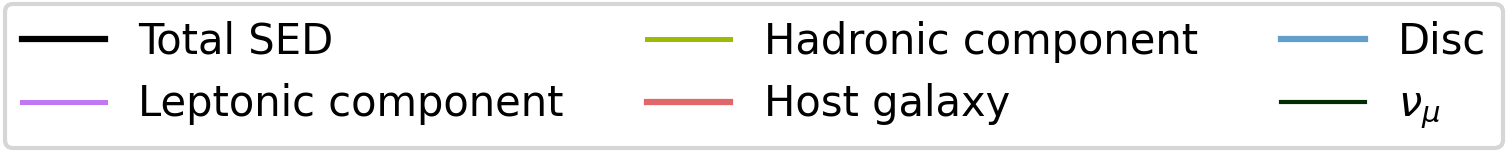}

\caption{Multiwavelength and neutrino emission from the G20 blazars previously identified as masquerading BL Lacs. We show the leptonic (purple) and hadronic (light green) components of the jet emission; the corresponding neutrino emission is shown in dark green. The error bands show the 1$\sigma$ uncertainty range. In red is the host contribution, and in blue the spectrum from the accretion disk, broad lines, and dust torus (cf.~\Fig\ref{fig:disc_and_host}). The green region represents the allowed energies and fluxes of the peak of the neutrino spectrum so as to satisfy the IceCube limits at the 68\% confidence level (cf.\Sec\ref{sec:icecube} and \App\ref{app:icecube}).}
\label{fig:all_masquerading}
\end{figure*}

\begin{figure*}[htpb!]\ContinuedFloat

\includegraphics[width=\sedsize,trim={4mm 5mm 4mm 4mm}, clip]{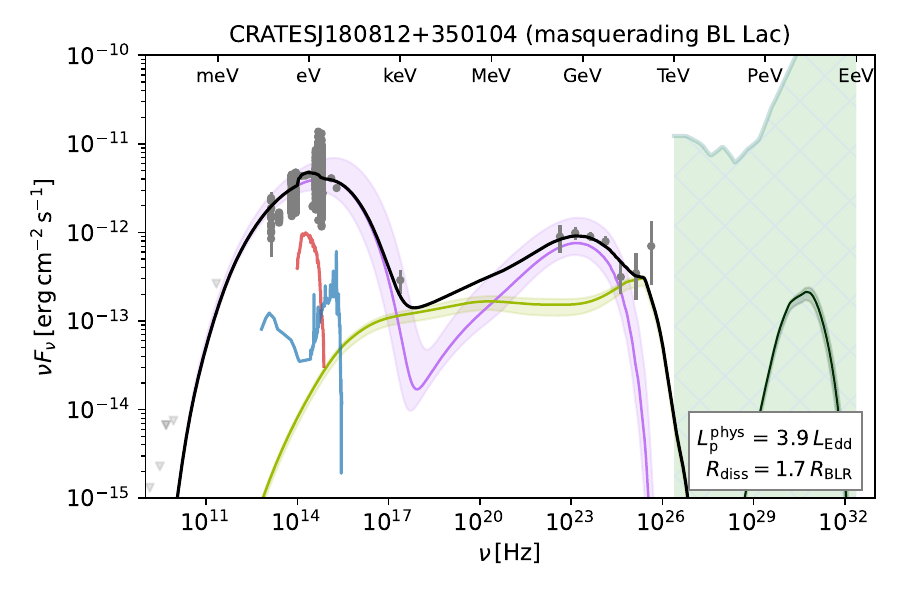}\includegraphics[width=\sedsize,trim={4mm 5mm 4mm 4mm}, clip]{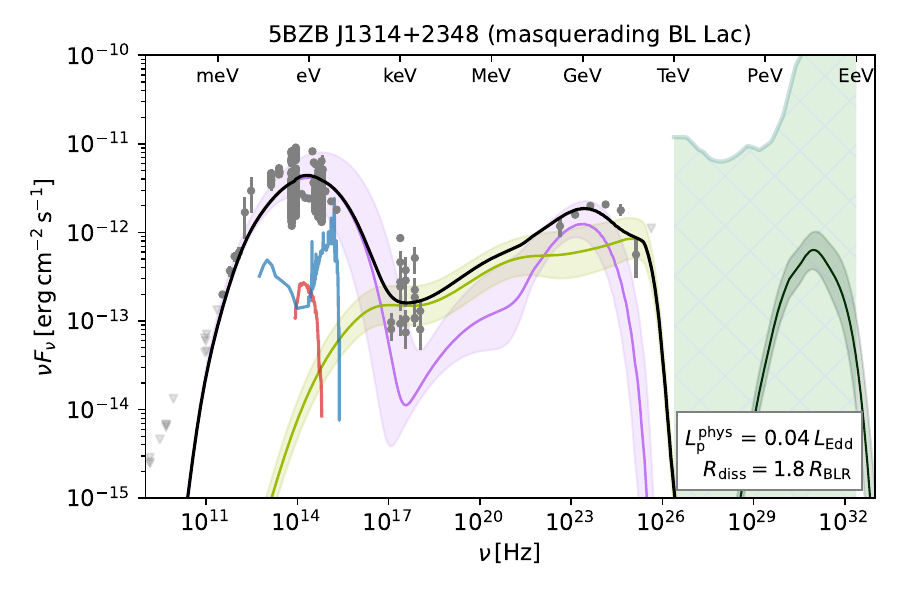}

\includegraphics[width=\sedsize,trim={4mm 5mm 0mm 4mm}, clip]{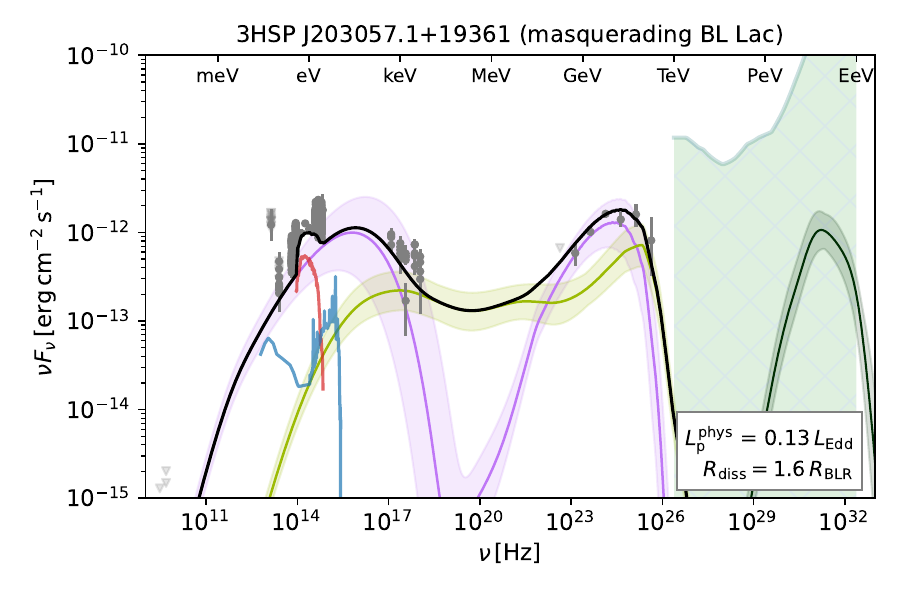}\includegraphics[width=3.6cm,trim={0 0 0 0}, clip]{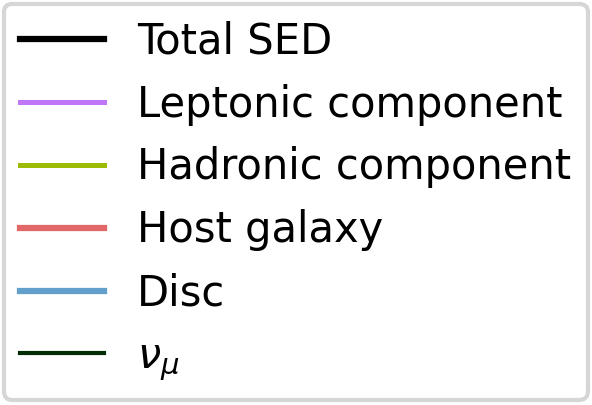}

\captionsetup{labelformat=empty}
\caption{Fig. 5. continued. Modeled multiwavelength emission and corresponding neutrino predictions for the blazars in the sample previously identified as masquerading BL Lacs.}
\end{figure*}

For the masquerading BL Lacs shown in the upper five panels of \Fig\ref{fig:all_masquerading}, we can see that IceCube data provide a significant lower limit on the point source flux at the 68\% confidence level. For the sources in the two upper rows, this is the case at all analyzed energies; in the case of source CRATESJ232625+011147 (third row, left panel), the data are compatible with zero below 100~TeV, but that energy range can generally be neglected in this model due to the low neutrino production efficiency. As we can see, in four out of these five cases, the peak of the predicted neutrino spectrum lies inside the green region, which means the model describes the IceCube points source data at the 68\% confidence level. In the case of blazar 3HSP~J152835.7+20042 (second row, left panel) the peak of the predicted spectrum lies below the allowed region, which means the model is in tension with the IceCube data. 

Taking as an example the case of TXS~0506+056 (upper left panel of \Fig\ref{fig:all_masquerading}), we can see that in this model, hadronic cascade emission dominates both the X-ray and \textit{Fermi}-LAT fluxes. In previous leptohadronic studies of this source, the $\gamma$-ray spectrum has generally been suggested to be dominated by inverse Compton emission, while the hadronic cascades contribute to the X-ray flux \citep[e.g.,][]{Cerruti:2018tmc,Gao:2018mnu,Keivani:2018rnh,Petropoulou:2019zqp}. As a consequence, in those models, the neutrino flux scales with the X-ray flux, and is therefore limited by observations in the X-ray range, where often lies the ``valley''  between the two broadband SED features. In contrast, in the solution shown here, the hadronic cascade peaks in the gigaelectronvolt range, and therefore the emitted flux in neutrinos is comparable to the LAT flux. At the same time, the $\gamma$-ray data fitted in this work is time-averaged, and therefore lower in flux compared to the 2017. In that sense, the present model suggests that a hadronic component may dominate the average $\gamma$-ray flux, even with sub-Eddington proton injection (cf.~\Sec\ref{sec:discussion}); but this conclusion does not necessarily apply to flaring states. In \Sec\ref{sec:discussion} we discuss in greater detail the differences between this and previous models of TXS~0506+056.

For the remaining six masquerading BL Lacs the lower limit on the IceCube point source flux is compatible with zero. In these cases, the model is not constrained by neutrino data, except for the upper limits on the IceCube flux, which are generally high compared to the LAT fluxes and therefore do not effectively constrain the model. In those cases, if the solution shows a LAT flux that is dominated by hadronic cascades, such as in the upper-left panel of \Fig\ref{fig:all_gray_area}, that is because the fit to the multiwavelength data are as good, or better, than alternative electron-dominated solutions within the parameter space region being probed. It is worth emphasizing that for a given power in accelerated electrons $L_\mathrm{e}^\prime$, we tested a minimum value of $L_\mathrm{p}^\prime>10\,L_\mathrm{e}^\prime$ (cf.~\Tab\ref{tab:steps}). As explained in \Sec\ref{sec:optimization}, this choice is made in order to avoid solutions where the SED is explained solely by electron emission. This is the reason why none of the best-fit SEDs have purely leptonic origin, even for sources where IceCube data do not constrain the minimum neutrino flux.

\begin{figure*}[htpb!]
\centering

\includegraphics[width=\textwidth,trim={0 1cm 0 0}, clip]{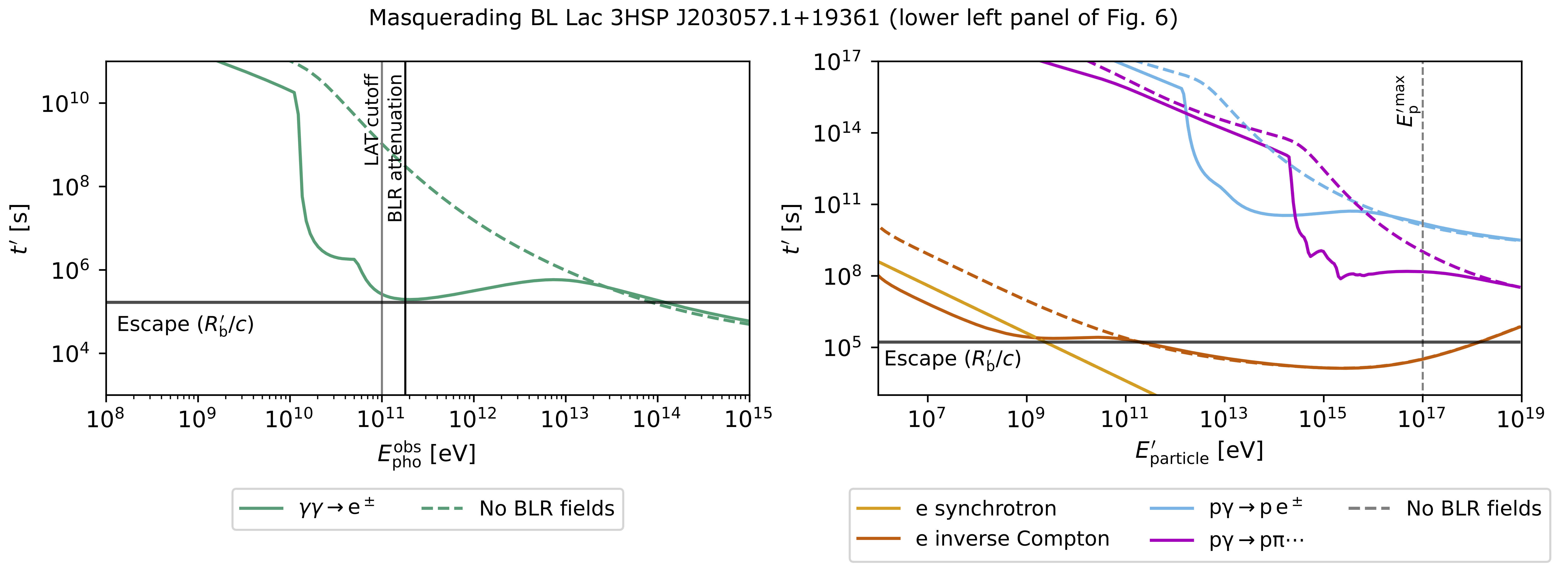}
\caption{Interaction timescales of photons \textit{(left)} and of electrons and protons \textit{(right)}, for one of the masquerading BL Lacs in the sample. The solid curves show the total interaction timescales in the best-fit scenario (cf.~\Fig{\ref{fig:all_masquerading}}). The two peaky features in the magenta curve result from resonant photo-meson interactions between protons and broad line photons, due to the proximity between the dissipation region and the BLR. The dashed curves show how the interaction rates would behave without the presence of external photon fields, as in the case of true BL Lacs.} 
\label{fig:timescales}
\end{figure*}

As indicated in the lower right corner of each panel, the best-fit value for the distance between the dissipation region in the jet and the supermassive black hole in masquerading BL Lacs is generally between one and three times the BLR radius. In the case of a source such as VOU~J150720-370902, shown in the upper right panel, the model favors a dissipation region located on the BLR, $R_\mathrm{diss}=R_\mathrm{BLR}$, which maximizes the density and the Doppler boost of BLR radiation in the jet frame. In this case, the resulting $\gamma$-ray spectrum predicted by the model contains a series of spectral features between ten and 100~GeV, due to enhanced photo-hadronic interactions as well as attenuation of the emitted $\gamma$ rays. Further investigation of this effect would require a dedicated analysis and modeling of time-selected $\gamma$-ray data, which lies beyond the scope of this work.

To better illustrate the effect of external fields in masquerading BL Lacs, we show in \Fig\ref{fig:timescales} the energy-dependent interaction timescales in source 3HSP~J203057.1+19361 (cf. respective panel in \Fig\ref{fig:all_masquerading}). The source has a best-fit value of $R_\mathrm{diss}/R_\mathrm{BLR}=1.6$, which means that the interaction zone is located relatively close to the BLR. In this case, the BLR photons can play a role as targets for photon annihilation, external Compton scattering, and hadronic interactions. In the left-hand plot we show as a solid curve  the interaction timescale for photon-photon annihilation in this source (given in the jet rest frame) as a function of the photon energy (given in the observer's frame). The dashed curve represents the annihilation timescale if the BLR photons were not present in the source, which would be the case if the dissipation region were located far from the BLR. We can see that for $\gamma$ rays between 10~GeV and 1~TeV, the presence of external photons, mainly from broad line emission, increases the photon annihilation efficiency by a factor as large as $10^3$. At about 200~GeV, the photon annihilation timescale matches that of the physical escape from the source, leading to a cutoff in the emitted photon spectrum. This energy roughly matches the cutoff observed in the LAT fluxes. If the interaction region were considerably closer to the BLR, the external photons would receive a larger relative boost, lowering the annihilation curve and therefore the cutoff frequency. In this sense, the fact that the LAT observes emission up to $\sim100$~GeV implies a lower limit on the distance between the interaction region and the BLR in this source. This constraint also applies to sources VOU~J150720-370902, 3HSP J152835.7+20042, and 5BZU~J0158+0101, where the BLR emission results in a well-defined cutoff in the $\gamma$-ray spectrum (shown in \Fig\ref{fig:all_masquerading} in the upper right panel, left panel of the second row, and lower right panel).

In the right panel of \Fig\ref{fig:timescales}, we show the electron and proton interaction timescales in the same source. The two peaks visible in the photo-meson timescale (solid magenta curve) are due to the helium and hydrogen Lyman $\alpha$ lines. By comparing this curve with the dashed one, we can see that external broad line photons play a major role as interaction targets for protons between 100~TeV and 1~EeV.  Above this energy, the interaction timescale remains approximately constant up to the best-fit maximum energy, $E_\mathrm{p}^{\prime\,\mathrm{max}}=100~\mathrm{PeV}$. 
On the contrary, in the case of true BL Lacs, photo-meson interactions are less effective at these energies because synchrotron photons are the only targets present, as represented by the dashed magenta curve. Regarding Bethe-Heitler pair production (cyan curves), we can see that this process is enhanced by the presence of BLR photons for proton energies between 1~TeV and 10~PeV. This leads to increased emission of $\gamma$ rays in the LAT range through synchrotron-supported cascades, as discussed above and shown in greater detail in \App\ref{app:components}.

\begin{figure*}[htpb!]

\includegraphics[width=\sedsize,trim={0 5mm 0 0}, clip]{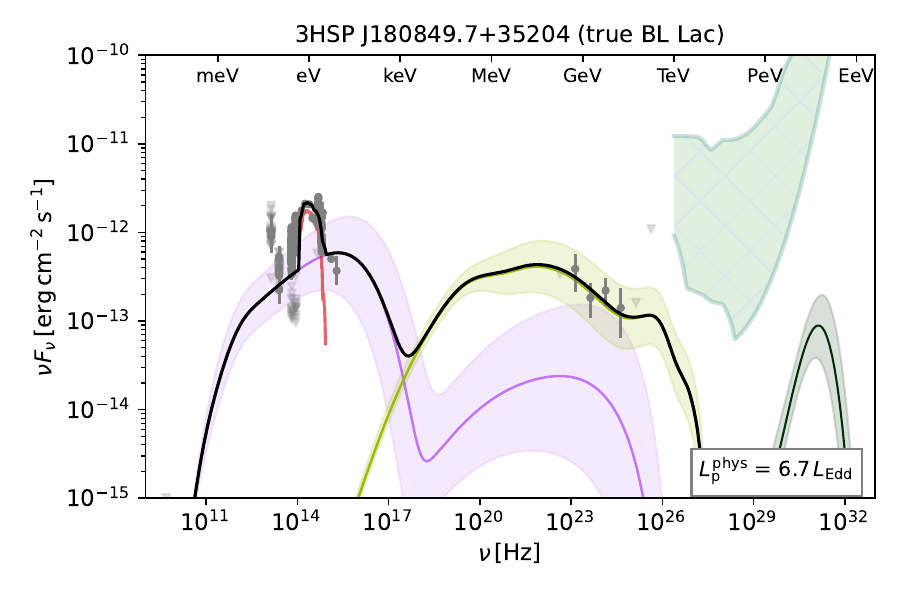}\includegraphics[width=\sedsize,trim={0 5mm 0 0}, clip]{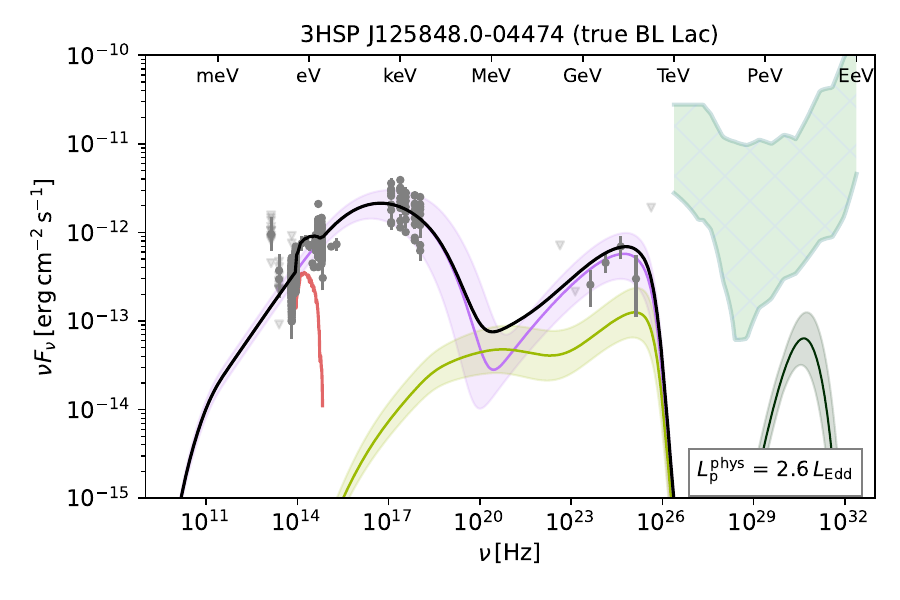}

\includegraphics[width=\sedsize,trim={0 5mm 0 0}, clip]{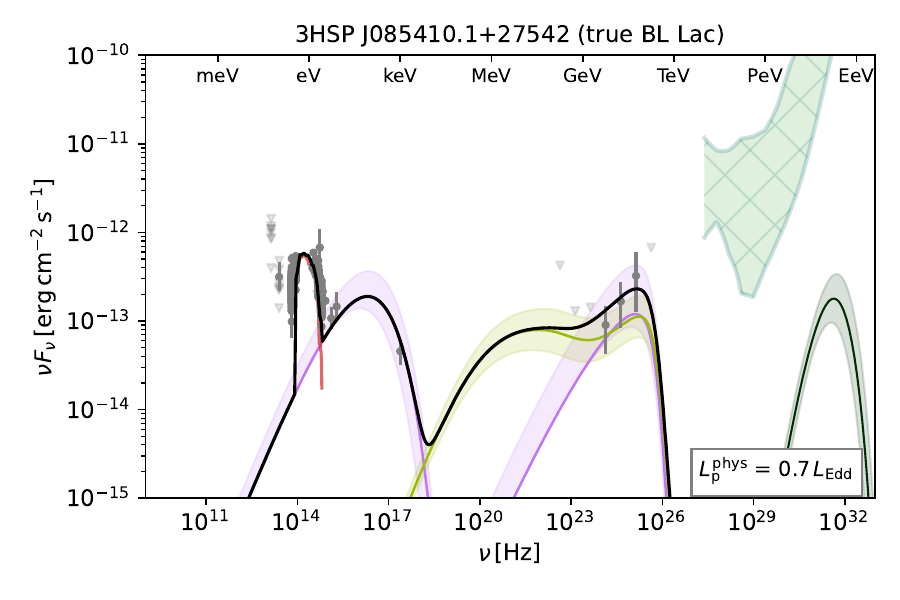}\includegraphics[width=\sedsize,trim={0 5mm 0 0}, clip]{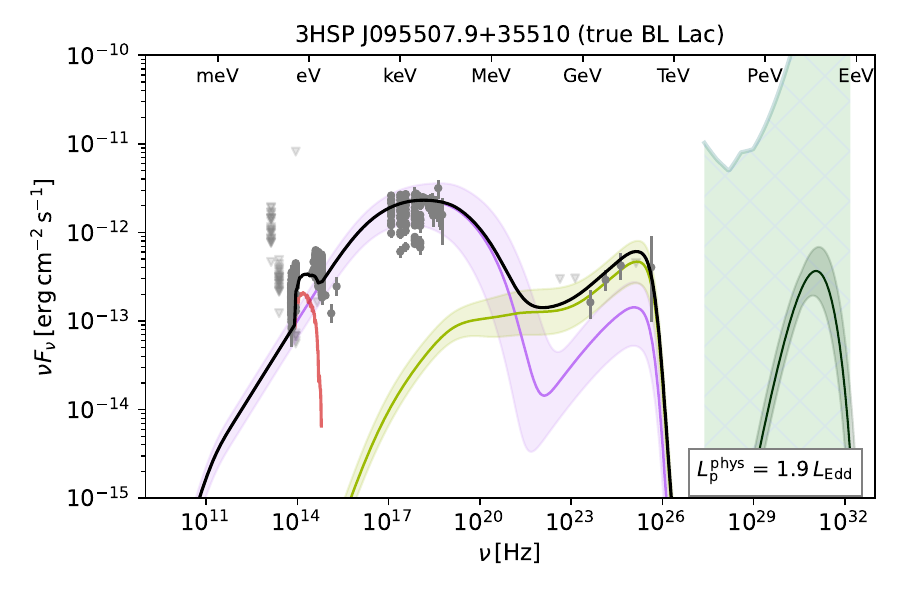}

\includegraphics[width=\sedsize,trim={0 5mm 0 0}, clip]{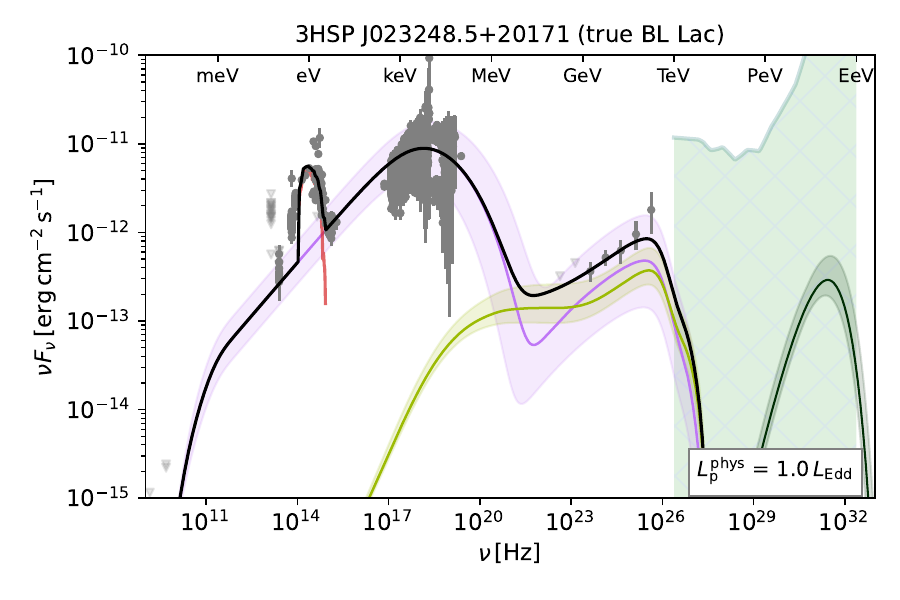}\includegraphics[width=\sedsize,trim={0 5mm 0 0}, clip]{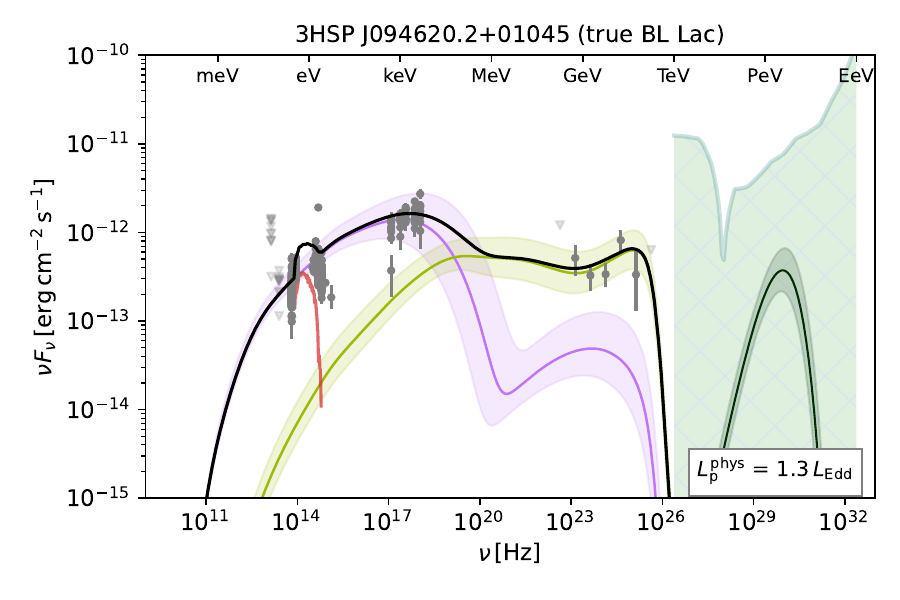}

\includegraphics[width=\sedsize,trim={0 5mm 0 0}, clip]{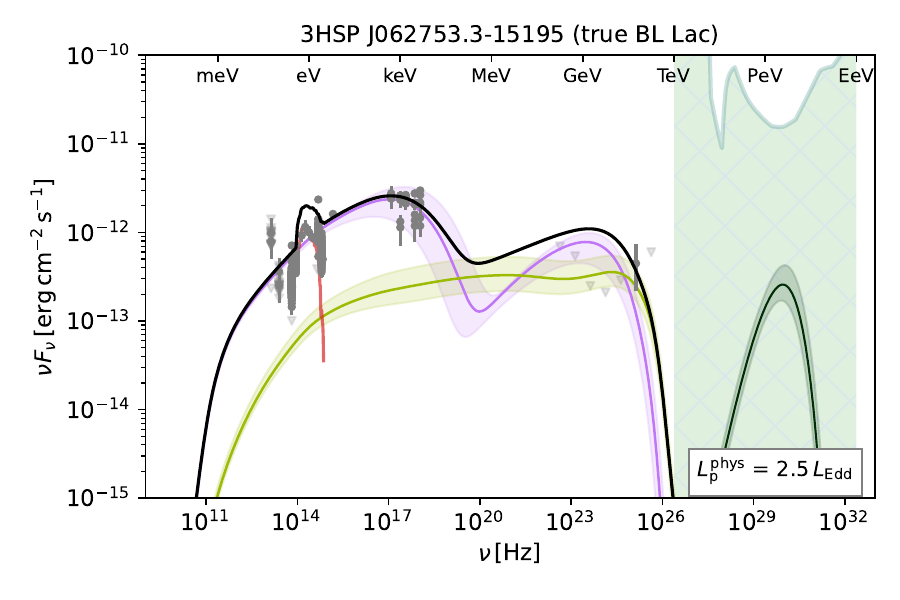}\hspace{1cm}\includegraphics[width=0.2\textwidth,trim={0 0 0 0}, clip]{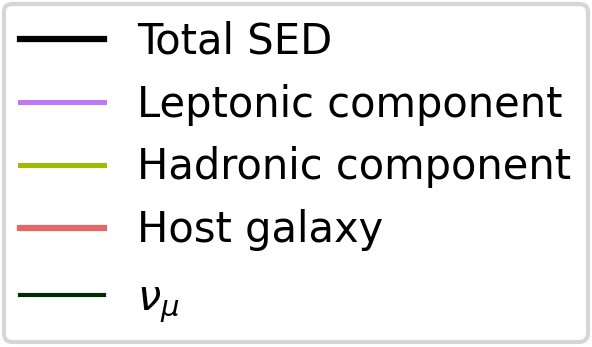}
\caption{Modeled multiwavelength emission and corresponding neutrino predictions for the blazars in the sample previously identified as true BL Lacs. The multiwavelength data are shown in black and the IceCube point source fluxes as a green band. In the cases where there is a lower limit on the neutrino flux (upper row), the model undershoots this lower limit, due to the absence of target photons from a BLR in these sources.} 
\label{fig:all_nonmasquerading}
\end{figure*}

We now move from the masquerading BL Lac objects to the true BL Lacs, for which we show in \Fig\ref{fig:all_nonmasquerading} the best-fit multimesssenger spectra. In the cases where the IceCube point source fluxes are incompatible with zero at the 68\% confidence level (top three sources), we see that it is more challenging to explain these fluxes than in the case of masquerading BL Lacs. Physically, this is due to the lack of external photon fields from a BLR, leading to sources that are extremely optically thin up to the highest proton energies. We can see that the best-fit neutrino spectrum peaks at 10-100~PeV, a regime where interactions between protons and synchrotron photons become efficient. This also leads to cascades that contribute significantly to the LAT flux. On the contrary, at the petaelectronvolt level photo-meson production is inefficient, making neutrino production less viable in these sources compared to masquerading BL Lacs, as can be seen by comparing the magenta curves in \Fig\ref{fig:timescales}.

Finally, we discuss the best fits for those blazars whose nature as masquerading or true BL Lacs is undetermined. Those results are shown in \Fig\ref{fig:all_gray_area}. In these cases, we allowed for the presence of a BLR. The $R_\mathrm{diss}$ parameter, which is optimized based on the multiwavelength and neutrino data, ultimately determines the level of contribution of the BLR fields to the interaction model. The value of this parameter in each source is shown in the respective SED  plot, as well as in \Tab\ref{tab:parameters}.

In the case of a source such as 3HSP~J125821.5+21235, shown in the left panel on the third row of \Fig\ref{fig:all_gray_area}, we can see that the model predicts a spectral break in the $\gamma$-ray spectrum at $\sim10$~GeV, due to interactions with the BLR photons. Observations of such a feature would favor the masquerading nature of this source. Since current LAT data
do not allow us to constrain a cutoff in the spectrum, it is challenging to draw a more definitive conclusion in this direction for the sources in this sample. 

\begin{figure*}[htpb!]
\centering
\includegraphics[width=\sedsize,trim={4mm 5mm 4mm 4mm}, clip]{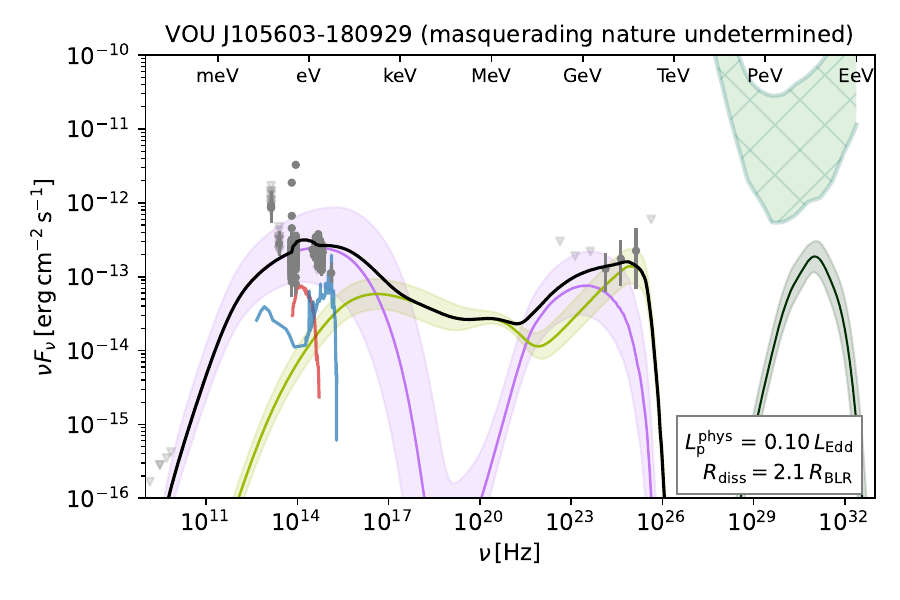}\includegraphics[width=\sedsize,trim={4mm 5mm 4mm 4mm}, clip]{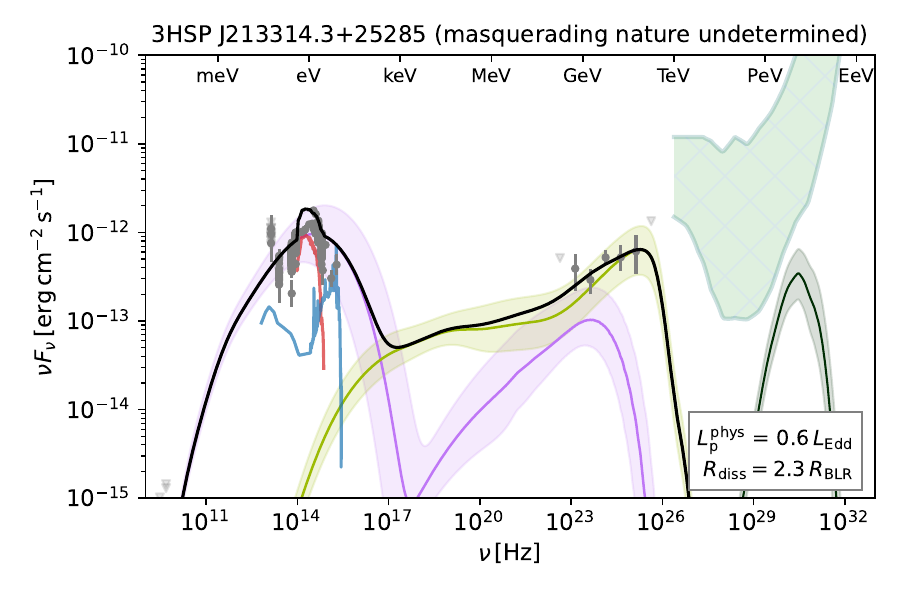}

\includegraphics[width=\sedsize,trim={4mm 5mm 4mm 4mm}, clip]{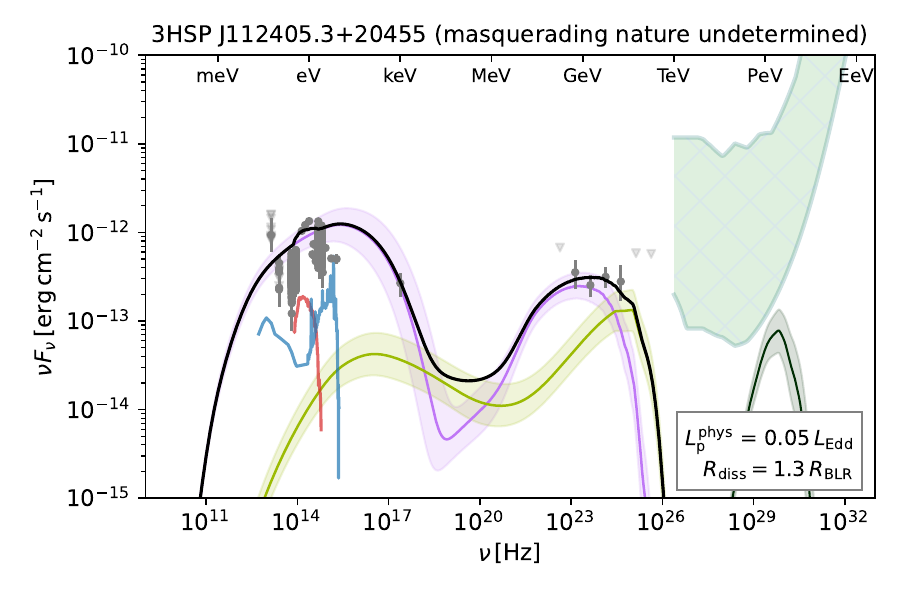}\includegraphics[width=\sedsize,trim={4mm 5mm 4mm 4mm}, clip]{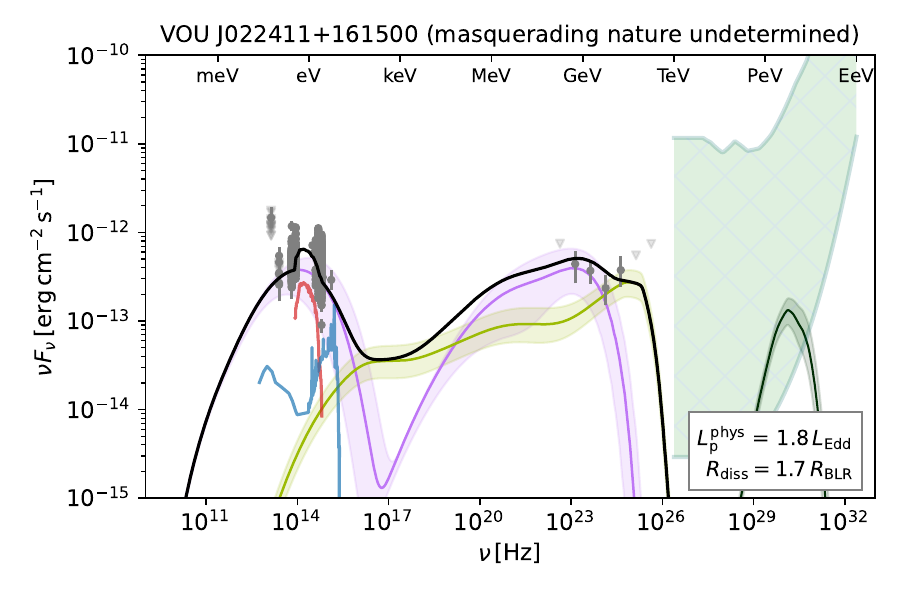}

\includegraphics[width=\sedsize,trim={4mm 5mm 4mm 4mm}, clip]{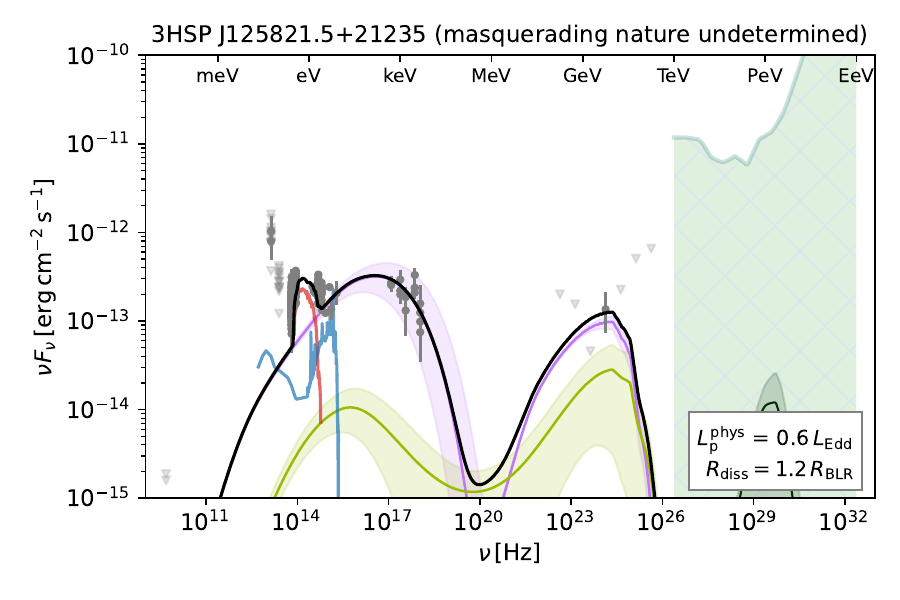}\includegraphics[width=\sedsize,trim={4mm 5mm 4mm 4mm}, clip]{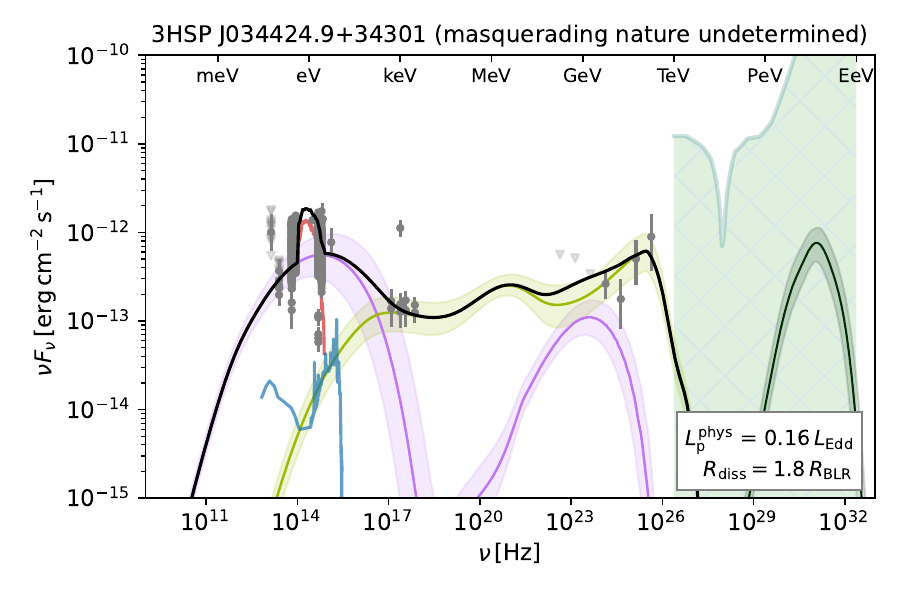}

\includegraphics[width=\sedsize,trim={4mm 5mm 4mm 4mm}, clip]{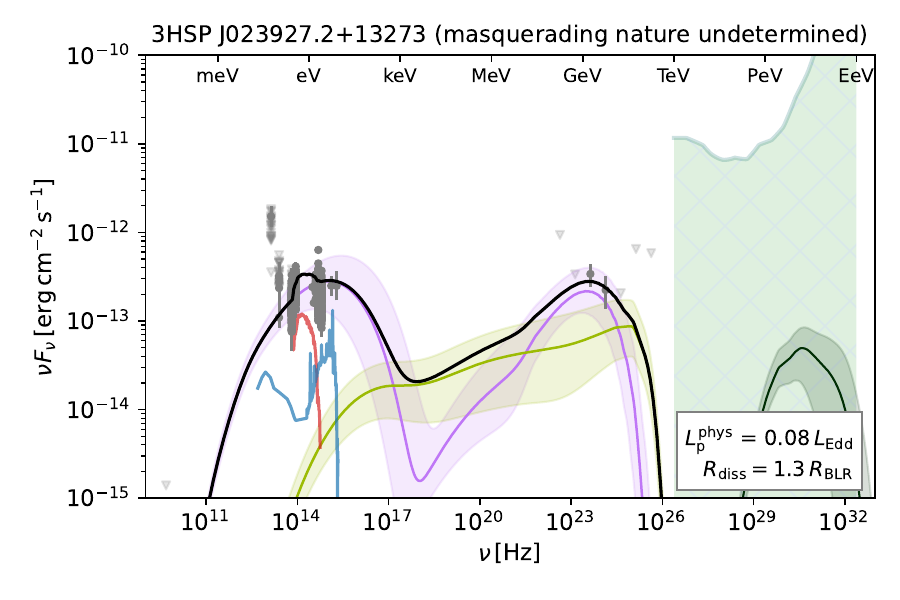}\includegraphics[width=\sedsize,trim={4mm 5mm 4mm 4mm}, clip]{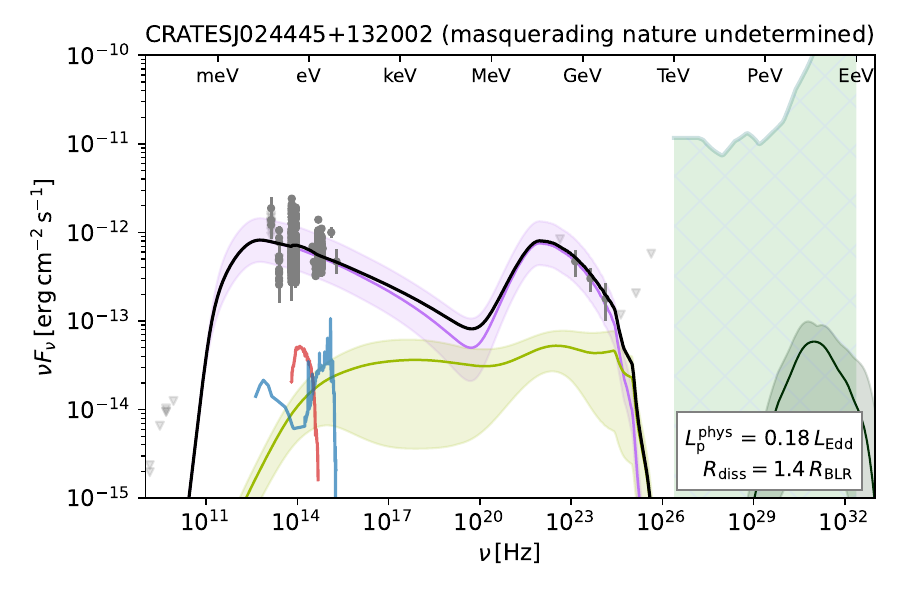}

\includegraphics[height=9mm,trim={0 0 0 0}, clip]{sed_legend_horizontal.png}

\caption{Modeled multiwavelength emission and corresponding neutrino predictions for the blazars in the sample whose nature as masquerading or true BL Lacs is undetermined. In these cases, we included the possibility of a BLR powered by an efficient accretion disk. The best-fit result can be similar to a masquerading BL Lac scenario if the production region lies close to the putative BLR, or to a true BL Lac scenario, if it is far from the BLR and no external photons are involved.} 
\label{fig:all_gray_area}
\end{figure*}

\begin{figure*}[htpb!]\ContinuedFloat
\centering
\includegraphics[width=\sedsize,trim={4mm 5mm 4mm 4mm}, clip]{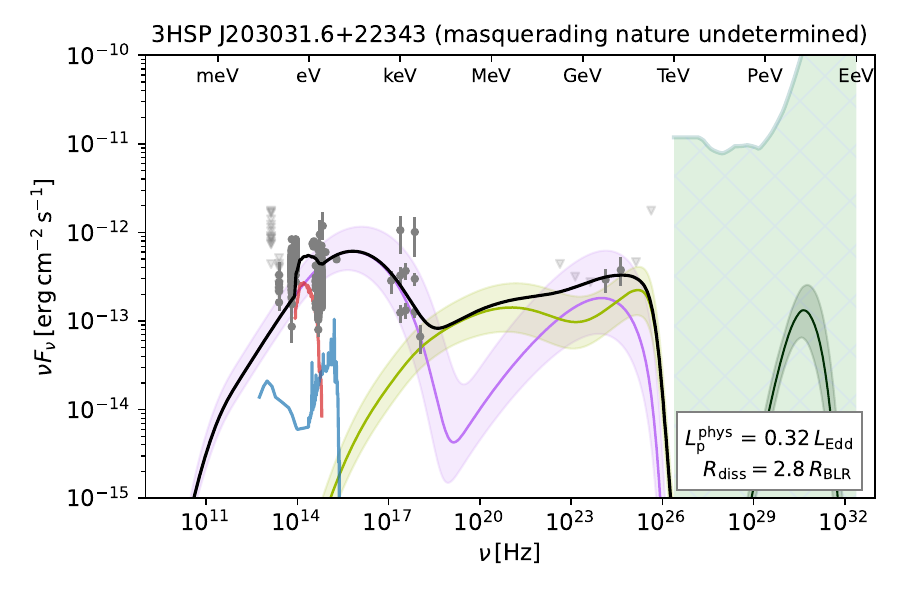}\includegraphics[width=\sedsize,trim={4mm 5mm 4mm 4mm}, clip]{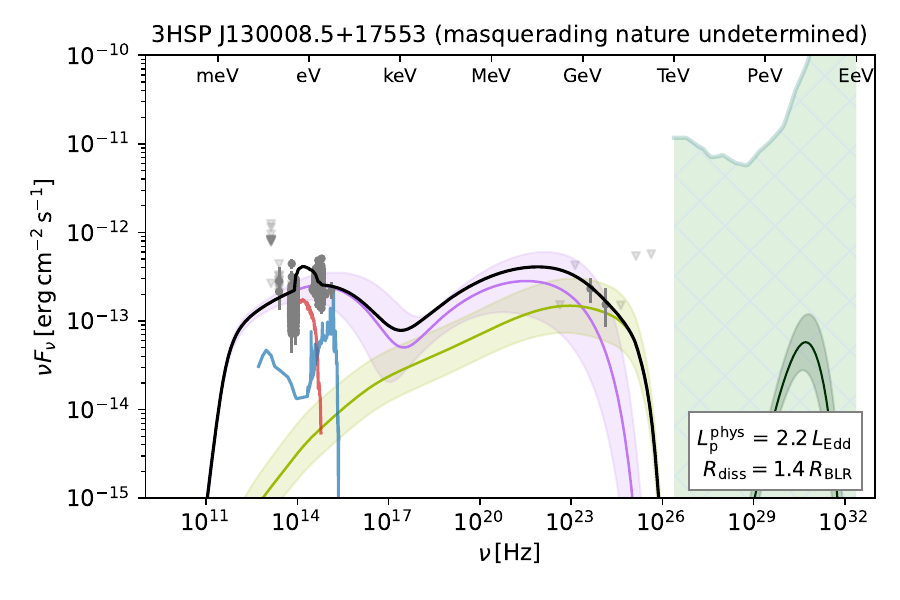}

\includegraphics[width=\sedsize,trim={4mm 5mm 4mm 4mm}, clip]{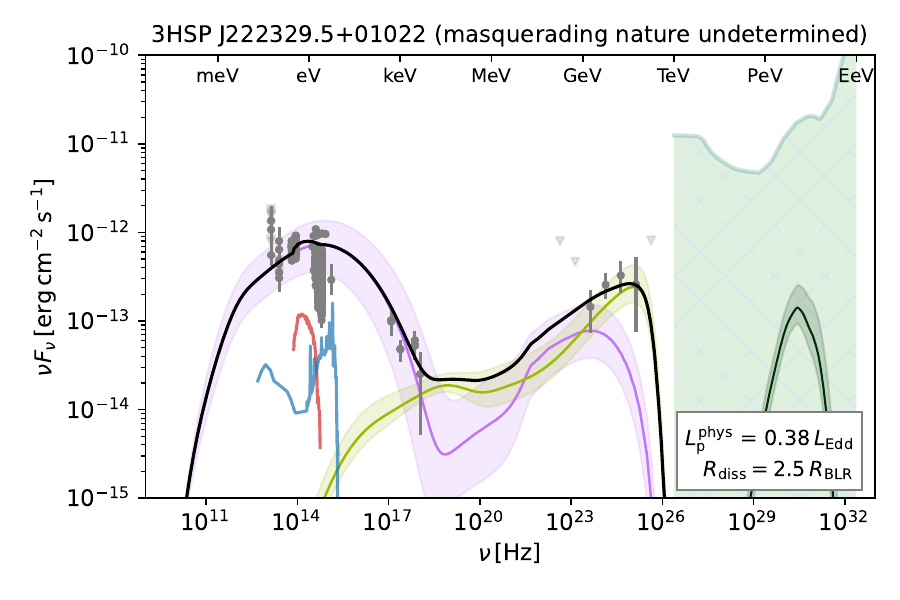}\includegraphics[width=\sedsize,trim={4mm 5mm 4mm 4mm}, clip]{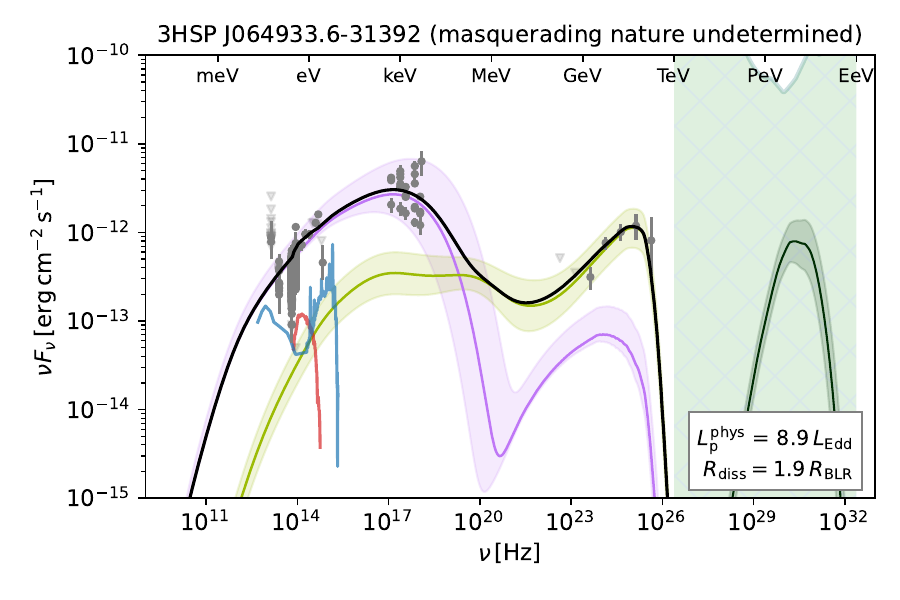}

\includegraphics[width=\sedsize,trim={4mm 5mm 4mm 4mm}, clip]{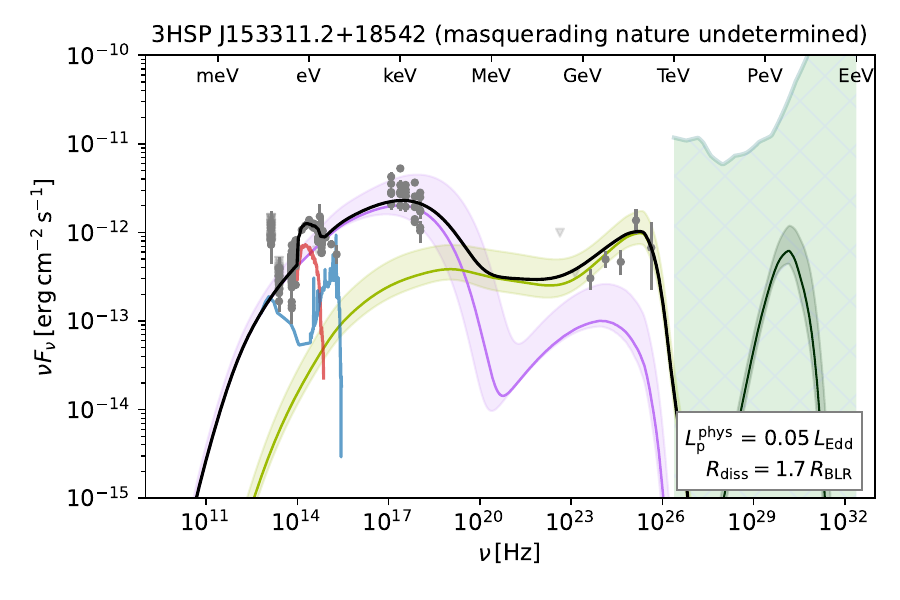}\includegraphics[width=\sedsize,trim={4mm 5mm 4mm 4mm}, clip]{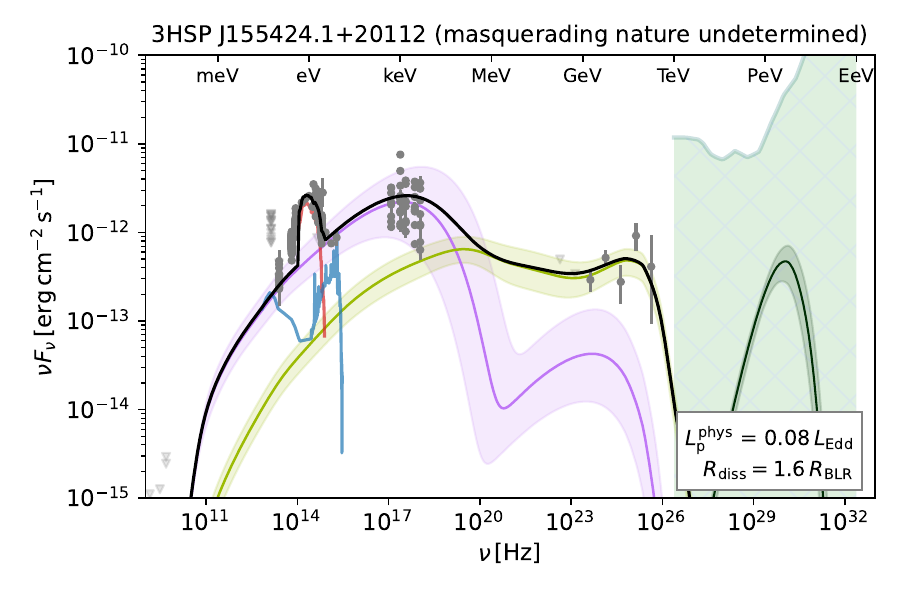}

\includegraphics[width=\sedsize,trim={4mm 5mm 4mm 4mm}, clip]{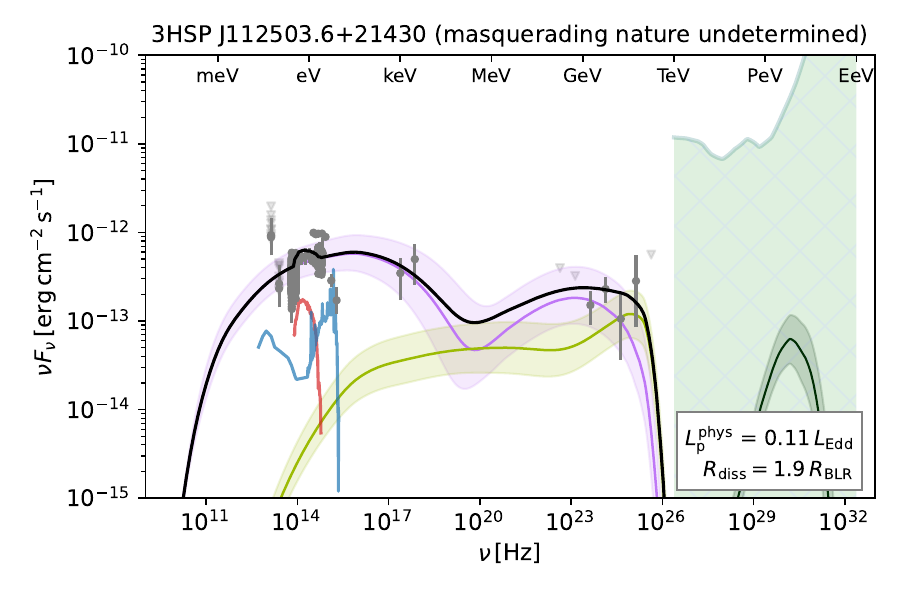}\includegraphics[width=\sedsize,trim={4mm 5mm 4mm 4mm}, clip]{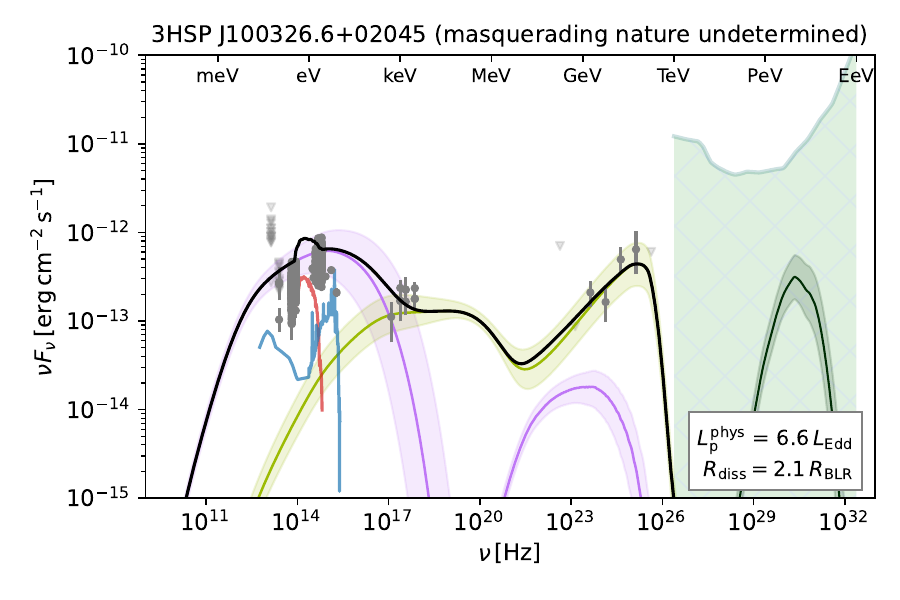}

\includegraphics[height=9mm,trim={0 0 0 0}, clip]{sed_legend_horizontal.png}

\captionsetup{labelformat=empty}
\caption{Fig. 8. continued. Modeled multiwavelength emission and corresponding neutrino predictions for the blazars in the sample whose nature as masquerading or true BL Lacs is undetermined.} 
\end{figure*}

\section{Discussion}
\label{sec:discussion}

Based on the individual simulation results, we now discuss the most prominent features of the model, place it in the context of previous leptohadronic studies, and analyze statistical trends within the sample. Starting with the predicted SED features, we can see that for the majority of the sources, the emission in the LAT range has a considerable contribution from hadronic interactions. More specifically, it consists of synchrotron emission by secondary photo-pairs produced either by protons through the so-called Bethe-Heitler process, or by the annihilation of very-high-energy hadronic $\gamma$ rays and low-frequency target photons. The contribution of these different processes to the overall fluxes is detailed for each source in \Fig\ref{fig:app_components_1} (\App\ref{app:components}). For two sources (TXS~0506+056 and 3HSP 180849.7+35204), the $\gamma$-ray flux in the LAT range is heavily dominated by synchrotron emission from Bethe-Heitler photo-pairs, as discussed in \App\ref{app:components}. This type of solution was recently shown by~\citet{Karavola:2024uog} to exist in certain regions of parameter space of IBLs. Our results support this prediction, given that both these sources are in fact IBLs.

For a few sources, the model also predicts a contribution from proton synchrotron emission to the $\gamma$-ray flux at 0.1-10~MeV (maroon curves in \Fig\ref{fig:app_components_1}). This feature results from the combination of a high proton energy and a moderate strength of the homogeneous magnetic field, and could in some cases be probed by proposed megaelectronvolt $\gamma$-ray missions such as ASTROGAM~\citep{e-ASTROGAM:2017pxr} or AMEGO-X~\citep{Fleischhack:2021mhc}. As shown with other models in the literature, in order for proton synchrotron to completely dominate the LAT fluxes, a magnetic field strength of   $B>10~\mathrm{G}$ is typically required~\citep[e.g.,][]{Muecke:2002bi, Cerruti:2014iwa,Petropoulou:2015swa,Liodakis:2020dvd,Rodrigues:2020fbu}. In comparison, the present model suggests an average value of $\langle B \rangle=2.6~\mathrm{G}$; this corresponds to an intermediate zone in parameter space between the proton synchrotron model and other leptohadronic models with typical proton energies $E_\mathrm{p}^{\prime\mathrm{max}}\lesssim$PeV (cf. literature reviewed in~\Sec\ref{sec:intro}). The latter class of model can also predict a feature at the 1-100~MeV range, but typically due to the Bethe-Heitler process rather than proton synchrotron emission \citep[e.g.,][]{Petropoulou:2015upa}.

To close the discussion on the source-by-source modeling, we now compare the result obtained for TXS~0506+056 with some of the abundant literature on this source. Following the association of the source with a high-energy IceCube event~\citep{icfermi}, leptohadronic models have typically described the system by evoking protons accelerated to maximum energies of up to a few teraelectronvolt, whose interactions generate electromagnetic cascades in the source. The bulk of the $\gamma$-ray flux in the LAT range is generally attributed to synchrotron self-Compton \citep[e.g.,][]{2018ApJ...863L..10A,2018ApJ...866..109S,Cerruti:2018tmc,Gao:2018mnu,Petropoulou:2019zqp,Oikonomou:2019djc} or external Compton emission~\citep[e.g.,][]{Keivani:2018rnh,Rodrigues:2018tku}. In the current model, as we can see in the upper left panel of \Fig\ref{fig:all_masquerading}, the time-averaged LAT SED is described exclusively by proton emission, a unique feature compared to other models of this source. As detailed in \Fig\ref{fig:app_components_1}, this emission comes specifically from Bethe-Heitler pair production, with an additional contribution from proton synchrotron below the gigaelectronvolt range. This is possible due to the high maximum energy of the accelerated protons, $E_\mathrm{p}^{\prime\mathrm{max}}\sim100~\mathrm{PeV}$. 

We can also compare the neutrino efficiency of TXS 0506-056 in this and other models. We resort for this to the $Y_{\nu\gamma}$ parameter, which is the ratio between
the all-flavor neutrino flux and the $\gamma$-ray flux in the LAT range. Based on simple energetic considerations, we know that in general, $Y_{\nu\gamma}\leq3$~\citep[e.g.,][]{Petropoulou:2014rla}. If we consider the models describing the 2017 neutrino event from TXS~0506+056 by \citet{Gao:2018mnu} and \citet{Keivani:2018rnh}, they predict $Y_{\nu\gamma}\approx 0.03$. In comparison, the present model predicts a value of $Y_{\nu\gamma}\approx 0.3$ for this source, a factor of ten larger. This is possible firstly due to the high proton energies in this model, as discussed above: we predict a neutrino spectrum peaking at $E_\nu=132.7\,\mathrm{PeV}$, while the two aforementioned models predict $\sim10\,\mathrm{PeV}$ and $\sim100\,\mathrm{TeV}$, respectively. At the same time, in this work we fitted a time-averaged $\gamma$-ray spectrum, which is a factor of 5 lower than that observed in 2017. This allows for a larger fraction of the $\gamma$-ray flux to be explained by hadronic cascades without overshooting X-ray observations.

It is also important to note two aspects regarding the X-ray data being fitted. Firstly, unlike $\gamma$ rays, the X-ray data do not have high cadence, but result from sporadic observations, often triggered within target-of-opportunity programs. This is likely to introduce a bias toward high fluxes in this frequency band, which propagates to the model results through the fitting procedure. A future solution to mitigate this bias could lie in time-domain data selection, accompanied by time-dependent modeling. This amounts to a more sophisticated procedure, challenged on the one hand by a the larger number of source model parameters and on the other hand by the need for simultaneous data in the infrared, optical, ultraviolet, and X-ray bands, which are likely to originate in the same particle population, as demonstrated by our results.

In the case of sources that have been monitored in the X-ray band, such as TXS~0506+056, the average fluxes considered are possibly more representative of the baseline emission; however, in these cases the data often display high flux variability. This translates into large error bars in this frequency band, which means the data are less constraining. In the specific example of TXS~0506+056, we can see in the upper left panel of \Fig\ref{fig:app_components_1} that the X-ray fluxes predicted by the model are slightly higher than the observed average, but still allowed within the large uncertainty range. This suggests that the proposed scenario, where X-ray and $\gamma$-ray fluxes have a strong contribution from hadronic cascades, may be put to a stricter test with a time-dependent data fitting method such as the one suggested above, which lies outside the scope of this work.

Finally, it is instructive to compare our prediction for blazar TXS 0506+056 with the study by \citet{Cerruti:2018tmc}, who described 2017 data with a model that suggests maximum proton energies between 60 PeV and 2.5~EeV. That model predicts a neutrino-to-$\gamma$-ray ratio of up to $Y_{\nu\gamma}\approx0.7$, more than twice than of the present model. Both predict a comparable level of $\gamma$-ray flux from hadronic interactions. The two main differences are \textit{1)} the model by \citet{Cerruti:2018tmc} predicts a higher inverse Compton emission from primary electrons, which is necessary in order to explain the enhanced $\gamma$-ray flux observed during the 2017 flare, and \textit{2)} the authors consider a higher power in relativistic protons, resulting also in a higher predicted neutrino flux.

It is interesting to note that in the cases where the IceCube analysis provides nonzero lower limits on the point source flux, as is the case of TXS~0506+056, those lower limits generally have their minimum value at around $\sim1$~PeV. This is a feature of the IceCube effective area, and it means that the requirement on the minimum neutrino flux is lowest when the predicted spectrum peaks at $\sim1$~PeV. Given that in the last optimization step we aimed at describing the IceCube constraints, it would be natural to expect solutions with a maximum proton energy of $E_\mathrm{p}^{\prime\,\mathrm{max}}\sim\mathrm{PeV}$, which would yield neutrinos with a similar energy in the observer's frame, $E_\mathrm{\nu}^{\mathrm{obs}}\sim2\,(\Gamma_\mathrm{b}/10)\,\mathrm{PeV}$, reducing the required proton luminosity and the corresponding cascade emission. However, in that case the cascade emission would peak at lower frequencies, overshooting more easily the observed X-ray fluxes. This would make the neutrino flux be limited by the X-ray observations, where the average flux is lower than in the LAT range. Higher proton energies, which lead to neutrino emission that peaks at tens to hundreds of petaelectronvolt, therefore help avert this constraint. On the other hand, these high proton energies also lead to high neutrino energies, which make it challenging to self-consistently explain the IceCube alert events, as discussed later in this section.

We now use the individual source results to infer some general properties of the sample. We start with the effect of the external fields on $\gamma$-ray attenuation. As listed in \Tab\ref{tab:parameters}, in this model the dissipation region is located on the outside the BLR, at a distance to the central black hole of approximately 1-3$\,R_\mathrm{BLR}$. The picture of a $\gamma$-ray-emitting region lying outside the BLR is generally consistent with previous results. For example,~\citet{Costamante:2018anp} analyzed the LAT spectra of a large FSRQ sample and concluded that 2/3 of the sources were optically thin to $\gamma$-ray attenuation, $\tau_\mathrm{\gamma\gamma}^\mathrm{max}<1$, and only 10\% were very optically thick ($\tau_\mathrm{\gamma\gamma}^\mathrm{max}>5$). Our best-fit results are consistent with these numbers: considering $\gamma$ rays up to 300~GeV in the observer's frame, we find $\tau_\mathrm{\gamma\gamma}^\mathrm{max}<1$ for 59\% of the sources where a BLR may be present (i.e., excluding true BL Lacs), with 11\% of these having $\tau_\mathrm{\gamma\gamma}^\mathrm{max}>5$. For the remaining sources, the maximum optical thickness is of order unity, as in the example shown in the left panel of \Fig\ref{fig:timescales}. For those sources, the resulting SED can display in some cases a spectral break due to BLR attenuation (e.g. right panels on the first and fourth rows of \Fig\ref{fig:all_masquerading}), but no cutoff is predicted in the LAT range, in agreement with the data.

\begin{figure*}[htpb!]
\includegraphics[width=\textwidth,trim={0 0 0 0}, clip]{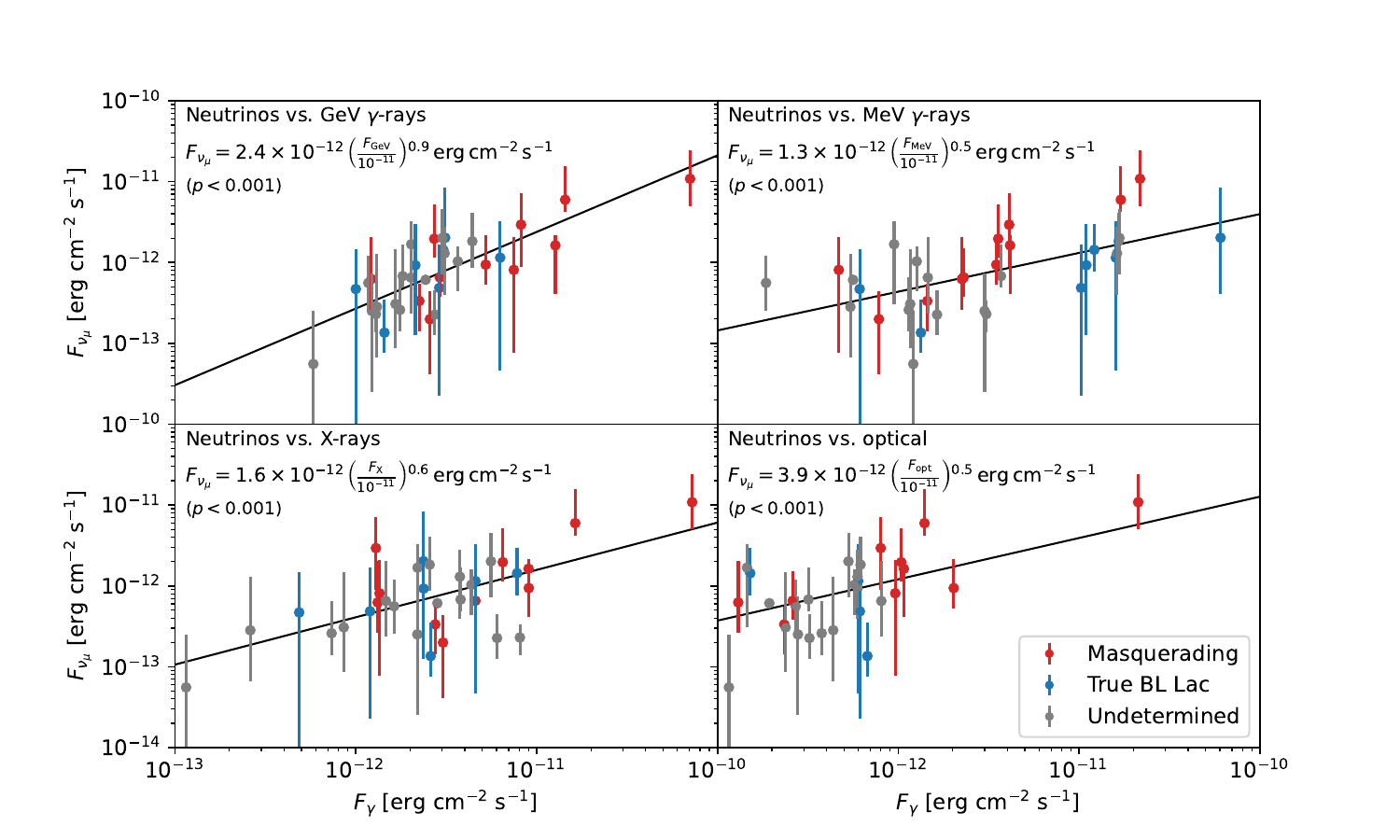}
\caption{Empirical relations between the neutrino and photon fluxes derived from the model results. We show the total predicted muon neutrino flux from each source in the sample, plotted against the total photon flux in the range 0.1-$100\,\mathrm{GeV}$ \textit{(upper left)}, 0.1-$10\,\mathrm{MeV}$ \textit{(upper right)}, 0.1-$100\,\mathrm{keV}$ \textit{(lower left)}, and in the optical range, 380-$750\,\mathrm{nm}$ \textit{(lower right)}. The best-fit power-law relations are shown as black lines, and the corresponding relations are given in the upper right. Below, we report the p-value of the corresponding Pearson correlation test.}
\label{fig:flux_relations}
\end{figure*}

We can also derive relationships between the photon flux in different wavelength bands and the predicted neutrino emission. In   \Fig\ref{fig:flux_relations} we show the total muon neutrino flux predicted by the model as a function of the total flux in the LAT range (upper left), megaelectronvolt $\gamma$ rays (upper right), X-rays (lower left) and optical (lower right). As we can see, the neutrino flux displays the strongest scaling with $\gamma$ rays in the LAT range, with a best-fit power-law relation of $F_{\nu_\mu}\sim F_\mathrm{GeV}^{1.1}$. 
The neutrino flux also correlates significantly with the flux in megaelectronvolt $\gamma$ rays, X-rays, and optical, but with a softer power-law index than for gigaelectronvolt $\gamma$ rays (0.5, 0.6, and 0.5, respectively).

The reason for the strong, almost linear correlation between neutrino and LAT fluxes is two-fold: \textit{1)} the high neutrino fluxes derived from the IceCube point source data favor solutions that maximize neutrino production. As shown in \Fig\ref{fig:neutrino_peak_vs_pl}, this leads to a stronger constraint than in previous literature, given this is the first study to self-consistently consider the neutrino spectral shape in the calculation of the point source flux.
In a scenario where neutrino production efficiency is maximized, the LAT fluxes tend to be dominated by hadronic emission, leading to the observed correlation; \textit{2)} the high maximum proton energies lead to secondary synchrotron emission that peaks in the LAT range rather than in the X-ray band. In contrast, previous models of IceCube blazar candidates typically assume proton energies in the sub-PeV to $\sim1$~PeV range, leading to hadronic cascades with a bright X-ray component that is more severely limited by observations (see e.g., the literature on TXS~0506+056 referenced in \Sec\ref{sec:intro}).

We can quantify this using $Y_{\nu\gamma}$, the ratio between the neutrino and $\gamma$-ray flux, as defined previously. From the results in the upper-left panel of \Fig\ref{fig:flux_relations}, and accounting for a factor of three between the observed flux in muon neutrinos and in all-flavor neutrinos, we can derive an average value of $\langle Y_{\nu\gamma}\rangle\approx0.8$ for the G20 sample. This is in agreement with the results by \citet{Petropoulou:2015upa}, who modeled six other IHBLs. This consistency reflects the fact that in both models the high-energy emission has a strong contribution from hadronic cascades, with predicted neutrino spectra that peak above the petaelectronvolt range. Using the results by \citet{Petropoulou:2015upa} and their extrapolation to the blazar population by \citet{Padovani:2015mba}, \citet{2016PhRvL.117x1101A} showed that the neutrino-to-$\gamma$-ray ratio is in fact limited to $Y_{\nu\gamma}<0.13$, based on the IceCube limits above 10~PeV at the time. The present model, like the model by \citet{Petropoulou:2015upa}, overshoots this limit by a factor of 6. This implies that sources in our sample have a value of $Y_{\nu\gamma}$ that is necessarily higher than the average BL Lac object, meaning they are exceptional neutrino emitters. This is particularly the case for the masquerading BL Lacs in the sample, including TXS~0506+056. Three considerations are in order when interpreting this number: \textit{1)} blazars in the G20 sample have been selected based on associations with high-energy IceCube events. It is therefore natural to expect that this sample has an over-representation of neutrino-efficient IHBLs that do not reflect the characteristics of the general population; \textit{2)} at the individual source level, the model attempts to describe neutrino flux limits estimated at a 68\% confidence level. This suggests that the lower bounds on the neutrino flux might be overestimated in certain instances, resulting in an overestimation of the modeled neutrino flux from those sources; \textit{3)} the relatively low sample size leads to large uncertainties when extrapolating the neutrino spectrum to the entire BL Lac population. This is particularly important given the wide range of values of neutrino peak energy ($1\lesssim E_\nu\lesssim100~\mathrm{PeV}$) and neutrino-to-$\gamma$-ray ratio ($0.1<Y_{\nu\gamma}<2.2$) predicted by the model.

It is important to note that even in the optimistic scenario where BL Lac objects have the maximally allowed value of $Y_{\nu\gamma}=0.13$ suggested by \citet{2016PhRvL.117x1101A}, they should still undershoot the IceCube diffuse flux in this model, because the predicted BL Lac flux peaks at higher energies~\citep[cf. e.g., \Fig1 of][]{Padovani:2015mba}. This underlines the idea that the IceCube diffuse flux is in fact dominated by another, or multiple other, source populations. As we know given the current constraints, such sources should either be numerous and with low intrinsic luminosity~\citep[cf. e.g.,][]{Palladino:2018lov}, or their emission must be obscured in the \textit{Fermi}-LAT band~\citep[][]{Murase:2015xka,Fang:2022trf}.

As also argued recently by \citet{Padovani:2024tgx}, jetted AGN may in fact produce a considerable diffuse flux, but peaking above the IceCube energy range. 
Our results support this scenario, given that the predicted neutrino spectra peak consistently above the petaelectronvolt range. Non-jetted AGN, on the other hand, could contribute a flux below the energy range of the IceCube diffuse flux. This flux was estimated by \citet{Padovani:2024tgx} by exploiting our knowledge of the AGN X-ray luminosity function and evolution and taking the source NGC 1068 as a benchmark for the neutrino emission level \citep{doi:10.1126/science.abg3395f}.

\begin{figure*}[htpb!]
\includegraphics[width=\textwidth,trim={0 0 0 0}, clip]{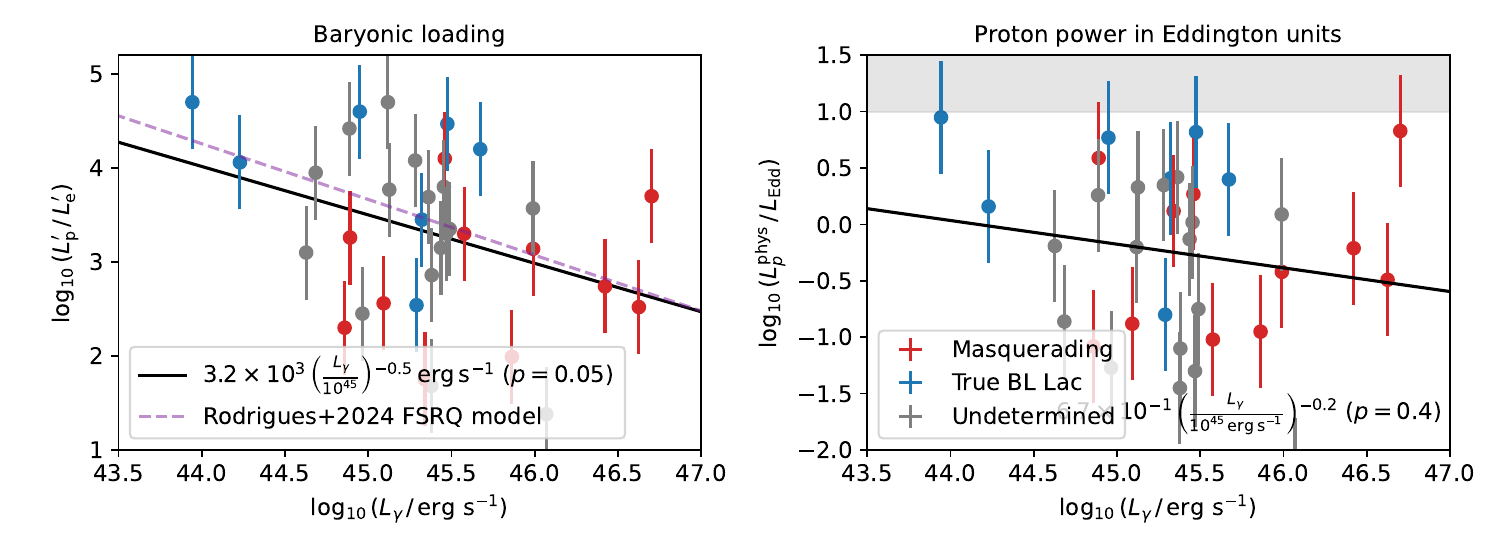}
\caption{Best-fit proton power, given on the left in terms of the electron power (baryonic loading), and on the right in terms of the Eddington luminosity of the supermassive black hole. The black lines show the empirical relations obtained with a Pearson test, revealing a slightly negative correlation at a low significance level ($p>0.05$ in both cases). As a point of comparison, on the left panel we show as a dashed purple line the relation obtained by \citet{Rodrigues:2023vbv} for a sample of FSRQs using a qualitatively similar BLR model.}

\label{fig:baryonic_loading}
\end{figure*}

We can now quantify the power in accelerated protons required by the model to simultaneously explain the multiwavelength and neutrino data across the sample. In \Fig\ref{fig:baryonic_loading} we show in the left panel the baryonic loading, defined as the proton-to-electron luminosity ratio, as a function of the $\gamma$-ray luminosity of the source in the LAT energy range. As shown in the figure, a Pearson correlation test shows a weak negative correlation between the baryonic loading with the $\gamma$-ray luminosity, albeit at a low significance level (solid black line). The fact that all sources have a baryonic loading $L^\prime_\mathrm{p}/L^\prime_\mathrm{e}>10$ results directly from the parameter search criteria, as discussed in \Sec\ref{sec:optimization}, in order to avoid leptonic solutions. For most sources, the baryonic loading lies above $10^3$ and can be as high as  $5\times10^4$. This shows that a physical scenario involving a proton-dominated jet is necessary to explain the neutrino fluxes derived here from public IceCube data. When comparing with a leptohadronic model of FSRQs \citep[][dashed pink line]{Rodrigues:2023vbv}, we see that the predicted baryonic loading ranges are roughly compatible, a consequence of the similar nature of the source geometry underlying the two models.

We can then compare the required proton power to the Eddington luminosity of each source, calculated using the black hole mass estimates from Papers II and III. In the right panel of \Fig\ref{fig:baryonic_loading} we show the proton physical luminosity\textsuperscript{\ref{footnote:physical_luminosity}} in units of the source's Eddington luminosity. For about half the sources in the sample, the best-fit proton power is sub-Eddington, down to a few percent of $L_\mathrm{Edd}$. For the remaining blazars, the model requires super-Eddington proton powers. On the one hand, this level of proton injection may be challenging to maintain over long periods of time; on the other hand, the proton power never exceeds $10\,L_\mathrm{Edd}$, making the model more energetically viable compared to other leptohadronic frameworks where the proton injection power can exceed the Eddington limit by multiple orders of magnitude~\citep[e.g.,][]{Gao:2018mnu,Keivani:2018rnh,Petropoulou:2019zqp,Liodakis:2020dvd,Rodrigues:2020fbu,Rodrigues:2023vbv}.
This feature arises from the constraint imposed at the optimization stage (\Sec\ref{sec:optimization}), which ensures that the proton injection is not arbitrarily large compared to the Eddington limit.  

By comparing the IceCube alert energies, listed in \Tab\ref{tab:sample}, with the predicted peak energies in \Tab\ref{tab:parameters}, we see that their ranges do not overlap, with the former up to hundreds of teraelectronvolt and the latter starting from a few petaelectronvolt. This is because we did not include the alert energy in the model fitting process, but focused only on the IceCube point source constraints. As we discussed above, to obey those constraints, this model requires high maximum proton energies, leading to a high neutrino flux that peaks above $\sim1$~PeV and to hadronic cascades that peak in the LAT range. The model's predictions can potentially be tested with future neutrino experiments targeting energies above the current IceCube sensitivity range, such as IceCube-Gen2~\citep{2021JPhG...48f0501A}, GRAND~\citep{2020SCPMA..6319501A}, and RNO-G~\citep{2021JInst..16P3025A}.

Based on this, it is clear that the steady-state multimesssenger spectra predicted by this model cannot be directly used to interpret the IceCube alert associated with the respective source, which poses a potential limitation of the model. On the other hand, using simple energetic considerations, it is easy to argue that the high proton energies obtained here are, in fact, more in line with expectations than sub-PeV energies. Let us use the example of TXS~0506+056 and consider a generic acceleration mechanism with a timescale given by $t^\prime_\mathrm{acc}=\eta E^\prime/(eB^\prime c^2)$, where $\eta\geq10$~\citep{2002PhRvD..66b3005A} is a parameter inversely proportional to the acceleration efficiency. In a self-consistent scenario, the maximum energy of the electrons would result from balancing this acceleration timescale with the timescale of the leading energy loss mechanism, in this case synchrotron, $t^\prime_\mathrm{syn}=m_\mathrm{e}c^4/(E^\prime_\mathrm{e}B^{\prime2})$. Because  the jet is generally optically thin to proton interactions, the maximum proton energy would result from  balancing acceleration and escape, which in this model is purely advective, $t^\prime_\mathrm{esc}=R^\prime_\mathrm{b}/c$. Using the best-fit parameters for TXS~0506+056 (\Tab\ref{tab:parameters}), these two equilibrium relations would be written for electrons as $E_\mathrm{e}^{\prime\mathrm{max}}=5\,\mathrm{GeV}(B^\prime/{2~\mathrm{G}})^{0.5}(\eta/20)^{-1}$ and for protons as $E_\mathrm{p}^{\prime\mathrm{max}}=100\,\mathrm{PeV}\,(ct^\prime_\mathrm{esc}/10^{16.5}\mathrm{cm})(\eta/200)^{-1}$. We conclude that in our best-fit scenario, the value of $\eta$ for electrons and protons differs by only a factor of ten. In contrast, a scenario that predicts neutrino energies below the petaelectronvolt implies extremely low acceleration efficiencies for protons compared to electrons. For example, a maximum proton energy of $E_\mathrm{p}^{\prime\mathrm{max}}=390\,\mathrm{TeV}$ ~\citep[e.g.,][]{Gao:2018mnu}, implies a value of  $\eta\approx6\times10^4$ for proton acceleration. Although such scenarios can more naturally explain the association with a sub-PeV IceCube neutrino event, they require a radically different acceleration mechanism for electrons and protons, which may be challenging to explain in a one-zone framework.

\section{Conclusion}
\label{sec:conclusion}

We have presented a theoretical framework for leptohadronic interactions in IHBLs based on numerical, time-dependent single-zone modeling. We applied the model to a sample of 34 sources spatially associated with high-energy IceCube events. The model was shown to self-consistently describe the available public multiwavelength data. The model can describe the IceCube flux constraints for a large fraction of masquerading BL Lacs, but not for any of the true BL Lacs.

The neutrino flux from each source was  constrained using public IceCube point source data. To derive these limits, we employed for the first time a neutrino spectrum typical of leptohadronic models as the assumption for the signal shape, instead of assuming a power law as had been done in the literature to date. For 12 out of the 34 IHBLs in the sample, this analysis suggests a nonzero minimum neutrino point source flux at the 68\% confidence level. Out of these 12 IHBLs, five are masquerading, three are true BL Lacs, and four are of an undetermined nature. For the masquerading BL Lacs, the model can describe a neutrino flux compatible with 68\% of contours in four out of the five cases; for the three true BL Lacs, the model cannot explain the IceCube data. Physically speaking, this distinction is due to the strong broad line emission surrounding the central engine in masquerading BL Lacs, which is not present in true BL Lacs, and which provides a rich target for the production of neutrinos at and above the petaelectronvolt range. This is the first theoretical result from a systematic blazar sample study that supports masquerading BL Lacs as efficient neutrino sources.

Taking into consideration the results for the entire sample, we see that the predicted steady-state neutrino flux scales quasi-linearly with the photon flux in the LAT range. It also shows a significant positive correlation with the observed flux in the optical and X-ray bands, and the predicted flux in megaelectronvolt $\gamma$ rays. The average flux ratio predicted between neutrinos and $\gamma$ rays in the LAT range is $Y_{\nu\gamma}=0.8$, in agreement with previous leptohadronic models. Because the present sample was selected based on neutrino associations, it is likely that it over-represents neutrino-bright sources; these results therefore should not be directly extrapolated to characterize neutrino emission from the IHBL population at large. However, the model supports the IHBL sample by \citet{Giommi:2020hbx} as promising neutrino emitters, and suggests which sources have the highest potential as neutrino emitters (cf.~\Tab\ref{tab:parameters}, where the sources have been ordered in descending order of predicted neutrino flux). At the top of the ranking is masquerading BL Lac TXS 0506+056.

A novel feature of the model is the fact that the required proton power does not exceed the Eddington limit by a factor larger than ten. The main reason is that the maximum energy of the protons lies considerably above the petaelectronvolt range in the best-fit scenario, a regime where neutrino production efficiency is large. In contrast, leptohadronic models where the proton energies are limited to the sub-petaelectronvolt range often require proton luminosities in excess of the Eddington limit by several orders of magnitude in order to explain high a neutrino flux.

Another defining feature of the model is the prediction of neutrino spectra that peak above petaelectronvolt, up to $100~\mathrm{PeV}$ in the case of source TXS~0506+056. Although in masquerading BL Lacs broad line photons can provide abundant targets for neutrino emission in the petaelectronvolt and sub-petaelectronvolt range, at those energies the corresponding cascade emission typically leads to high X-ray fluxes, a result that is often excluded by observations. In the present model, the secondary emission tends to peak at higher energies, often up to the gigaelectronvolt range, which means that the extension of hadronically triggered cascades is not limited by X-ray observations, but rather by the LAT flux level. This allows the model to describe a larger neutrino flux that can explain the IceCube point source data. 

A potential limitation lies in the fact that the predicted neutrino peak energy lies above the range of reconstructed energies of the IceCube alert events associated with the sample because this information was not included in the model optimization process. In this context, it is important to note that our model describes steady-state blazar emission, given that the model is fitted to time-independent multiwavelength fluxes and time-integrated IceCube point source data. It is, therefore, conceivable in this scenario that a single $\sim100~\mathrm{TeV}$ event detected in a time span of over a decade may have been produced during a transient state of the source, during which parameters such as the proton luminosity and maximum energy would potentially differ compared to our best-fit values. This hypothesis can be verified by testing the model on multiwavelength data simultaneous with each alert event, an approach that is beyond the scope of this work. It is also worth noting that the reconstructed alert event energies result from an IceCube analysis that has the underlying assumption of a power-law signal \citep{abbasi2023icecat1}. In contrast, this model suggests that the steady-state neutrino flux from IHBLs should peak above the petaelectronvolt, and be approximately flat in the range from tera- to petaelectronvolt, where IceCube is most sensitive. Considering this model-based signal might potentially affect the reconstruction of the alert energies. This is particularly relevant for through-going events, for which the original neutrino energy is less constrained.

The hadronic framework underlying our IHBL model can be tested with continued multiwavelength blazar monitoring, more sensitive neutrino instrumentation above the petaelectronvolt regime, and theory-driven stacking searches. These efforts are necessary to shed further light on the neutrino associations with the G20 sample, the connection between individual detections and steady-state neutrino emission from blazars, and ultimately the nature of AGN as cosmic accelerators.

\begin{acknowledgements}
We thank Aldo Treves for valuable comments on the manuscript and Martin Wolf for his assistance with the open-source framework SkyLLH. This work is supported by the Deutsche Forschungsgemeinschaft (DFG, German Research Foundation) through grant SFB 1258 ``Neutrinos and Dark Matter in Astro- and Particle Physics'' and by the Excellence Cluster ORIGINS which is funded by the DFG under Germany's Excellence Strategy - EXC 2094 - 390783311. X.R. also acknowledges support by Institut Pascal at Université Paris-Saclay during the Paris-Saclay Astroparticle Symposium 2023, with the support of the P2IO Laboratory of Excellence (program ``Investissements d’avenir'' ANR-11-IDEX-0003-01 Paris-Saclay and ANR-10-LABX-0038), the P2I axis of the Graduate School of Physics of Université Paris-Saclay, as well as IJCLab, CEA, IAS, OSUPS, and the IN2P3 master project UCMN. M.P. acknowledges support from the Hellenic Foundation for Research and Innovation (H.F.R.I.) under the ``2nd call for H.F.R.I. Research Projects to support Faculty members and Researchers'' through the project UNTRAPHOB (Project ID 3013). 

\end{acknowledgements}



\begin{appendix}

\section{IceCube point source analysis}
\label{app:icecube}

We now describe in greater detail the method mentioned in \Sec\ref{sec:icecube} for the analysis of the IceCube data. This novel method allows us to estimate the point source signal contribution to the data while taking into account the specific spectral shape expected in a one-zone leptohadronic framework.

\subsection{Method description}

\begin{figure}[htpb!]
\begin{minipage}{0.5\textwidth}
    \includegraphics[width=\textwidth]{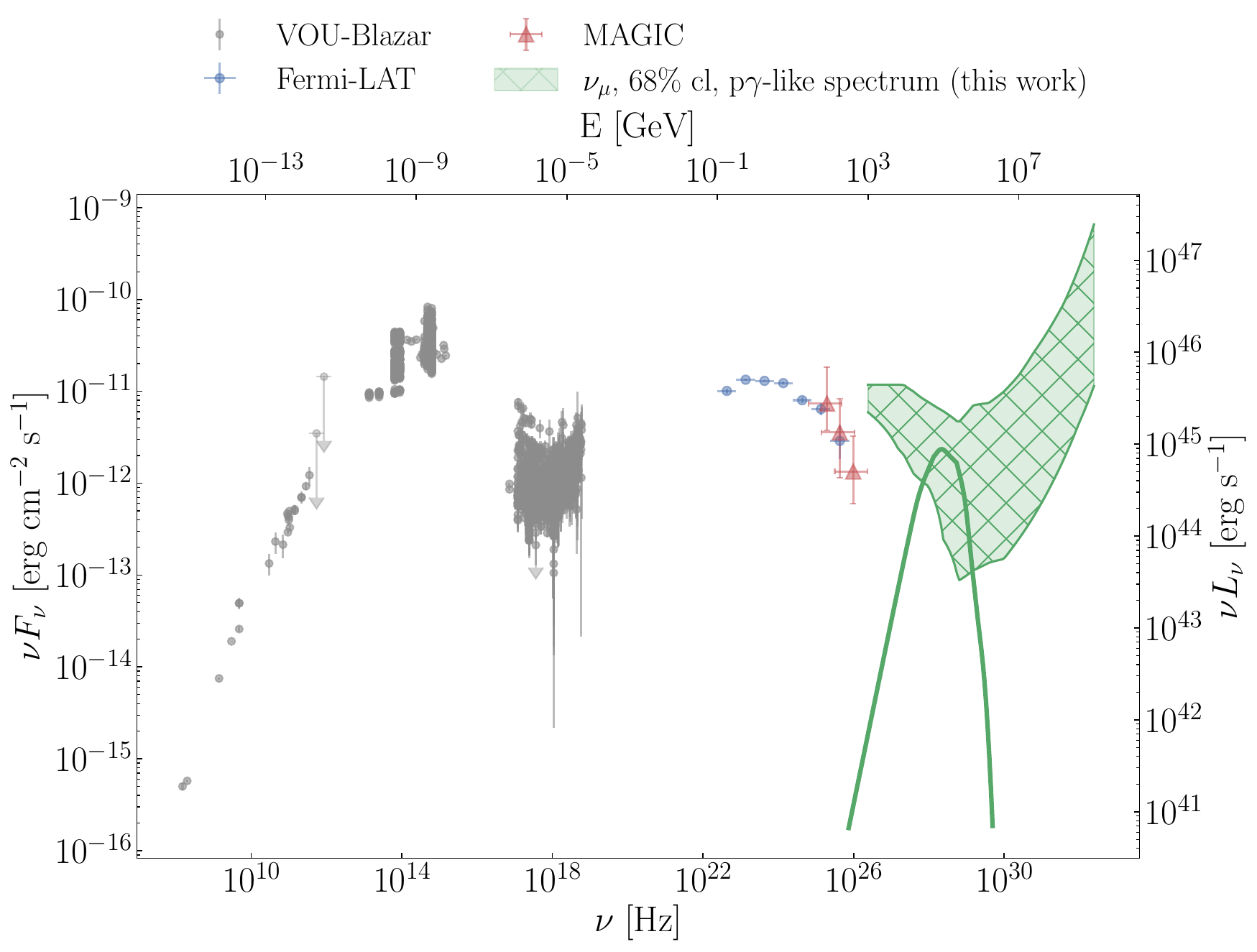}
    \end{minipage}
    \begin{minipage}{0.5\textwidth}
    \includegraphics[width=\textwidth]{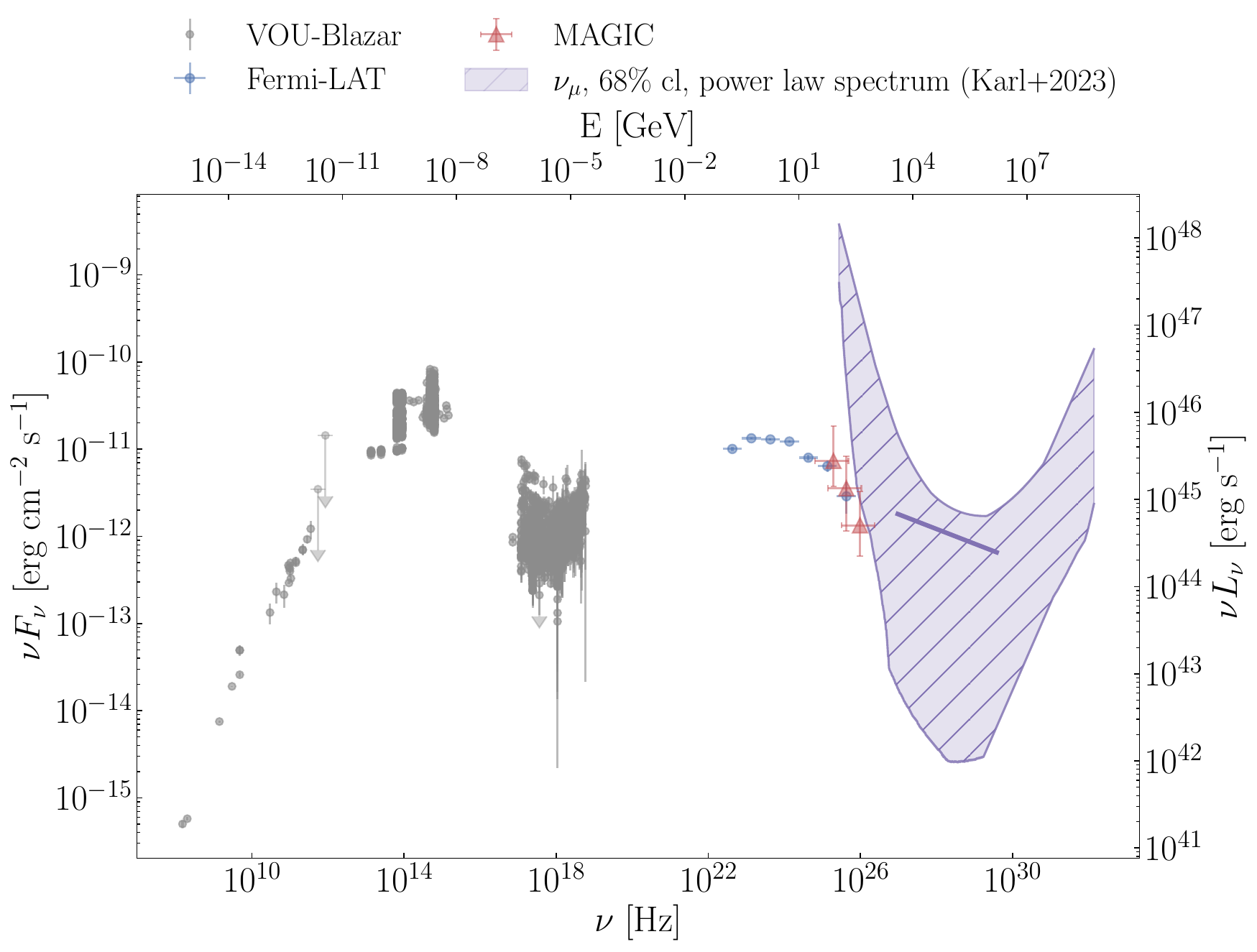}
    \end{minipage}
    \caption{multimesssenger fluxes from TXS~0506+056. The gray, blue, and red data points show the multiwavelength flux observations, according to the legend shown above. The colored bands show the range of the best-fit point source neutrino flux at the 68\% confidence level, derived from public IceCube data, assuming two different signal spectral shapes. \textit{Top:} The signal spectrum is assumed to be peaky, as predicted by leptohadronic blazar models; the template shape considered for all sources is shown as a green curve. The green band shows the allowed peak positions at the 68\% confidence level. These are the data used in this work to constrain the model. \textit{Bottom:} The signal spectrum is assumed to be a power law, as exemplified by the purple line. The purple band shows the point source neutrino fluxes from this blazar compatible with the data at the 68\% confidence level. This was the analysis performed in Paper IV.}
    \label{fig:neutrino_peak_vs_pl}
\end{figure}

The basic procedure consists of applying unbinned maximum likelihoods \citep{Braun:2008bg} to test two different hypotheses (background and signal): 
\begin{itemize}
    \item background hypothesis $H_B$: the neutrino emission is composed of atmospheric background and diffuse astrophysical neutrino emission;
    \item signal hypothesis $H_S$: an additional signal component originates from the source and clusters around it. The total observed neutrino emission then consists of both the signal and background components. In this work, we assume that the signal follows a peaked energy spectrum typical of p$\gamma$ interactions. As an approximation, we neglect differences between different sources and consider a single spectral template adopted from a previous leptohadronic study \citep{Rodrigues:2023vbv}. That template is shown as a green curve in the upper panel of \Fig\ref{fig:neutrino_peak_vs_pl}. We refer henceforth to this spectral shape assumption as the ``p$\gamma$ spectrum,'' in contrast to the power-law spectrum used as the signal assumption in Paper IV (e.g., purple line in the lower panel of \Fig\ref{fig:neutrino_peak_vs_pl}).
\end{itemize}

We fit the source signal by optimizing the likelihood ratio of the background and signal hypotheses. Because the shape of the signal is fixed to the assumed p$\gamma$ spectrum, the signal is fully described by only two parameters: the energy of the peak, $E_{\rm{peak}}$, which specifies its position along the x-axis, and the mean number of signal neutrinos, $n_\mathrm{s}$, which determines the normalization of the spectrum and therefore specifies its position along the y-axis~\citep[cf. e.g.,][]{Padovani:2015mba}. These are the only parameters of our fit, since the background is fixed. We set bounds for the fit parameters. For $E_\mathrm{peak}$ the lower bound is 10~TeV (below which the data are generally dominated by atmospheric background) and the upper bound is 1~EeV (the maximum energy for which we have information on the detector). Values of $n_\mathrm{s}$ have a lower bound of 0 and an upper bound of $10^3$ neutrinos. As mentioned in the main text, this fitting procedure, as well as the neutrino flux simulations explained below, are performed using the open-source software SkyLLH~\citep{Bellenghi:20230u}. 

We then use the likelihood ratio test result optimized in the previous step as a threshold test statistic (TS) value to calculate the 68\% confidence intervals for each blazar, following the method by \citet{Feldman_1998}. Essentially, the confidence limits are given by the possible fluxes for which 68\% of their weighted TS distribution are compatible with our threshold TS value. To determine the compatible fluxes, we simulate a neutrino signal with different $E_\mathrm{peak}$ values ranging from 10~TeV to 10~EeV and with different $n_\mathrm{s}$ between 0 and 150 detected neutrinos per ten years. For $E_\mathrm{peak}$, we adopt a step size of $\approx 0.2$ dex; for $n_\mathrm{s}$, we adopt a step size that depends on the signal strength: for $n_\mathrm{s}$ between 0 and 1 detected neutrinos per ten years, we adopt a step size of $\approx 0.025$; for $n_\mathrm{s}$ higher than 1 neutrino per ten years, we adopt and larger step size, ranging between 0.5 and ten. The reason for the varying step sizes is that smaller steps are generally necessary near the limits of the confidence band in order to define it precisely. Whenever the lower bound of the neutrino flux is compatible with no neutrino emission, we only consider the 68\% upper flux limit. Typically, a few hundreds of simulations for each flux realization are necessary to gather enough statistics; the exact number of simulations depends on the smoothness of the TS distribution for that particular signal spectrum. 

The result of this procedure is shown in the top panel of \Fig\ref{fig:neutrino_peak_vs_pl} for TXS ~0506+056. The green curve represents the template spectral shape, shown here for a given value of $E_\mathrm{peak}$ and $\nu F_{\nu,\mathrm{peak}}$. As argued above, the signal can be fully characterized by the value of $E_\mathrm{peak}$ and $\nu F_{\nu,\mathrm{peak}}$, since the spectral shape is fixed. The result of the above analysis can therefore be summarized as a region of ($E_\mathrm{peak}$, $\nu F_{\nu,\mathrm{peak}}$) values compatible with the data at the 68\% confidence level, which is shown as a green band in \Fig\ref{fig:neutrino_peak_vs_pl}. This is also the definition of the green band shown in the plots in \Sec\ref{sec:results}, where the fit results were presented for each source. This plotting choice allows for an easy visual evaluation of the goodness of the model: the predicted neutrino spectrum is compatible with the point source data at the 68\% confidence level if its peak lies within the flux band.

It is important to note that in this representation, the flux bands show a constraint on the neutrino energy flux and not the number flux. This is partly responsible for the increasing confidence bands at exaelectronvolt energies. Let us take as an example the case of TXS~0506+056, shown in the figure. For a spectrum peaking at 10~TeV, the flux band corresponds to a mean number of detected signal neutrinos between 8 and 39 per ten years. For a spectrum peaking at 1~EeV, the bounds correspond to a number between 0.18 and 8 neutrinos per ten years. 

\subsection{Method limitations}

The first limitation of the method is the adoption of the same spectral shape template when deriving the IceCube limits for all sources. On the one hand, this choice is justified by the fact that the blazar neutrino spectrum is expected to be highly peaked in a $\nu F_\nu$ representation, because the interactions are threshold-dominated. The total neutrino flux is therefore highly dominated by the peak, thus minimizing the effect of the specific spectral shape compared to the case of a broad spectrum, such as a power-law signal. On the other hand, the exact shape of the neutrino spectrum can in fact vary, since it depends on the proton spectral index and the target photon spectrum, which differ from source to source. To test the impact of this variation, for every best-fit result presented in \Sec\ref{sec:results}, we have retroactively checked the consistency of our approach by comparing the neutrino spectrum resulting from the model resulting to the template neutrino spectrum in the upper panel of \Fig\ref{fig:neutrino_peak_vs_pl}. We did this by taking each best-fit result, rescaling the x- and y-axes to match the peak position of the template, and then calculating the ratio between the energy-integrated neutrino flux for the best-fit spectrum and for the template spectrum. In a scenario where the template were perfectly representative, this ratio would be 1 for every source. We obtained an average of 1.3, with 70\% of the results lying below this value. This indicates that this method introduces an intrinsic error of at most 30\% when comparing each source model result with the IceCube limits calculated with the template. In theory, this could be further mitigated by employing a fully self-consistent method where the neutrino spectrum resulting from each source simulation is fed into the data analysis to determine the exact constraints for that spectral shape. However, because the IceCube analysis involves computationally expensive simulations, it cannot realistically be performed after every individual source model. Notwithstanding this uncertainty in the method, the assumption of a peaked signal remains a more accurate approximation of blazar emission than a power-law signal.

An additional limitation of the method is the fact that IceCube public data covers the years 2008 to 2018, which means that not all detected alert events in the G20 sample are part of this data sample. \citet{Abbasi_2024} found no additional lower-energy neutrino component from the direction of alert events (apart from the case of TXS~0506+056), and we expect the IceCube alert events to contribute significantly to the neutrino flux. Hence, the neutrino flux fits and constraints for G20 sources where the alert event is not part of the IceCube public data are most likely very conservative and underestimate the actual neutrino flux.

\subsection{Comparison with the assumption of a power-law signal}

We now compare the results obtained with the assumption of a p$\gamma$-like signal, used in this work, and a power-law signal, used in Paper IV. The lower panel of \Fig\ref{fig:neutrino_peak_vs_pl} shows the result for TXS~0506+056 obtained assuming a power-law signal as in Paper IV. The band represents the overlapping possible realizations of the source's differential flux that are compatible with the data at the 68\% confidence level.

The procedure used in Paper IV to obtain this constraint is explained in detail in Appendix A of that reference. The general method is the same as that explained above; however, the different assumptions on the signal spectral shape (i.e., the energy pdf $S_{\rm{energy}}$ in equation A2 of Paper IV) translate into some technical differences. Given that a power-law spectrum is not intrinsically limited in energy, the two appropriate parameters to characterize the signal in this case are the spectral index $\gamma$ (unlike $E_\mathrm{peak}$ in the present approach) and the signal strength, $n_\mathrm{s}$. This also means that for each realization of $\gamma$ and $n_\mathrm{s}$, the simulated events are not limited in energy, owing to the fact that they are sampled from a power-law spectrum. The energy range in which the derived flux limit is valid is estimated by \textit{1)} folding the flux with the effective area and reconstruction properties of the detector to estimate the distribution of detected events, \textit{2)} sampling $\approx 10^5$ signal events from the distribution of detectable events, and \textit{3)} considering the true energy distribution of the detected events. The relevant energy range where the flux limit is valid is given by the central 90\% quantile of the distribution of the true energy of the detected events.

The valid energy range of each test result is an essential component underlying the purple band in the bottom panel of \Fig\ref{fig:neutrino_peak_vs_pl}: for each value of spectral index $\gamma$, the method by \citet{Feldman_1998} reveals the flux range that is compatible with data at the 68\% confidence level, which is then plotted in its corresponding valid energy range. For steeper power-law signals, the flux is dominated by low-energy events; therefore most detected events lie at lower energies, and the valid energy range where the result applies is therefore lower. Conversely, for hard spectra the valid energy range is generally higher.

\begin{figure}
    \centering
    \includegraphics[width=\columnwidth]{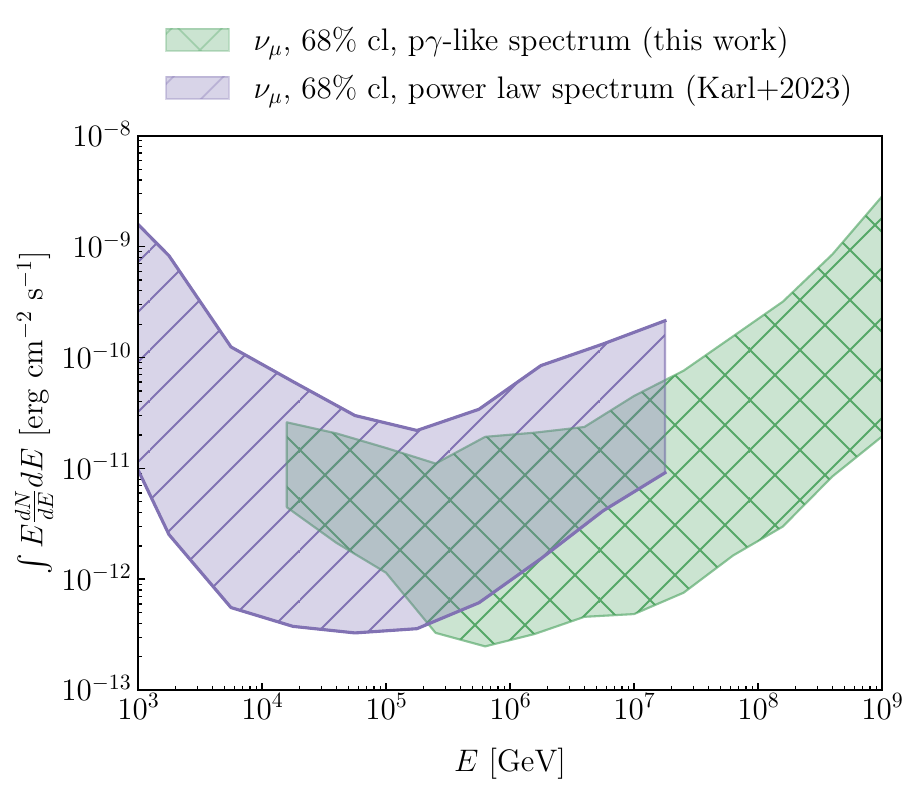}
    \caption{Integrated fluxes of the 68\% confidence limit bands as shown in \Fig\ref{fig:neutrino_peak_vs_pl} for the power-law assumption (purple diagonally hatched) and the p$\gamma$ spectrum (green cross-hatched), for source TXS~0506+056. For the p$\gamma$ spectrum, the flux is shown as a function of the peak energy. For the power-law spectrum, the flux is shown at the respective mean energy (cf. lower panel of \Fig\ref{fig:neutrino_discovery_potential}). This is the reason why the purple band covers a more limited energy range than in the lower panel of \Fig\ref{fig:neutrino_peak_vs_pl}, where the differential flux is shown in the full energy range of the analysis.}
    \label{fig:integrated_limits}
\end{figure}

While the representations of the 68\% confidence band in \Fig\ref{fig:neutrino_peak_vs_pl} allow for a comparison with the neutrino spectrum predicted by the model, a direct comparison between the two bands is hindered by the fact that the two assumed signals have drastically different spectral shapes, leading to different conversion factors between the differential neutrino flux, indicated by the bands, and the integrated flux, which characterizes the source's overall neutrino emission. To better compare the two approaches, in \Fig\ref{fig:integrated_limits} we show the allowed range of energy-integrated neutrino flux from TXS~0506+056. In the case of a p$\gamma$ spectrum (green band), the differential fluxes allowed at the 68\% confidence level are integrated over the entire energy range; for the power-law signal (purple band), the integration is performed over the respective valid energy range, estimated as discussed above.

\begin{figure}
\begin{minipage}{0.5\textwidth}
    \includegraphics[width=\textwidth]{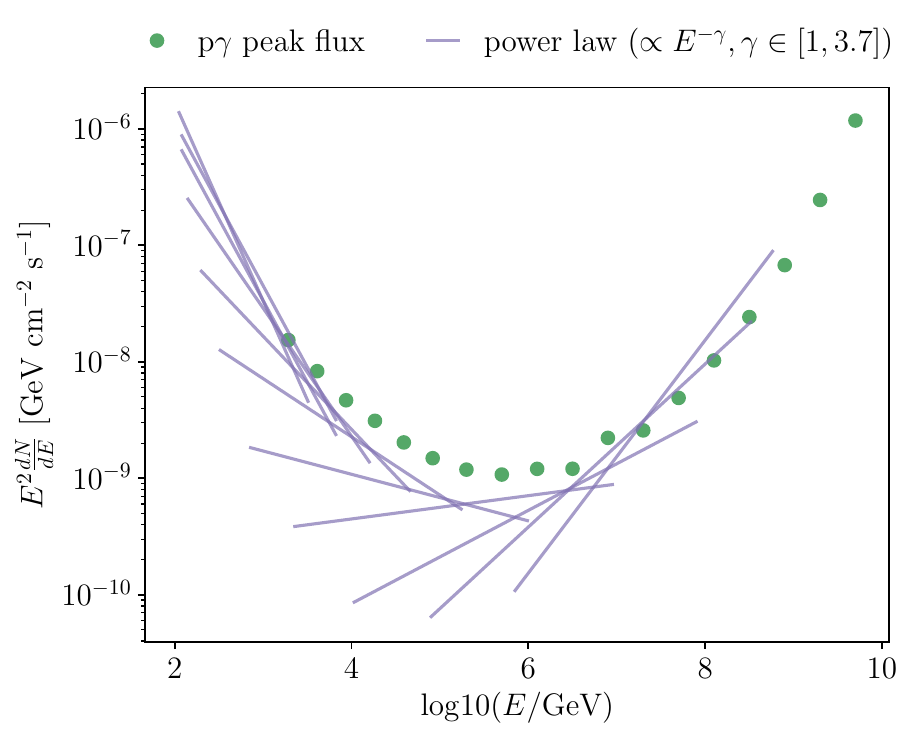}
    \end{minipage}
    \begin{minipage}{0.5\textwidth}
    \includegraphics[width=\textwidth]{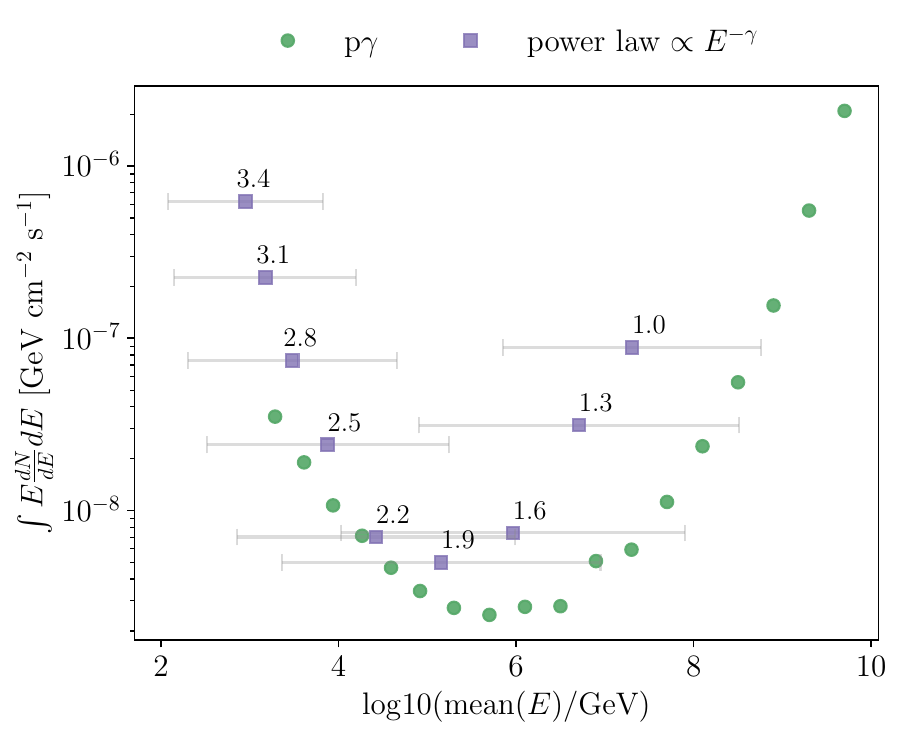}
    \end{minipage}
    \caption{ 3$\sigma$ neutrino discovery potential at the location of TXS~0506+056, assuming the source emits a power-law spectrum (purple) and a p$\gamma$-like spectrum (green). The 3$\sigma$ discovery potential is the flux a source would need to emit to have a 50\% probability of being detected with 3$\sigma$ significance. \textit{Top:} The purple lines show the differential discovery potential for different power-law spectra with spectral indices between 1 and 3.7. Each spectrum is shown extending across the corresponding valid energy range, estimated following the procedure described in the text. The green points show the peak position of the 3$\sigma$ discovery potential fluxes assuming a p$\gamma$-like spectrum (upper panel of \Fig\ref{fig:neutrino_peak_vs_pl}) with different $E_{\text{peak}}$ values. \textit{Bottom:} The integrated 3$\sigma$ discovery potential fluxes for the different spectra. The purple squares show the integrated power-law fluxes and the gray bars indicate the energy integration range. The numbers state the respective spectral index, $\gamma$, of the power law. The green dots show the integrated fluxes from 100~GeV to 10~EeV for the full p$\gamma$ spectra at their respective $E_{\text{peak}}$.}
    \label{fig:neutrino_discovery_potential}
\end{figure}

As we can see, in the common energy range of the two analyses\footnote{The energy range of the confidence limits in Paper IV (purple band in \Fig\ref{fig:neutrino_peak_vs_pl}) is determined by the sampled range of the spectral index $\gamma$,  fixed to the range $\gamma \in [1, 3.7]$, for all sources. For each $\gamma$, the valid energy range is then determined following the procedure described in this appendix (as an example, see \Fig\ref{fig:neutrino_discovery_potential}). This leads to an energy range that differs from source to source, unlike the present analysis where we sample a fixed range of $E_\mathrm{peak}$.}, there is a large overlap in the allowed range of integrated signal flux obtained with both analyses, but there are also significant differences. The upper limit on the integrated signal flux obtained under the assumption of a p$\gamma$ spectrum is more stringent than the power-law assumption by a factor of 2.5-3, at all overlapping energies. On the other hand, the lower limit on the integrated flux is higher for the p$\gamma$ spectrum for peak energies below 300~TeV, and higher for the power law assumption above 300~TeV. As an example, considering a signal p$\gamma$ spectrum with $E_\mathrm{peak}\approx20~\mathrm{TeV}$, the minimum flux required to explain the data from TXS~0506+056 at the 68\% confidence level under the assumption of a p$\gamma$-like signal is a factor of ten higher than under the assumption of a power law. Conversely, for $E_\mathrm{peak}\approx30~\mathrm{PeV}$, the minimum flux required under the p$\gamma$ assumption is 15 times lower compared to the power law assumption.

Finally, we compare the sensitivity of both analyses to neutrino emission. This can be done by estimating the so-called 3$\sigma$ discovery potential, that is, by simulating the neutrino flux necessary to have a 50\% probability of obtaining evidence of neutrino emission from the source at the 3$\sigma$\ level. We show the result in the upper panel of \Fig\ref{fig:neutrino_discovery_potential} for a power-law spectrum (purple) and a p$\gamma$ spectrum (green). As in \Fig\ref{fig:neutrino_peak_vs_pl}, the band shows the allowed range of differential neutrino flux for every energy analyzed. While \Fig\ref{fig:neutrino_peak_vs_pl} shows the flux that best fits the point source data, here we show the flux that satisfies the discovery potential. 

In the lower panel of \Fig\ref{fig:neutrino_discovery_potential}, we show the energy-integrated discovery potential flux for the two assumptions. We can see that below 100~TeV, both p$\gamma$ and power-law spectra require a comparable level of integrated signal neutrino flux to satisfy the discovery potential criterion. Above 100~TeV, the power-law spectrum assumption is associated with a higher integrated discovery potential flux compared to the p$\gamma$ assumption. 
This difference can be explained by the fact that a p$\gamma$ spectrum differs strongly from the spectral shape of the atmospheric background and the diffuse neutrino component, both of which can be well described by power-law spectra in the energy range of interest \citep[for example][]{Abbasi_2022}. Hence, a source emitting a signal with a p$\gamma$ spectral shape requires a lower flux for a significant detection compared to a signal that follows a power-law spectrum, even if it is harder than the background distribution.

\section{Origin of the multiwavelength photon emission}
\label{app:components}

In \Fig\ref{fig:app_components_1} we show the best-fit SEDs decomposed into the process where the emission originates. This is obtained by tracking the photons produced through synchrotron and inverse Compton emission by the different particle populations: primary electrons (yellow), secondary electron-positron pairs from proton photo-pair production (i.e., Bethe-Heitler, blue), secondary pairs from photon annihilation (green), secondary pairs from the decay of charged pions produced in p$\gamma$ interactions (magenta), and primary protons (orange).

The shaded areas below the emission curves above $\sim100$~GeV represent the extent of photon attenuation during propagation, due to interactions with the extragalactic background light (EBL). The attenuation length for each source is calculated using the Gammapy software package~\citep{gammapy:2023,gammapy:zenodo-1.1} and assuming the EBL model by \citet{2011MNRAS.410.2556D}. The final SED, shown as solid black curves, is that resulting from this attenuation calculation.

We do not show explicitly the $\gamma$ rays produced in neutral pion decay, which have typical energies between 10~TeV and a few petaelectronvolt in this model. A fraction of this emission is attenuated in the source through annihilation with lower-frequency photons, leading to pair production as mentioned above; this flux is therefore included in the green curves, which originate in pair production   The remaining fraction escapes the source and is attenuated through EBL interactions. 

\begin{figure*}
    \centering

\includegraphics[height=42mm, trim={5mm 11mm 4.01mm 4mm}, clip]{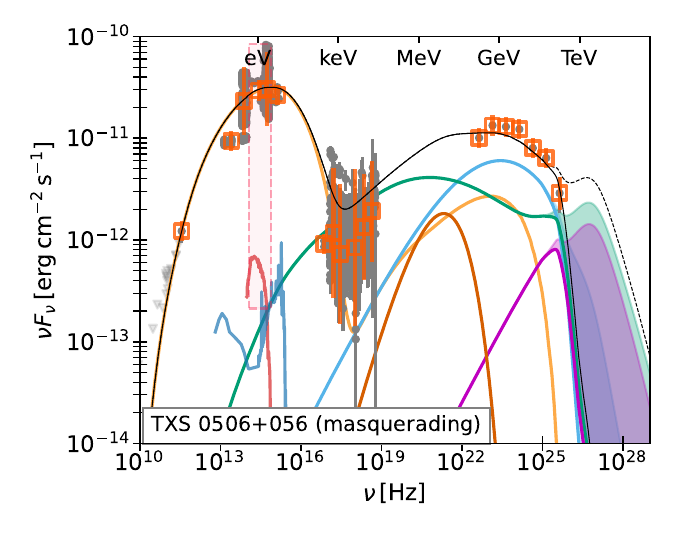}\includegraphics[height=42mm, trim={10mm 11mm 4mm 4mm}, clip]{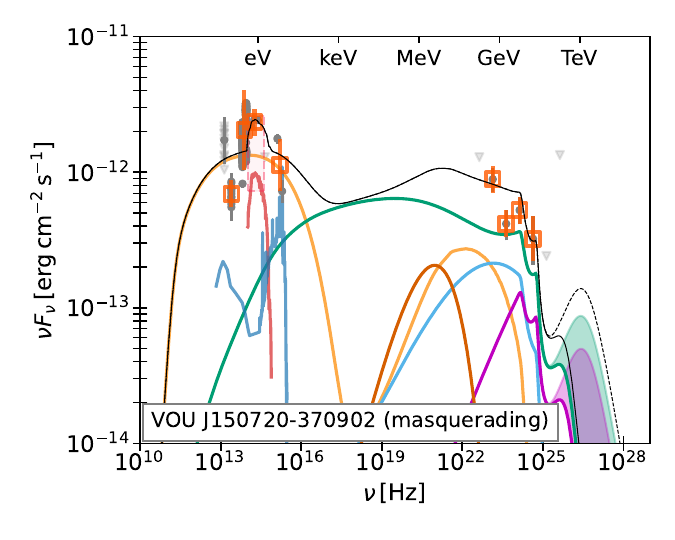}\includegraphics[height=42mm, trim={10mm 11mm 4mm 4mm}, clip]{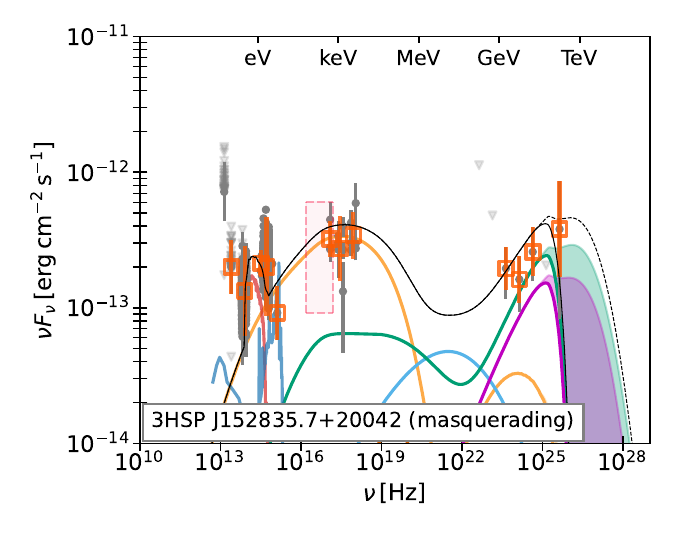}
    
\includegraphics[height=42mm, trim={5mm 11mm 4mm 4mm}, clip]{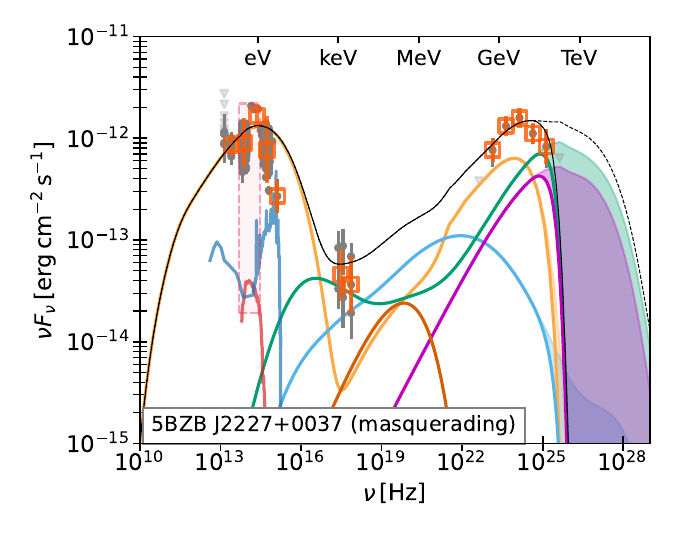}\includegraphics[height=42mm, trim={10mm 11mm 4mm 4mm}, clip]{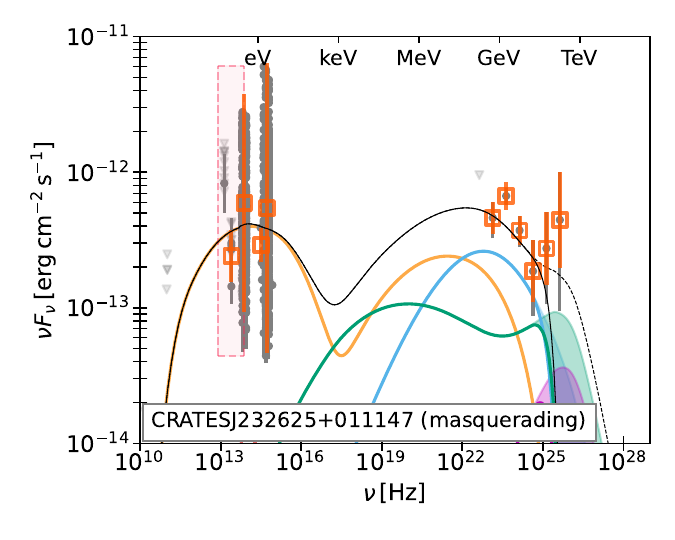}\includegraphics[height=42mm, trim={10mm 11mm 4mm 4mm}, clip]{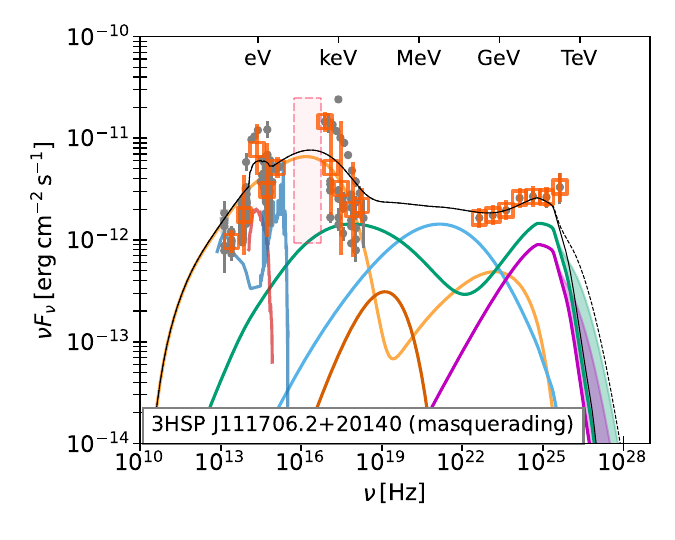}

\includegraphics[height=42mm, trim={5mm 11mm 4.01mm 4mm}, clip]{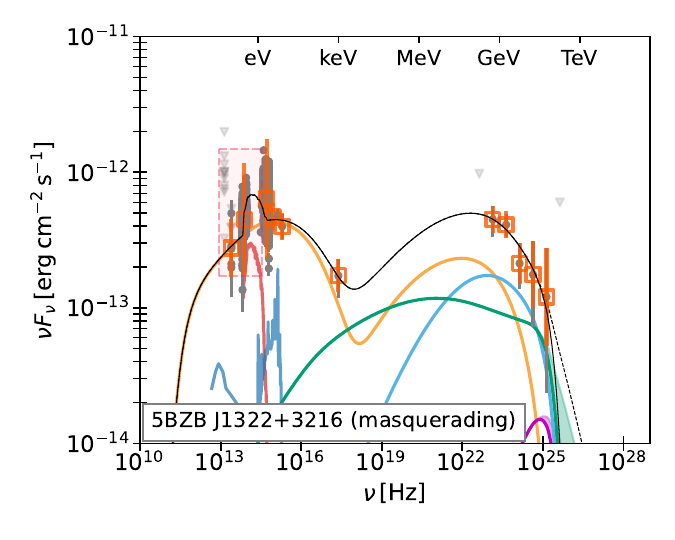}\includegraphics[height=42mm, trim={10mm 11mm 4mm 4mm}, clip]{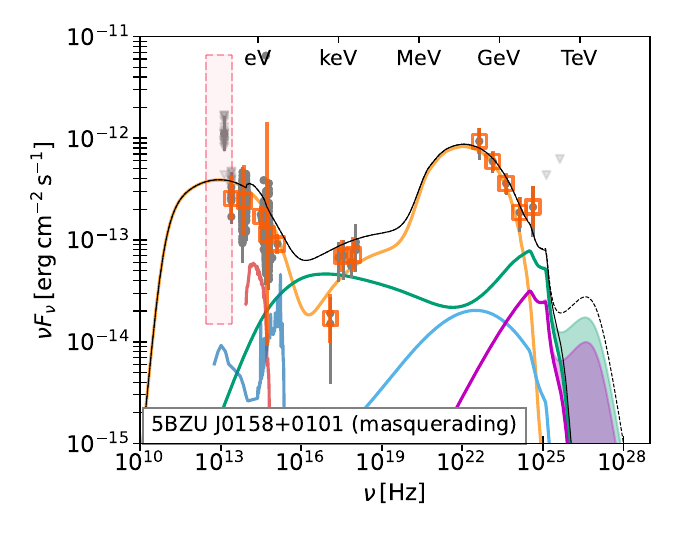}\includegraphics[height=42mm, trim={10mm 11mm 4mm 4mm}, clip]{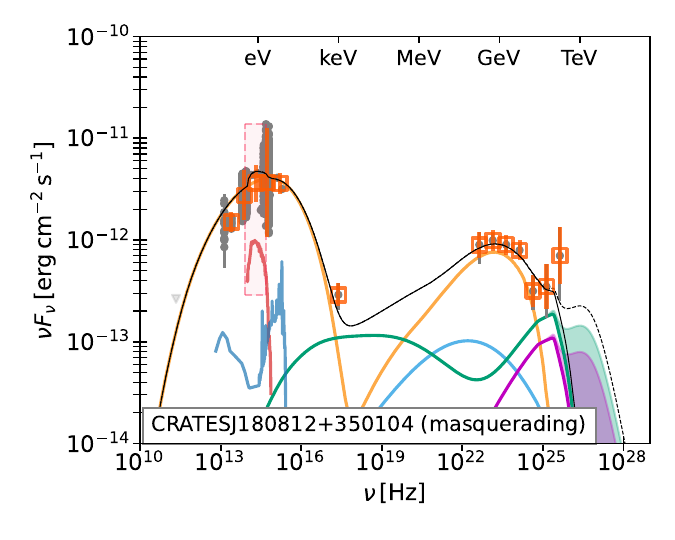}

\includegraphics[height=42mm, trim={5mm 11mm 4.01mm 4mm}, clip]{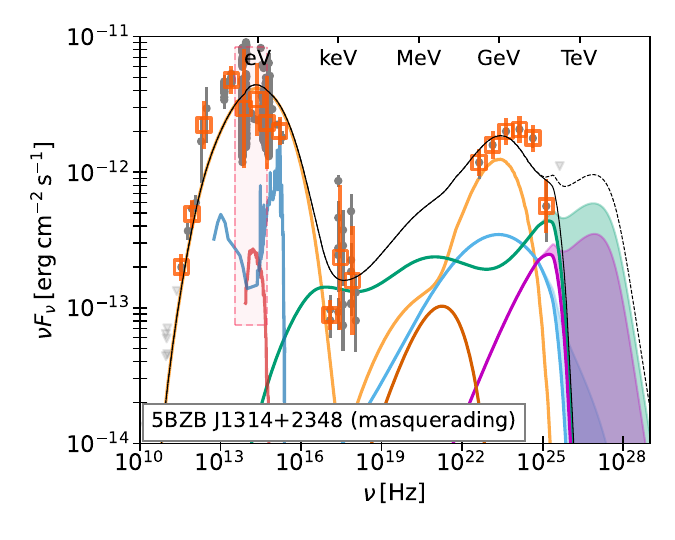}\includegraphics[height=42mm, trim={10mm 11mm 4mm 4mm}, clip]{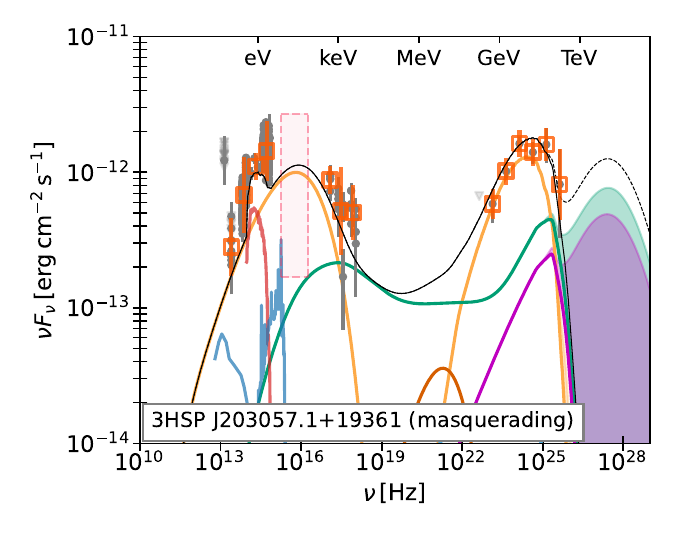}\includegraphics[height=42mm, trim={10mm 10mm 4mm 4mm}, clip]{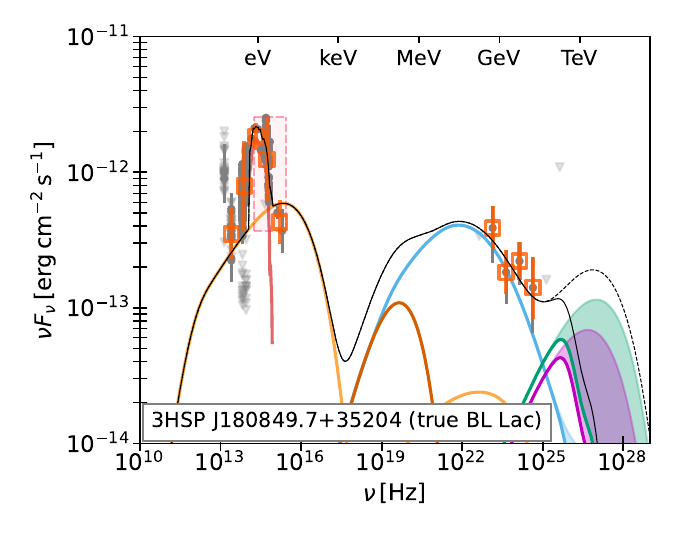}

\includegraphics[height=46.3mm, trim={5mm 3mm 4.01mm 4mm}, clip]{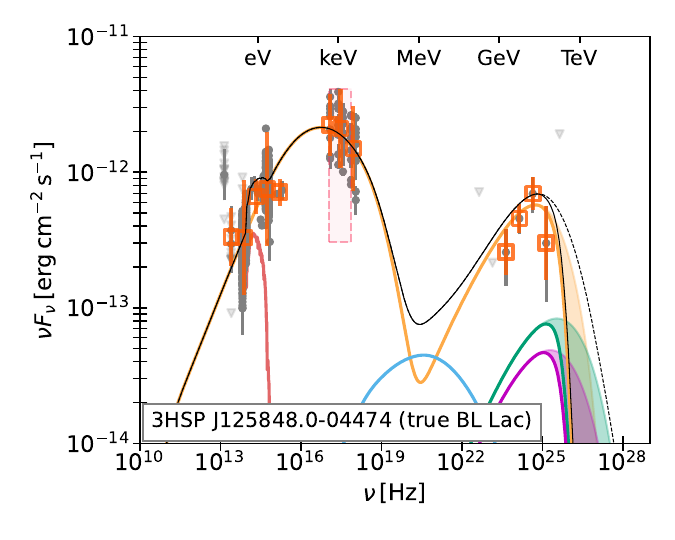}\includegraphics[height=45.7mm, trim={10mm 3mm 4mm 4mm}, clip]{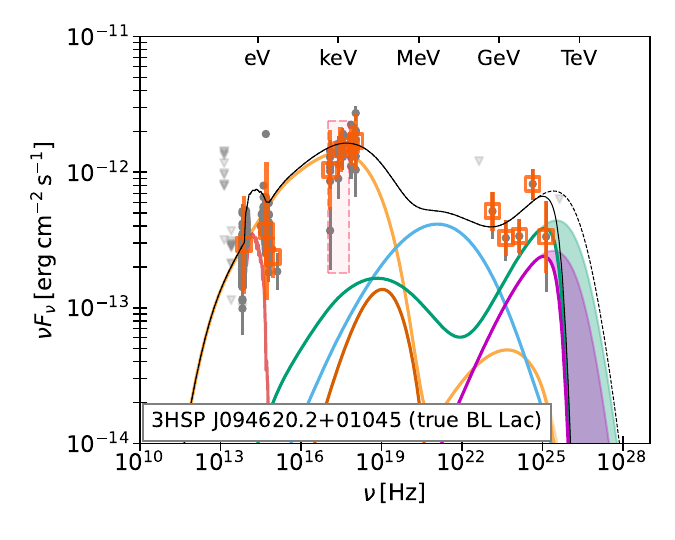}\includegraphics[height=45.7mm, trim={10mm 3mm 4mm 4mm}, clip]{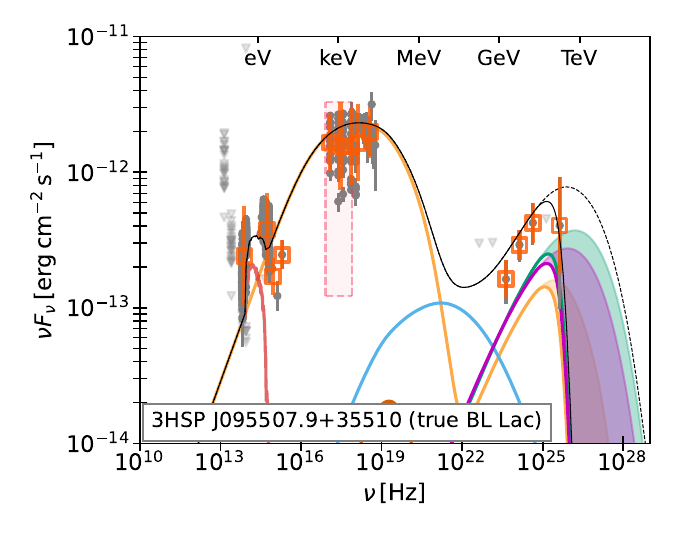}

    \hspace{15mm}\includegraphics[height=13mm, trim={0 0 0 0}, clip]{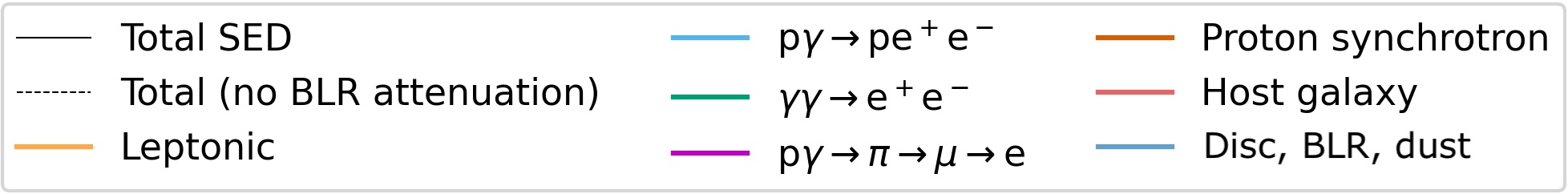}
    
    \caption{Detailed multiwavelength spectral energy distributions predicted by the best-fit leptohadronic model for each source. The colored curves represent the photon fluxes at the observer, separated by the different processes from which they originate, according to the caption. The colored bands represent the extent of the attenuation on the EBL of each of the components. The rectangular area inside the dashed red lines represents the uncertainty range of the synchrotron peak position, as plotted in \Fig\ref{fig:synchrotron_peak} for the entire sample. The gray points show the respective SED. In orange, we show the binned fluxes (cf. \Sec\ref{sec:binning}).}
    \label{fig:app_components_1}
\end{figure*}

\begin{figure*}\ContinuedFloat
    \centering

\includegraphics[height=42mm, trim={5mm 11mm 4.01mm 4mm}, clip]{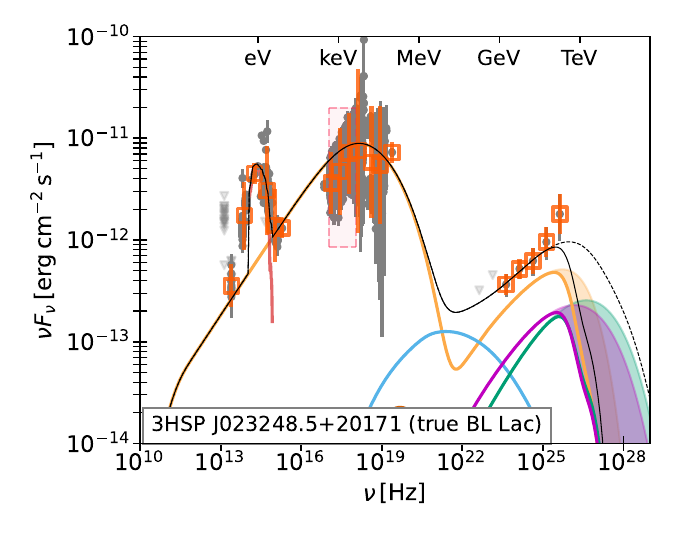}\includegraphics[height=42mm, trim={10mm 11mm 4mm 4mm}, clip]{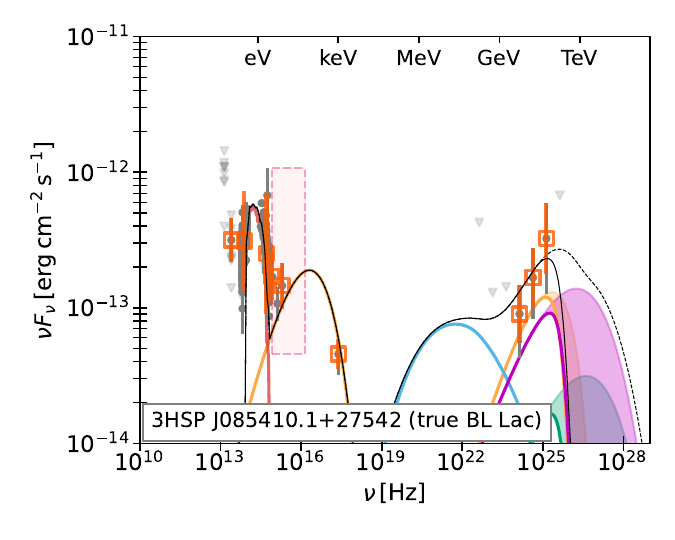}\includegraphics[height=42mm, trim={10mm 11mm 4mm 4mm}, clip]{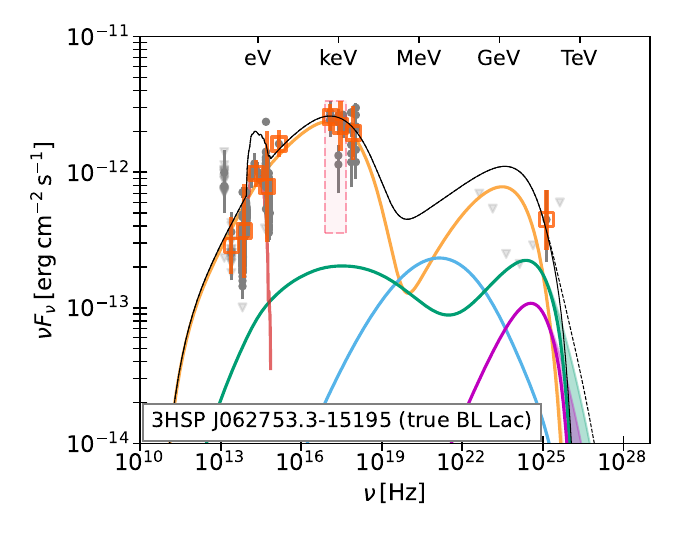}
    
\includegraphics[height=42mm, trim={5mm 11mm 4mm 4mm}, clip]{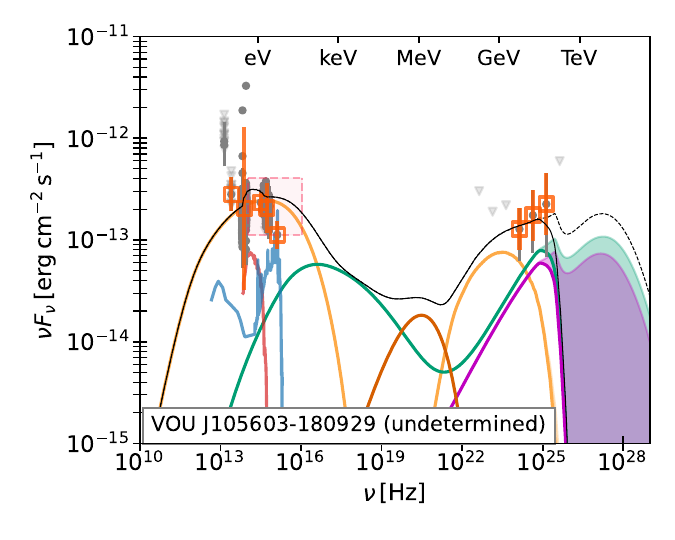}\includegraphics[height=42mm, trim={10mm 11mm 4mm 4mm}, clip]{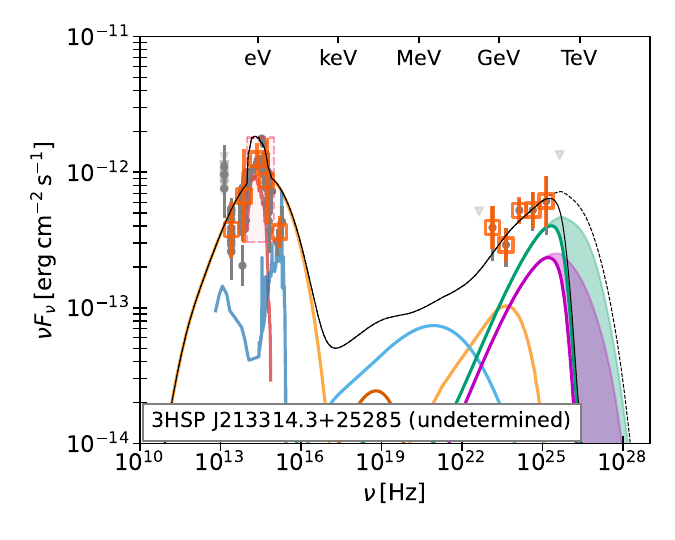}\includegraphics[height=42mm, trim={10mm 11mm 4mm 4mm}, clip]{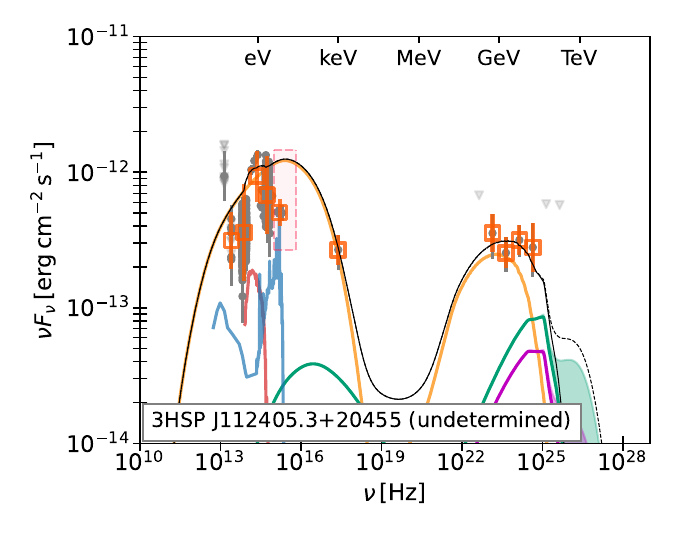}

\includegraphics[height=42mm, trim={5mm 11mm 4.01mm 4mm}, clip]{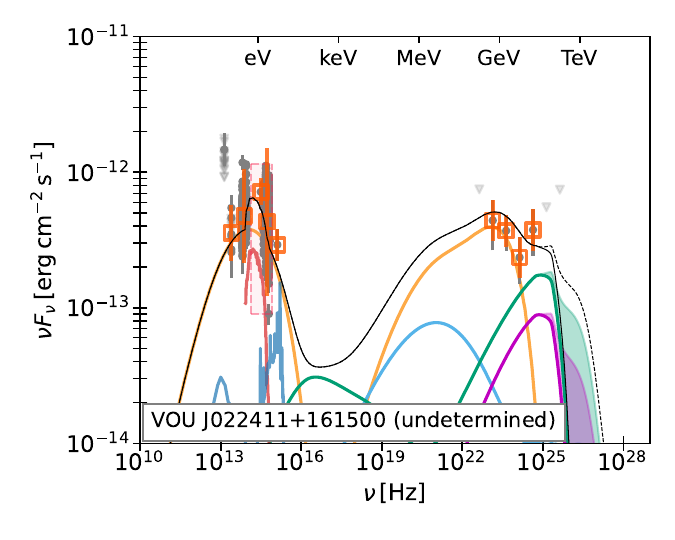}\includegraphics[height=42mm, trim={10mm 11mm 4mm 4mm}, clip]{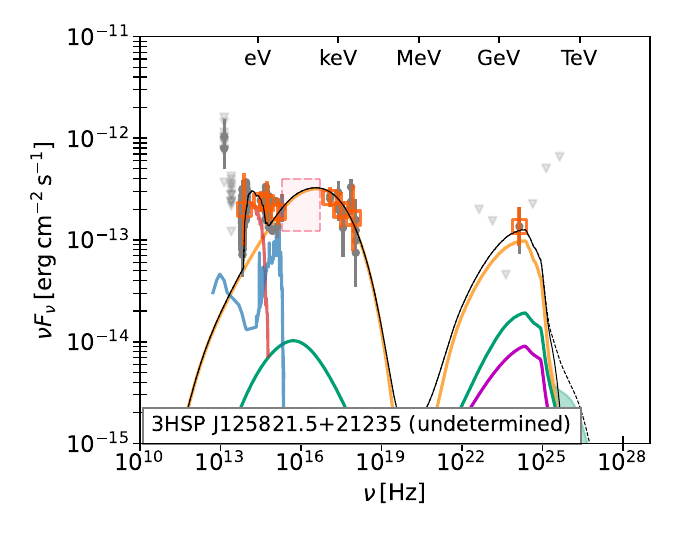}\includegraphics[height=42mm, trim={10mm 11mm 4mm 4mm}, clip]{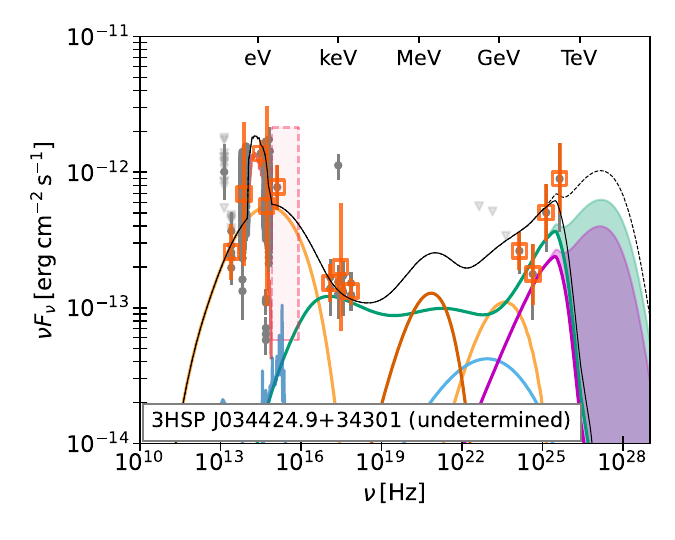}

\includegraphics[height=42mm, trim={5mm 11mm 4.01mm 4mm}, clip]{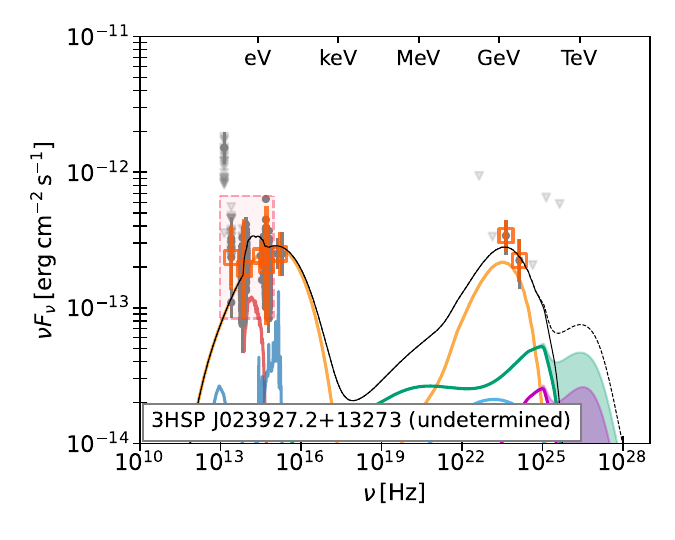}\includegraphics[height=42mm, trim={10mm 11mm 4mm 4mm}, clip]{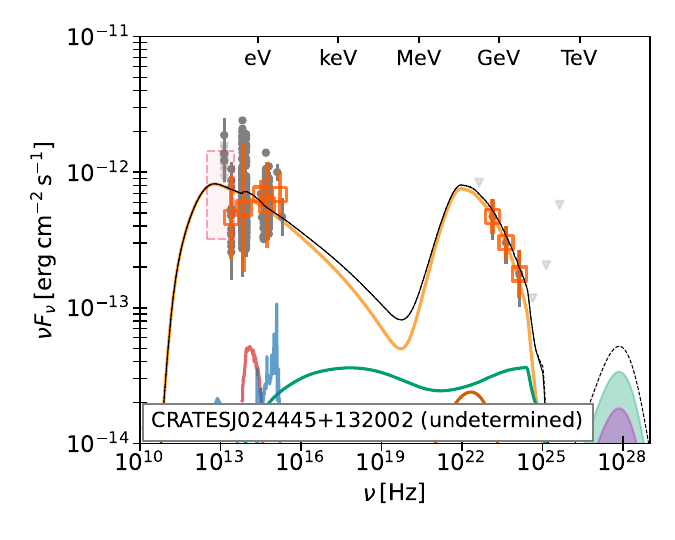}\includegraphics[height=42mm, trim={10mm 10mm 4mm 4mm}, clip]{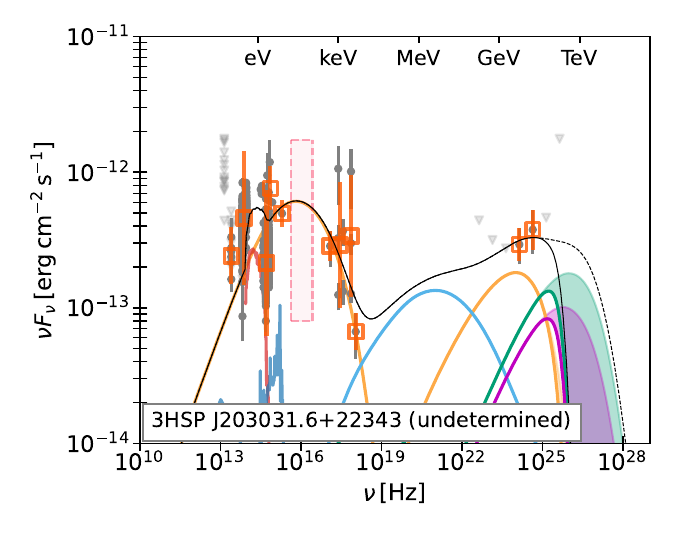}

\includegraphics[height=46.3mm, trim={5mm 3mm 4.01mm 4mm}, clip]{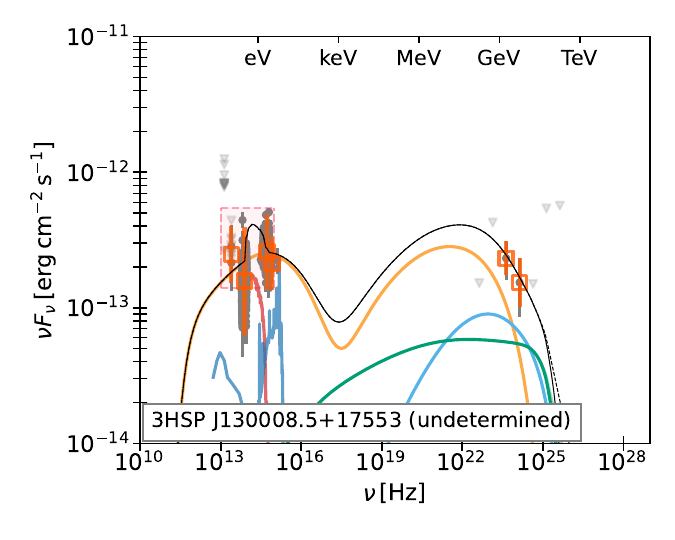}\includegraphics[height=45.7mm, trim={10mm 3mm 4mm 4mm}, clip]{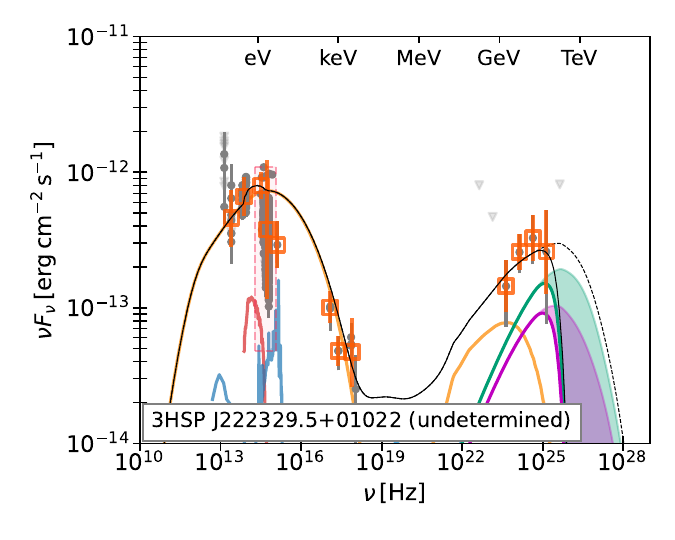}\includegraphics[height=45.7mm, trim={10mm 3mm 4mm 4mm}, clip]{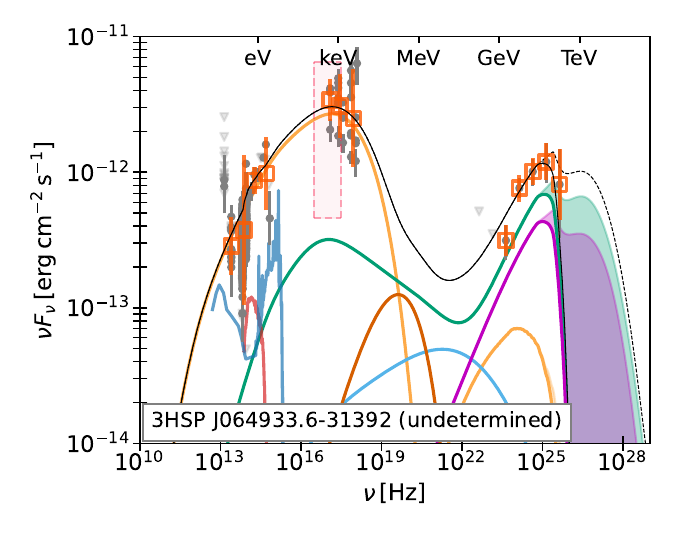}

\hspace{15mm}\includegraphics[height=13mm, trim={0 0 0 0}, clip]{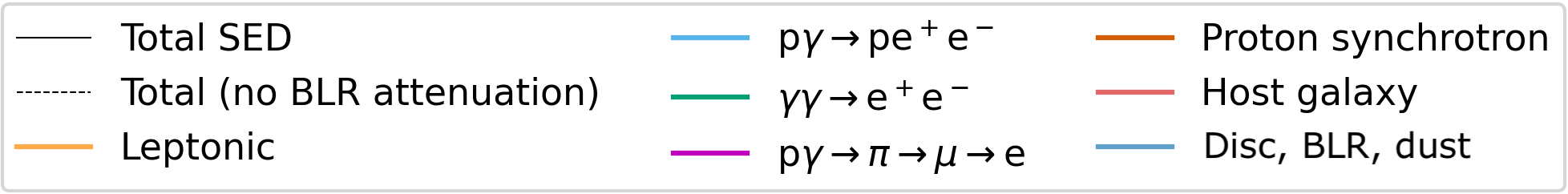}
    \captionsetup{labelformat=empty}
    \caption{Fig. B.1. continued. Detailed multiwavelength spectral energy distributions predicted by the best-fit leptohadronic model for each source.}
\end{figure*}

\begin{figure*}\ContinuedFloat
    \centering
    \includegraphics[height=45.7mm, trim={10mm 3mm 4mm 4mm}, clip]{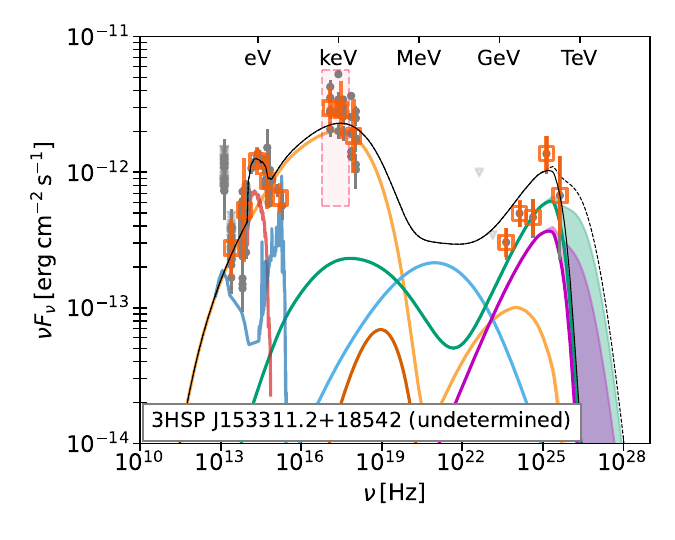}
    \includegraphics[height=45.7mm, trim={10mm 3mm 4mm 4mm}, clip]{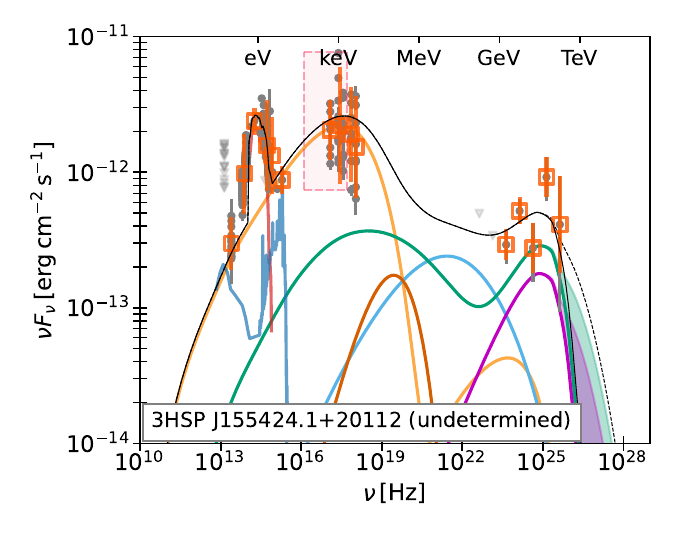}
    \includegraphics[height=45.7mm, trim={10mm 3mm 4mm 4mm}, clip]{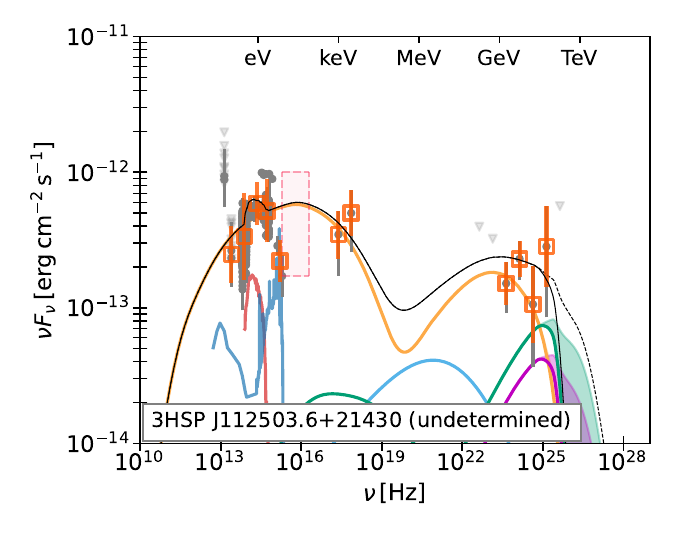}
    
    \includegraphics[height=45.7mm, trim={10mm 3mm 4mm 4mm}, clip]{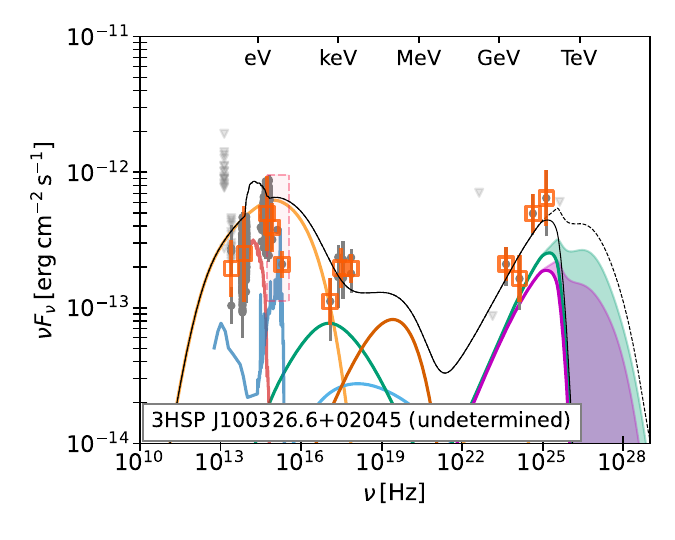}
    \hspace{15mm}\includegraphics[height=13mm, trim={0 0 0 0}, clip]{components_legend.jpg}
    \captionsetup{labelformat=empty}
    \caption{Fig. B.1. continued. Detailed multiwavelength spectral energy distributions produced by the best-fit leptohadronic model for each source.}
\end{figure*}

\section{Treatment of the host galaxy contribution}
\label{app:host}

We now report further details on the host galaxy spectrum considered for each BL Lac. Although this emission does not play a direct role in the jet radiation model, its contribution to the infrared SED is often significant in IHBLs. We therefore add this spectrum to the modeled nonthermal jet emission before fitting the model to the multiwavelength SED.

\begin{figure*}
    \centering
    \includegraphics[width=\textwidth, trim={5mm 5mm 5mm 18mm}, clip]{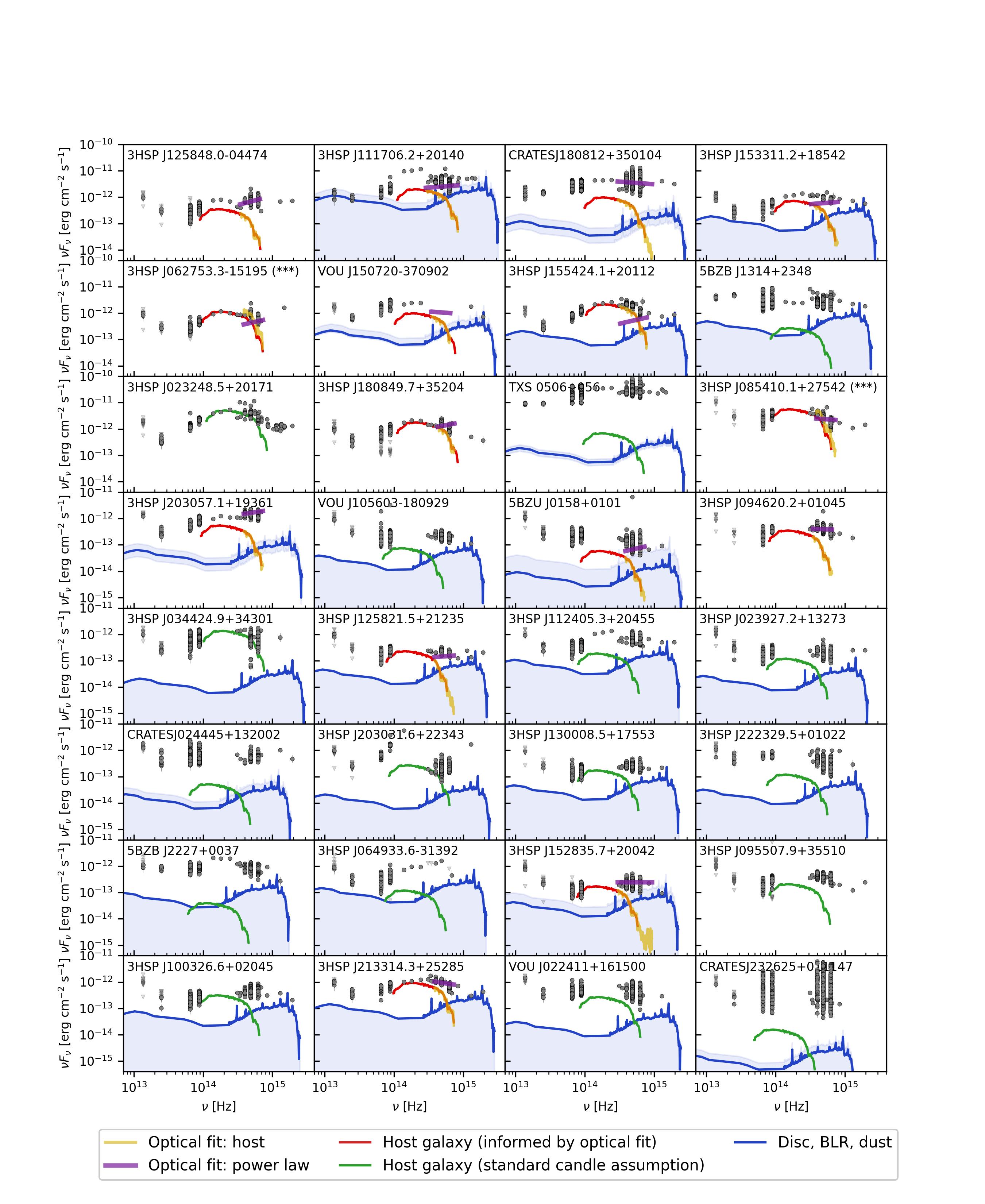}

    \caption{Results of the fitting of the host galaxy contribution for all sources in the sample. In all these cases, the host contribution was constrained based on optical spectroscopy, as reported in Paper III. The gray points show the broadband SED; the blue curve and the respective error band show the template spectrum for the disk, BLR, and dust torus emission, normalized to match the disk luminosity value; the red curve shows the fit to the host spectrum resulting from the spectral decomposition, and the purple line the power-law component.}
    \label{fig:app_host}
\end{figure*}

In \Fig\ref{fig:app_host} we show the fitted host galaxy templates, adopted from \citet{Mannucci:2001qa} and normalized following the procedure described in \Sec\ref{sec:host}. In the cases where optical spectroscopy data allowed us to perform a decomposition into the host contribution and a nonthermal continuum (cf. corresponding analyses in Papers I and III), those two components are shown in yellow and purple, respectively. The total host spectrum, shown in red, is obtained by renormalizing the host galaxy template to match the flux of the derived host galaxy component, as described in \Sec\ref{sec:host}. In the panels where the yellow and purple decomposition results are not shown, the normalization of the host template was obtained assuming a standard-candle luminosity (cf. \Fig\ref{fig:disc_and_host}).

When using the spectral decomposition in the optical range to normalize the host spectrum, the normalized template can overshoot the multiwavelength SED at infrared frequencies, which would lead to the total modeled spectrum to contradict the data. This is only the case for two sources, marked with three asterisks in \Fig\ref{fig:app_host}: 3HSP~J062753.3-15195 (second row, first panel) and 3HSP J085410.1+27542 (third row, fourth panel). We then reduced the normalization obtained originally so as to ensure that the template is consistent with the data. In both cases, the correction factor was lower than 2. As we can see in the figure, in these two instances the final host galaxy spectrum is assumed to dominate the infrared SED.

\section{Role of dust emission in hadronic models of masquerading BL Lacs}
\label{app:dust}

As explained in \Sec\ref{sec:model_BLR}, the infrared emission from the dust torus in masquerading BL Lacs was not accounted for when calculating the particle interactions in the jet. This was based on two considerations: \textit{1)} although the total luminosity of the dust emission may be comparable, or even superior, to that of broad lines, the large radius of the dust torus compared to the BLR implies a considerably lower photon energy density of the former compared to the latter; \textit{2)} the inner radius of the torus has been shown to display a large spread, and it is challenging to confidently establish a scaling relation with the disk luminosity~\citep[e.g.,][]{2011A&A...527A.121K,Burtscher:2013aza}. Given that the infrared photon energy density scales with $R_\mathrm{dust}^{-2}$, this introduces large uncertainties on the energy density of the dust radiation in the rest frame of the jet, and therefore on the role of this radiation as a target for particle interactions. 

In spite of the above considerations, dust emission may play a significant role in the leptohadronic model under certain conditions. Specifically, we show in this appendix that for a small fraction of our sources, the presence of bright infrared dust photons may lead to enhanced megaelectronvolt $\gamma$-ray emission, up to a factor of 1.8 compared to the baseline scenario, by boosting photo-pair production by relativistic protons.

Neglecting effects related to the thickness of the dust torus, which is generally unknown, the density of infrared photons in the rest frame of the jet is given by\footnote{This expression differs slightly from that suggested by \citet{Ghisellini:2009wa}, where the density remains constant for $0<R_\mathrm{diss}<R_\mathrm{dust}$. While we adopt that expression for the BLR photons, as explained in the main text, here we adapt it to represent a toroidal dust geometry and a jet that moves perpendicularly to its plane.}
\begin{equation}
u^\prime_\mathrm{IR}=\frac{f_\mathrm{cov}^\mathrm{dust}\,\delta_\mathrm{dust}^{\mathrm{rel}2}\,L_\mathrm{disk}}{4\pi c\left(R_\mathrm{dust}^2+R_\mathrm{diss}^2\right)},
\label{eq:dust}
\end{equation}
where $\delta_\mathrm{dust}^\mathrm{rel}$ is the relative Doppler factor between the direction of the dust radiation and the direction of motion of the jet. The expression for the distance that appears in the denominator, $d^2=R_\mathrm{dust}^2+R_\mathrm{diss}^2$, comes from the assumption that the jet is perpendicular to the plane of the torus.

For all our best-fit results (cf. \Tab\ref{tab:parameters}), we have $R_\mathrm{BLR}\leq R_\mathrm{diss} \leq3R_\mathrm{BLR}\ll R_\mathrm{dust}$, which means the dissipation region lies relatively close to the plane of the torus. This implies an angle between the torus radiation and the axis of motion of the jet of approximately $\theta=90\deg$ in the rest frame of the black hole. In the rest frame of the jet, this radiation is relativistically beamed, appearing at an angle of $\theta^\prime\approx1/\Gamma_\mathrm{b}$. In other words, although in the black hole frame the dust torus is surrounding our dissipation region in the jet, in the rest frame of the jet the dust appears in a more frontal position. The respective Doppler factor corresponding to this angle is $\delta^\mathrm{rel}_\mathrm{dust}\approx\Gamma_\mathrm{b}$. Our original \eq(\ref{eq:dust}) therefore simplifies to
\begin{equation}
u^\prime_\mathrm{IR}\approx\frac{f_\mathrm{cov}^\mathrm{dust}\,\Gamma_\mathrm{b}^2\,L_\mathrm{disk}}{4\pi cR_\mathrm{dust}^2}~~\left(R_\mathrm{diss}\ll R_\mathrm{dust}\right).
\label{eq:dust_simplified}
\end{equation}
We can now evaluate the impact of the dust emission on the jet model for each source. We assume the jet Lorentz factor $\Gamma_\mathrm{b}$ is given by the best-fit value in the leptohadronic model (\Tab\ref{tab:parameters}),  we fix the dust covering factor to $f_\mathrm{cov}=0.3$, and we consider a disk luminosity $L_\mathrm{disk}$ following \Sec\ref{sec:disk}. As an estimate of the dust torus radius, we utilize the results by \citet{Burtscher:2013aza}, based on mid-infrared observations of an AGN sample. In spite of a large statistical spread, the authors found that the $12\,\mu\mathrm{m}$ dust radius generally obeyed $R_\mathrm{dust}\approx R_\mathrm{B}/3$, where $R_\mathrm{B}=1.3\,(L_\mathrm{UV}/10^{46}\,\mathrm{erg}\,\mathrm{s}^{-1})^{0.5}(T/1500\,\mathrm{K})^{-2.8}\,\mathrm{pc}$ is the theoretical relation informed by the dust sublimation temperature~\citep{1987ApJ...320..537B}. Following~\citet{Burtscher:2013aza}, we assume a bolometric correction $L_\mathrm{disk}=1.5\,L_\mathrm{UV}$ \citep{Runnoe:2012rg,Elvis:1994us}, leading to the overall relation $R_\mathrm{dust}=15\,(L_\mathrm{disk}/10^{45}\,\mathrm{erg}\,\mathrm{s}^{-1})^{0.5}(T/300\,\mathrm{K})^{-2.8}\,\mathrm{pc}$.

Regarding the spectral shape of the infrared emission, we adopt the template spectrum by~\cite{SDSS:2001ros}. The exact shape of this spectrum is shown in the observer's frame in \Fig\ref{fig:optical_fit_example}, in the form of a low-frequency bump around $10^{13}~\mathrm{Hz}$ in the curve labeled ``Disk, BLR, dust.'' We then use \eq(\ref{eq:dust_simplified}) to normalize the dust spectrum in the rest frame of the jet, and include it in addition to the usual BLR spectrum, whose treatment is described in \Sec\ref{sec:model_BLR}.

\begin{figure}
    \centering
    \includegraphics[width=0.5\textwidth, trim={0 0 0 0}, clip]{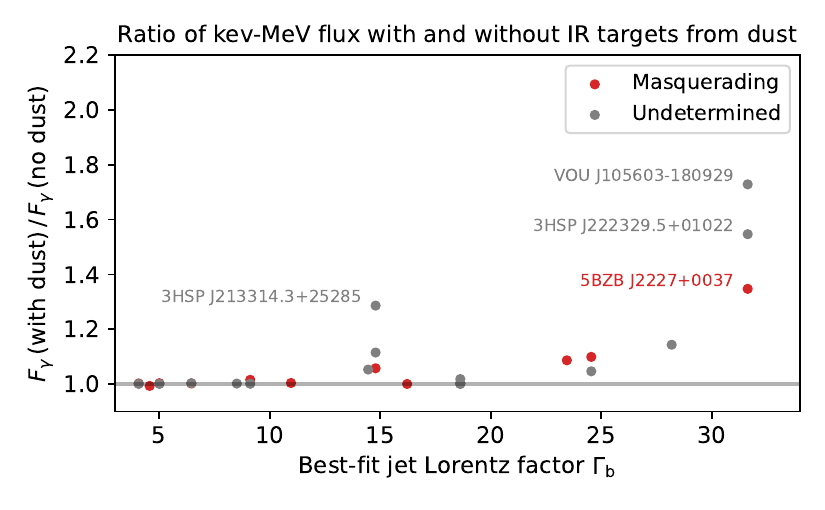}
    \caption{Ratio between the integrated observed flux in the kiloelectronvolt-megaelectronvolt range predicted by the model when accounting for, and neglecting, the presence of infrared dust photons surrounding the jet. The four sources labeled are those for which his effect leads to an enhancement of the flux higher than 1.2. For the remaining sources, the role of this emission in the model is minimal.}
    \label{fig:app_torus_fluxes}
\end{figure}

We have run the source simulations including the dust contribution, calculated as described above, and compared them with our baseline results, where it was not taken into account. We conclude that this effect can, under certain conditions, lead to increased nonthermal emission in the kilo- to megaelectronvolt range due to an enhanced proton interactions. This is quantified in \Fig\ref{fig:app_torus_fluxes} for all sources in the sample excluding true BL Lacs, which are not expected to contain a radiating dust torus. As we can see, for four of the sources, the presence of torus photons causes a non-negligible enhancement of the jet emission in the sub-megaelectronvolt range, with enhancement factors lying between 1.2 and 1.8. These sources are named explicitly in \Fig\ref{fig:app_torus_fluxes}. Three of these four sources have best-fit Lorentz bulk factors above 30, which is an exceptionally high value that leads to a strong boost of the infrared photons from the torus into the jet frame, and explains why the torus radiation plays some role in the leptohadronic model in these cases. For the remaining 18 sources, the net enhancement factor of the megaelectronvolt $\gamma$-ray flux is lower than 1.2.

\begin{figure*}
    \centering
    \includegraphics[width=0.5\textwidth, trim={0 5mm 0 0}, clip]{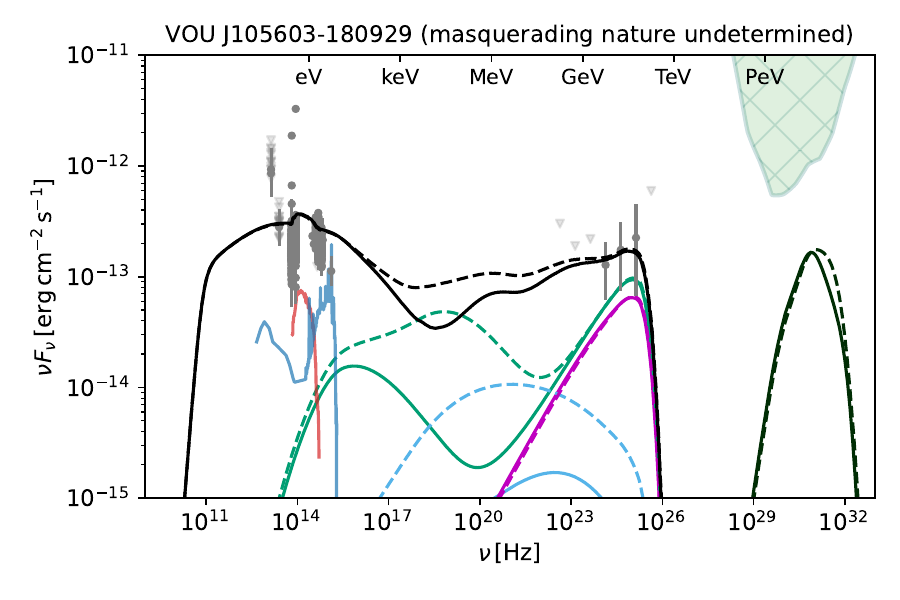}\includegraphics[width=0.5\textwidth, trim={0 5mm 0 0}, clip]{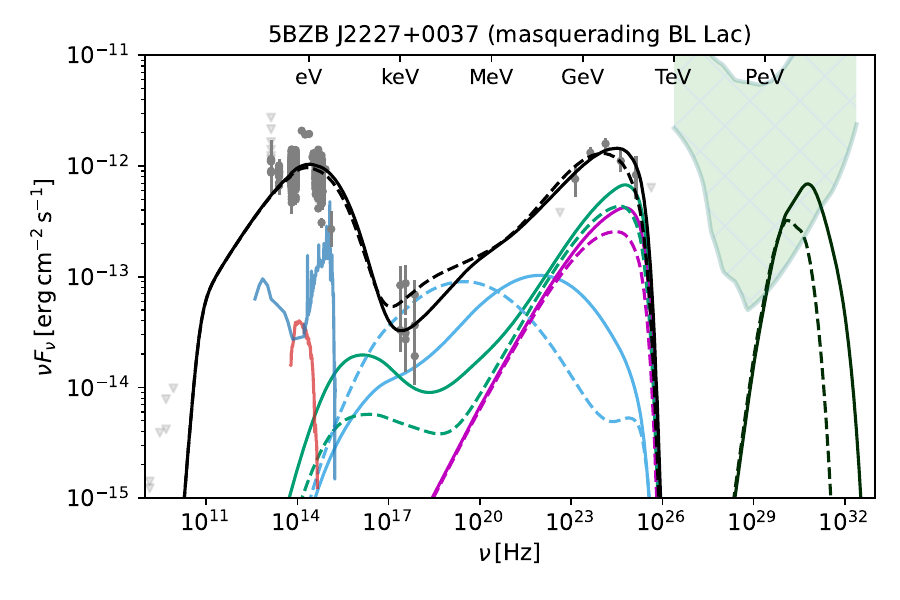}

    \includegraphics[width=0.5\textwidth, trim={0 5mm 0 0}, clip]{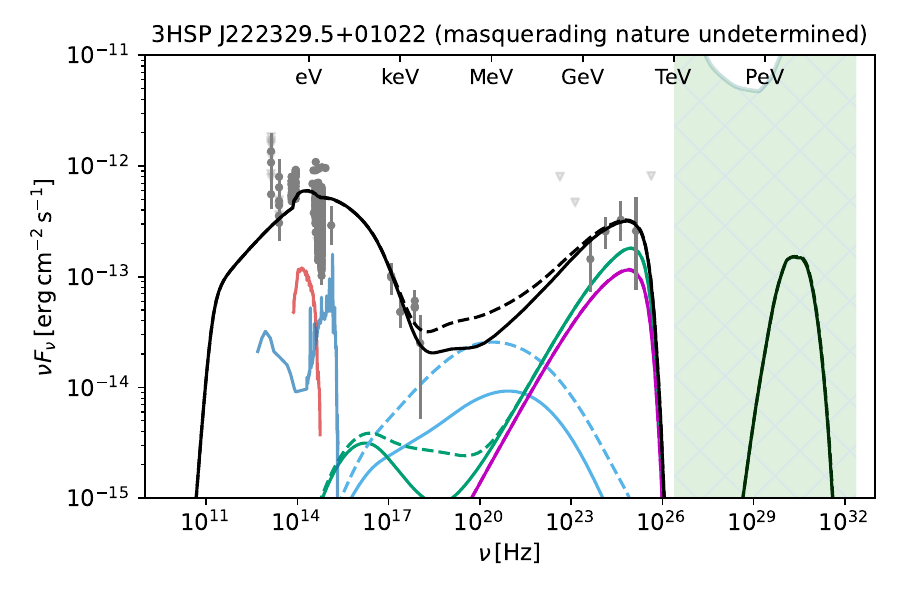}\includegraphics[width=0.5\textwidth, trim={0 5mm 0 0}, clip]{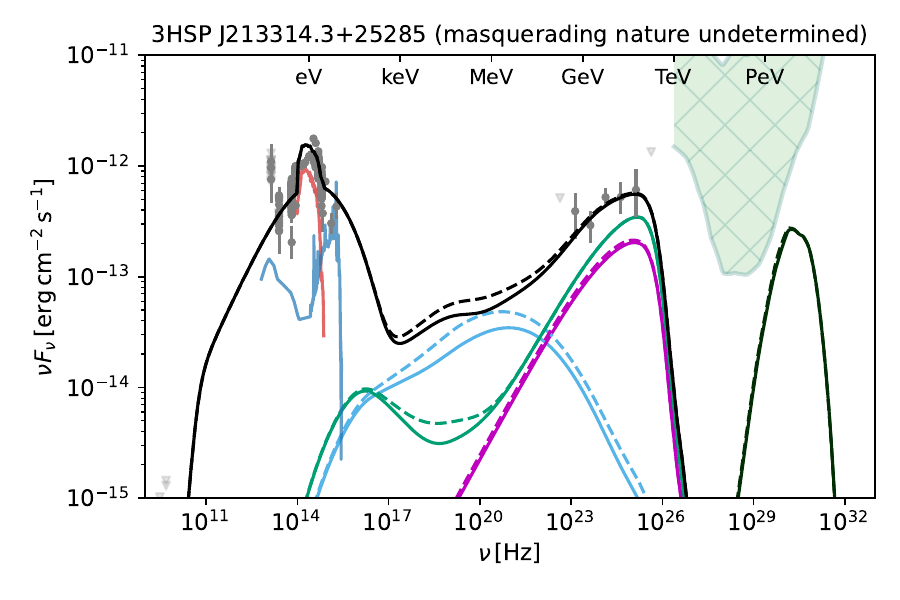}
    
    \includegraphics[width=0.67\textwidth, trim={0 0 0 0}, clip]{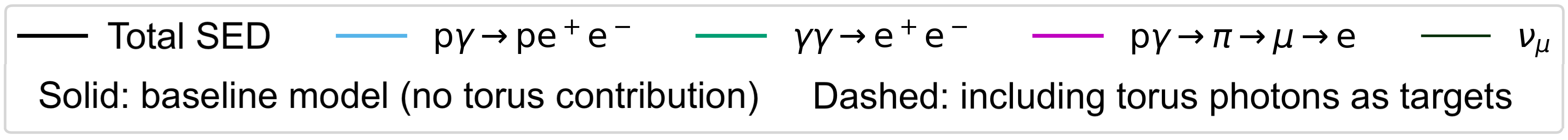}
    \caption{Effect of thermal infrared emission from a dust torus boosted into the jet frame. The solid curves show the fluxes from our our baseline result (\Tab\ref{tab:parameters}), without accounting for thermal dust emission. The dashed curves show the result of the model when accounting for the emission from a dust torus that is boosted into the jet. The net effect is an increase in photo-pair production and cascade emission, leading to an increased flux prediction in megaelectronvolt $\gamma$ rays. For all other sources in the sample, this effect does not change the predicted SED, as shown in \Fig\ref{fig:app_torus_fluxes}.}
    \label{fig:app_torus_sed}
\end{figure*}

To clarify the origin of this effect, we show in \Fig\ref{fig:app_torus_sed} the multiwavelength emission resulting from the jet modeling when ignoring the torus emission (solid curves, same as our baseline results), and when accounting for the torus emission (dashed curves) for the four sources labeled in \Fig\ref{fig:app_torus_fluxes}. As discussed above, these are the only four sources for which the dust emission affects the best-fit result.

Starting with the result shown in the upper left panel, we see that the dust photons boosted into the jet frame act as additional targets for photo-pair production by protons, leading to enhanced synchrotron emission from photo-pairs (compare dashed and solid cyan curves). Furthermore, considering that this source has a best-fit value of $\Gamma_\mathrm{b}=31.6$, the infrared dust emission appears in the jet frame as ultraviolet radiation, with a frequency $\nu^\prime=4(T/300\,\mathrm{K})(\Gamma_\mathrm{b}/31.6)\,\mathrm{eV}$. These photons can interact at threshold with 250~GeV photons from hadronic interactions, creating pairs that subsequently emit synchrotron radiation, leading to the enhanced flux that can be seen in the green dashed curve. This enhancement in photo-pair production and cascade emission results in a higher overall photon flux in the kilo- and megaelectronvolt range compared to the baseline model, as we can see by comparing the dashed and solid black curves.

In the case displayed in the upper right panel, we can see that the enhanced photo-pair production also leads to significant additional cooling of the steady-state proton distribution. This explains the slightly enhanced keV flux, which is due to synchrotron emission from Bethe-Heitler pairs, and to a steady-state peak neutrino flux that is a factor of $\sim2$ lower compared to the baseline model.

For the two sources shown in the lower panels, the only consequence of the inclusion of dust emission is an enhancement in the megaelectronvolt $\gamma$-ray flux by a factor of $\sim1.3$, as already shown in \Fig\ref{fig:app_torus_fluxes}, which is due to a slight increase in efficiency of photo-pair production.

Having discussed the possible effects of infrared dust emission, we should emphasize that its actual contribution to leptohadronic interactions is highly dependent on parameters that were considered fixed throughout this discussion but which, in reality, are not well constrained. For example, the disk luminosity is not well constrained for any of these four sources (cf.~\Fig\ref{fig:disc_and_host}, left panel). If the disk luminosity were considerably lower than we assume, that could drastically reduce the effect discussed here, in which case our baseline model would be the more accurate description of the system. 

On a final note of comparison, \citet{Blazejowski:2000ck} have described quasar data with a leptonic approach and shown that infrared photons from hot dust play a pivotal role as targets for external Compton scattering. In contrast, in our leptohadronic scenario $\gamma$-ray emission has a strong contribution from hadronic cascades and the model therefore relies less on efficient external Compton scattering. As we have shown in this section, infrared dust emission in IHBLs should play a significant role in hadronic interactions only for limited regions of parameter space, notably for high jet bulk Lorentz factors $\Gamma_\mathrm{b}\gtrsim30$.

\section{Best-fit model parameters}
\label{app:parameters}

Finally, we detail the best-fit leptohadronic parameter values for the G20 sample. The parameters were searched in the ranges listed in \Tab\ref{tab:steps} and the resulting best-fit multimesssenger predictions are those shown in \Figs\ref{fig:all_masquerading}, \ref{fig:all_nonmasquerading}, and \ref{fig:all_gray_area}.

In \Tab\ref{tab:parameters} we show the best-fit values of the ten leptohadronic parameters for each source, ordered by the predicted energy-integrated muon neutrino flux, $F_{\nu_\mu}$. We also report the values of the peak energy of the neutrino spectrum, which lie between 2.1~PeV and 172.9~PeV. As discussed in \Sec\ref{sec:methods}, the minimum electron and proton Lorentz factors are fixed to 100 in all cases.

\begin{figure*}[htpb!]
    \centering
    \includegraphics[width=\textwidth, trim={0 0 0 0}, clip]{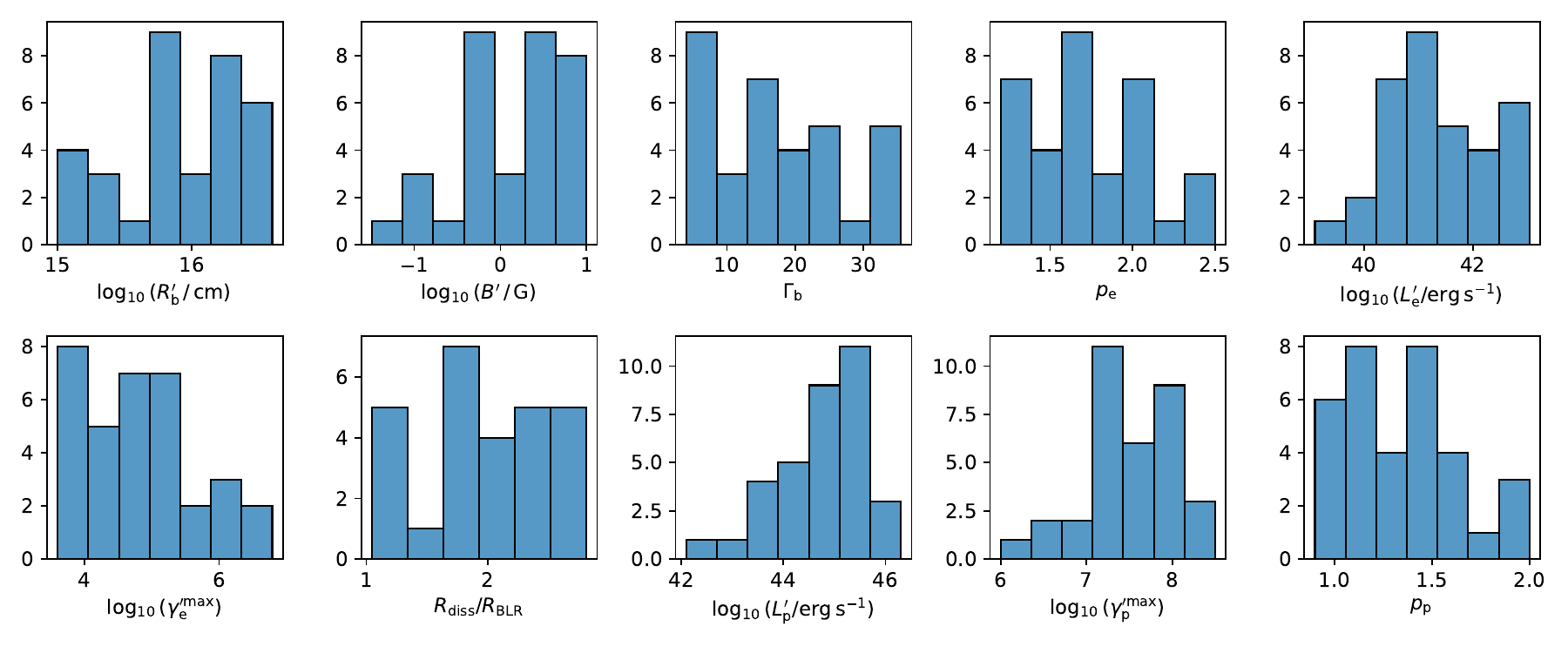}
    \caption{Distribution of each of the ten variable parameters of the leptohadronic IHBL model, as listed in \Tab\ref{tab:parameters}. We include all sources in the G20 sample, except in the case of the dissipation radius, which does not apply to true BL Lacs, leading us to exclude that subset.}
    \label{fig:parameter_histogram}
\end{figure*}

As a visual guide to the parameter distribution in this model, in \Fig\ref{fig:parameter_histogram} we plot a histogram of the best-fit values for each of the ten variable parameters of the model listed in \Tab\ref{tab:parameters}.

\begin{table*}[htpb!]
\setlength{\tabcolsep}{1.7pt}
\caption{Predicted muon neutrino flux, neutrino peak energy, and best-fit parameter values of the leptohadronic model for the blazars in the sample.}
\begin{center}
\small
\begin{tabular}{lrrrrrrrrrrrrr}
\hline\hline
Associated source & $\log_{10}(F_{\mu_\nu})$ & $E_\nu/\mathrm{PeV}$ & $\log_{10}(R_\mathrm{b}^\prime)$ & $B^\prime$~ & $\Gamma_\mathrm{b}$~  & $R_\mathrm{diss}/R_\mathrm{BLR}$ & $\log_{10}(\gamma_\mathrm{e}^{\prime\mathrm{max}})$ & $\log_{10}(\gamma_\mathrm{p}^{
\prime\mathrm{max}})$ & $p_\mathrm{e}$ & $p_\mathrm{p}$ & $\log_{10}(L^\prime_\mathrm{e})$ & $\log_{10}(L^\prime_\mathrm{p})$  &  $\chi_\mathrm{r}^{2^\ddagger}$ \\
\hline
TXS 0506+056$^\ast$ & -10.8 & 120.1 & 16.4 & 1.6 & 14.8 &  2.0 & 4.0 & 8.1  & 1.3 & 1.0 & 41.9  & 44.6  & 99.4 \\
3HSP J111706.2+20140$^\ast$ & -11.3 & 5.3 & 16.4 & 3.2 & 4.1 &  1.6 & 4.9 & 7.2  & 1.3 & 1.4 & 42.6  & 45.9  & 65.6 \\
3HSP J203057.1+19361$^\ast$ & -11.5 & 71.7 & 15.8 & 0.8 & 24.5 &  1.6 & 4.4 & 8.1  & 1.6 & 0.9 & 40.0  & 43.3  & 38.4 \\
VOU J150720-370902$^\ast$ & -11.6 & 9.9 & 16.2 & 8.1 & 4.6 &  1.0 & 4.0 & 7.9  & 1.6 & 1.4 & 42.3  & 44.6  & 110.5 \\
3HSP J064933.6-31392 & -11.6 & 8.8 & 16.3 & 2.3 & 18.6 &  1.9 & 5.2 & 7.3  & 1.3 & 1.5 & 41.2  & 44.8  & 111.8 \\
5BZB J2227+0037$^\ast$ & -11.7 & 44.7 & 16.5 & 0.2 & 31.6 &  2.7 & 4.1 & 7.9  & 2.1 & 1.7 & 41.3  & 45.0  & 92.8 \\
5BZB J1314+2348$^\ast$ & -11.7 & 37.8 & 15.8 & 1.9 & 23.4 &  1.8 & 3.7 & 8.2  & 1.7 & 1.7 & 40.9  & 44.0  & 44.0 \\
3HSP J034424.9+34301 & -11.7 & 47.2 & 16.0 & 1.3 & 14.8 &  1.8 & 4.1 & 7.9  & 1.6 & 1.0 & 40.0  & 44.0  & 244.6 \\
3HSP J153311.2+18542 & -11.8 & 6.2 & 15.8 & 3.1 & 9.1 &  1.7 & 5.4 & 6.9  & 1.4 & 1.1 & 41.5  & 45.2  & 33.6 \\
3HSP J155424.1+20112 & -11.9 & 4.4 & 16.1 & 6.3 & 4.1 &  1.6 & 5.3 & 7.2  & 1.2 & 1.3 & 42.5  & 45.6  & 38.3 \\
3HSP J094620.2+01045$^\dagger$ & -12.0 & 3.7 & 15.2 & 10.0 & 14.5 &  -- & 5.2 & 6.7  & 1.4 & 1.4 & 41.2  & 45.4  & 60.5 \\
3HSP J095507.9+35510$^\dagger$ & -12.0 & 46.6 & 16.1 & 0.2 & 24.5 &  -- & 6.7 & 7.5  & 1.9 & 1.0 & 40.8  & 45.2  & 54.1 \\
3HSP J023248.5+20171$^\dagger$ & -12.0 & 110.9 & 15.7 & 0.1 & 31.6 &  -- & 6.5 & 7.8  & 2.2 & 1.6 & 40.3  & 44.4  & 57.6 \\
3HSP J100326.6+02045 & -12.1 & 9.4 & 16.3 & 3.2 & 28.2 &  2.1 & 4.0 & 7.2  & 1.5 & 2.0 & 39.4  & 44.1  & 91.6 \\
3HSP J213314.3+25285 & -12.1 & 12.4 & 16.1 & 0.6 & 14.8 &  2.3 & 4.0 & 7.2  & 1.8 & 1.6 & 40.7  & 45.1  & 57.9 \\
3HSP J152835.7+20042$^\ast$ & -12.1 & 11.9 & 15.2 & 3.2 & 16.2 &  1.6 & 5.8 & 7.0  & 1.6 & 1.0 & 40.5  & 44.6  & 18.1 \\
3HSP J062753.3-15195$^\dagger$ & -12.1 & 3.9 & 15.7 & 4.7 & 5.1 &  -- & 5.2 & 7.1  & 1.4 & 1.5 & 42.6  & 45.2  & 141.8 \\
CRATESJ180812+350104$^\ast$ & -12.2 & 23.3 & 16.3 & 1.4 & 9.1 &  1.7 & 4.1 & 7.8  & 1.4 & 1.6 & 41.7  & 44.3  & 24.0 \\
VOU J105603-180929 & -12.3 & 41.5 & 16.6 & 1.0 & 31.6 &  2.1 & 4.1 & 7.7  & 1.7 & 1.1 & 39.6  & 42.9  & 64.3 \\
3HSP J085410.1+27542$^\dagger$ & -12.3 & 172.9 & 15.6 & 0.0 & 35.5 &  -- & 4.8 & 7.9  & 1.3 & 1.2 & 40.5  & 45.1  & 57.2 \\
5BZB J1322+3216$^\ast$ & -12.5 & 25.9 & 15.7 & 4.9 & 6.5 &  1.6 & 4.5 & 7.9  & 1.6 & 1.3 & 42.6  & 44.6  & 29.9 \\
3HSP J203031.6+22343 & -12.5 & 17.3 & 15.9 & 0.4 & 14.5 &  2.8 & 5.0 & 7.3  & 1.9 & 1.2 & 41.1  & 45.2  & 116.4 \\
3HSP J222329.5+01022 & -12.5 & 11.1 & 15.9 & 0.8 & 31.6 &  2.5 & 4.7 & 6.9  & 2.4 & 1.2 & 40.4  & 44.2  & 208.3 \\
VOU J022411+161500 & -12.5 & 5.7 & 16.3 & 0.5 & 6.5 &  1.7 & 3.8 & 7.3  & 1.2 & 1.3 & 42.3  & 45.5  & 48.9 \\
CRATESJ232625+011147$^\ast$ & -12.5 & 45.9 & 16.2 & 1.1 & 11.0 &  2.4 & 4.6 & 8.1  & 1.9 & 1.4 & 42.6  & 45.1  & 76.5 \\
5BZU J0158+0101$^\ast$ & -12.6 & 15.2 & 16.6 & 0.8 & 5.0 &  1.4 & 3.9 & 8.1  & 2.0 & 1.6 & 42.8  & 44.5  & 59.5 \\
CRATESJ024445+132002 & -12.6 & 40.4 & 16.3 & 4.3 & 24.5 &  1.4 & 6.3 & 8.5  & 2.4 & 1.5 & 40.8  & 42.2  & 61.7 \\
3HSP J023927.2+13273 & -12.7 & 16.1 & 15.1 & 3.4 & 18.6 &  1.3 & 3.9 & 8.3  & 1.6 & 2.0 & 40.6  & 43.9  & 45.0 \\
3HSP J180849.7+35204$^\dagger$ & -12.7 & 57.6 & 15.0 & 0.7 & 24.5 &  -- & 4.2 & 7.3  & 2.4 & 1.1 & 39.9  & 44.6  & 126.2 \\
3HSP J112405.3+20455 & -12.8 & 2.9 & 15.4 & 7.1 & 18.6 &  1.3 & 4.2 & 6.4  & 1.6 & 1.5 & 40.7  & 43.6  & 191.0 \\
3HSP J112503.6+21430 & -12.8 & 6.4 & 16.3 & 0.9 & 8.5 &  1.9 & 5.7 & 7.5  & 2.0 & 2.0 & 42.1  & 45.8  & 66.8 \\
3HSP J125848.0-04474$^\dagger$ & -12.8 & 18.1 & 16.3 & 0.2 & 15.8 &  -- & 6.0 & 7.2  & 2.0 & 1.2 & 41.4  & 44.9  & 43.5 \\
3HSP J130008.5+17553 & -12.8 & 20.6 & 15.4 & 5.8 & 5.0 &  1.4 & 4.3 & 7.9  & 1.6 & 1.4 & 42.7  & 44.4  & 44.3 \\
3HSP J125821.5+21235 & -13.6 & 2.1 & 15.4 & 4.7 & 18.6 &  1.2 & 4.7 & 6.0  & 1.3 & 1.0 & 40.1  & 42.6  & 84.5 \\
\hline
Sample mean & -12.2 & 16.2 & 15.9 & 2.6 & 16.8 & 1.8 & 4.7 & 7.5 & 1.7 & 1.4 & 41.2  & 44.5 & - \\
Sample spread (max-min) & 2.7 & 1.9 & 1.6 & 10.0 & 31.4 & 1.7 & 3.0 & 2.5 & 1.2 & 1.1 & 3.4  & 3.7 & - \\
Average model uncertainty & \phantom{-1}0.4 & \phantom{0}<0.1 & 0.2 & 0.2 & 0.3 & 0.1 & \phantom{0}0.3 & <0.1 & <0.1 & \phantom{0}0.1 & 0.1 & 0.3 & - \\
\hline\hline
\multicolumn{14}{l}{Note: All variables are given in CGS units, except the predicted neutrino peak energy, given in petaelectronvolt.}\\
\multicolumn{14}{l}{$^\ddagger$ Reduced chi-squared value, assuming a number of degrees of freedom of $N-10$, where $N$ is the number of data points in the multiwavelength}\\
\multicolumn{14}{l}{SED and 10 is the number of model parameters. This value does not directly represent the goodness of fit of the model because it includes the}\\
\multicolumn{14}{l}{effect of the intrinsic variability of the data. As noted in \Sec\ref{sec:binning}, we do not directly use the chi-squared value to optimize the model, but the}\\
\multicolumn{14}{l}{derived quantity $\chi^2_\mathrm{log}$, defined in \eq(\ref{eq:logchi}).}\\
\multicolumn{14}{l}{$^\ast$ Source previously identified as a masquerading BL Lac object (cf. \Tab\ref{tab:sample}).}\\
\multicolumn{14}{l}{$^\dagger$ Source previously identified as a true BL Lac object.}\\
\end{tabular}
\end{center}

\label{tab:parameters}
\end{table*}

\clearpage

\end{appendix}

\end{document}